\pdfoutput=1 
\documentclass[smallextended,numbook]{svjour3}

\journalname{Journal of Statistical Physics}

\usepackage{graphicx,epsfig,color}

\usepackage{mathrsfs} 
\usepackage{amsmath}
\usepackage{amssymb}
\usepackage{amsbsy}

\usepackage{mathabx} 

\usepackage[utf8]{inputenc}  
\usepackage[english,francais]{babel}

\usepackage{url,hyperref}  

\usepackage{enumitem}

  \definecolor{blue}{rgb}{0,0,1}
  \definecolor{green}{rgb}{0,.6,0}
  \definecolor{red}{rgb}{1,0,0}
  \definecolor{vio}{rgb}{1,0,1}
  \definecolor{uv}{rgb}{0.5,0,0.5}
  \definecolor{ama}{rgb}{0.3,0.3,0.3}

\definecolor{M_Beige}         {rgb}{0.96 , 0.96 , 0.86}

\definecolor{M_Brown}         {rgb}{0.65 , 0.16 , 0.16}

\definecolor{M_Gold}          {rgb}{1.00 , 0.84 , 0.00}

\definecolor{M_LemonChiffon}  {rgb}{1.00 , 0.98 , 0.80}

\definecolor{M_Orange}        {rgb}{1.00 , 0.60 , 0.00}

\definecolor{M_Pink}          {rgb}{0.80 , 0.55 , 0.60}

\definecolor{M_Violet}          {rgb}{0.83 , 0.21 , 0.93}

\definecolor{M_Green}          {rgb}{0.2 , 0.6 , 0.2}

\definecolor{M_Gray}          {rgb}{0.7 , 0.7 , 0.7}

\definecolor{M_BluPal}          {rgb}{0.7 , 0.7 , 0.9}

\def\Xint#1{\mathchoice
{\XXint\displaystyle\textstyle{#1}}%
{\XXint\textstyle\scriptstyle{#1}}%
{\XXint\scriptstyle\scriptscriptstyle{#1}}%
{\XXint\scriptscriptstyle\scriptscriptstyle{#1}}%
\!\int}
\def\XXint#1#2#3{{\setbox0=\hbox{$#1{#2#3}{\int}$}
\vcenter{\hbox{$#2#3$}}\kern-.5\wd0}}

\def\dashint{\Xint-}

\renewcommand{\leq}{\leqslant}
\renewcommand{\geq}{\geqslant}

\newcommand{\EXP}[1]{\mathrm{e}^{#1}}         

\def\eqdef{\stackrel{\mbox{\tiny def}}{=}}     
\newcommand{\ket}[1]{|\kern.3ex#1\kern.3ex\rangle}
\newcommand{\bra}[1]{\langle\kern.3ex #1 \kern.3ex|}

\newcommand{\braket}[2]{\langle\kern.3ex #1 \kern.3ex|\kern.3ex #2 \kern.3ex \rangle}
\newcommand{\braketYves}[2]{\left(\kern.3ex #1 \kern.3ex|\kern.3ex #2 \kern.3ex \right)}

\newcommand{\mean}[1]{\left\langle #1 \right\rangle} 
\newcommand{\smean}[1]{\langle #1 \rangle} 

\newcommand{\cotan}{\mathop{\mathrm{cotan}}\nolimits}

\newcommand{\re}{\mathop{\mathrm{Re}}\nolimits}      
\newcommand{\im}{\mathop{\mathrm{Im}}\nolimits}      
\newcommand{\sign}{\mathop{\mathrm{sign}}\nolimits}  
\renewcommand{\min}[2]{\mathop{\mathrm{min}}\nolimits\left( #1 , #2\right)}  
\renewcommand{\max}[2]{\mathop{\mathrm{max}}\nolimits\left( #1 , #2\right)}

\def\I{{\rm i}}                  
\def\D{{\rm d}}                  

\newcommand{\deriv}[2]{\frac{\mathrm{d}#1}{\mathrm{d}#2}}
\newcommand{\derivp}[2]{\frac{\partial #1}{\partial #2}}

\newcommand\antiddots{\mathinner{\mkern2mu\raise1pt\hbox{.}\mkern2mu
    \newline \raise4pt\hbox{.}\mkern2mu\raise7pt\hbox{.}\mkern1mu}}

\def\levy{\mathcal{L}}
\def\mass{m}

\def\covma{\boldsymbol{\sigma}}
\def\btheta{\overline{\theta}}
\def\bu{\overline{u}}
\def\bv{\overline{v}}
\def\bw{\overline{w}}

\def\Duu{D_{uu}}
\def\Dtt{D_{\theta\theta}}

\renewcommand{\vec}[1]{\boldsymbol{#1}}

\def\tvmu{ \vec{\tilde{\mu}} }
\def\tcovma{ \boldsymbol{\tilde{\sigma}} }

\def\IDoS{\mathcal{N}}

\def\Casimir{\mathscr{C}}

\begin{document}

\renewcommand{\labelitemi}{$\bullet$}
\renewcommand{\labelitemii}{$\star$}

\selectlanguage{english}

\title{Fluctuations of the product of random matrices and generalized Lyapunov exponent}

\author{
Christophe Texier 
}

\institute{
LPTMS, CNRS, Univ.~Paris-Sud, Universit\'e Paris-Saclay, F-91405 Orsay, France
\\
\email{christophe.texier@u-psud.fr}
}

\date{August 31, 2020}

\maketitle

\begin{abstract}

I present a general framework allowing to carry out explicit calculation of the moment generating function of random matrix products $\Pi_n=M_nM_{n-1}\cdots M_1$, where $M_i$'s are i.i.d..
Following Tutubalin [Theor. Probab. Appl. {\bf 10}, 15 (1965)], the calculation of the generating function is reduced to finding the largest eigenvalue of a certain transfer operator associated with a family of representations of the group.
The formalism is illustrated by considering products of random matrices from the group $\mathrm{SL}(2,\mathbb{R})$ where explicit calculations are possible.
For concreteness, I study in detail transfer matrix products for the one-dimensional Schr\"odinger equation where the random potential is a L\'evy noise (derivative of a L\'evy process). 
In this case, I obtain a general formula for the variance of $\ln||\Pi_n||$
and for the variance of $\ln|\psi(x)|$, where $\psi(x)$ is the wavefunction, in terms of a single integral involving the Fourier transform of the invariant density of the matrix product. 
Finally I discuss the continuum limit of random matrix products (matrices close to the identity ). In particular, I investigate a simple case where the spectral problem providing the generalized Lyapunov exponent can be solved exactly.


\end{abstract}



%
%





\section{Introduction}
\label{sec:Intro}

The study of random matrix products was initiated by Bellman in 1954 \cite{Bel54} and was later developed by Furstenberg and Kesten \cite{FurKes60,Fur63}, Guivarc'h and Raugi \cite{GuiRau85}, Le~Page \cite{LeP83} and others \cite{Oco75} (see the monograph \cite{BouLac85} or the recent one \cite{BenQui16book}).
Among the vast mathematical literature on this topic, one of the key problems is the derivation of sufficient conditions for the central limit theorem to hold. 
In most of these works, the emphasis is not placed on concrete calculations of the moments of the distribution of the random matrix product. 
Exception is Ref.~\cite{Pol10}, where a method for the calculation of the largest Lyapunov exponent is proposed.  
Some analytic results can also be obtained by making further assumptions on the matrices.
In Ref.~\cite{New86}, Newman has derived the spectrum of Lyapunov exponents for real Ginibre random matrices (strictly speaking, this paper has considered a $N$-dimensional stochastic linear dynamical system, corresponding to the continuum limit of random matrix products).
In the Physics literature, transport properties of disordered waveguides have been studied within transfer matrix approach. Some analytical results were obtained by making some isotropy assumption, i.e. studying phenomenological rather than microscopic models \cite{Dor83,Mel86,MelPerKum88} (see the review \cite{Bee97}). 
In this spirit, a classification of the possible Lyapunov spectra for the various symmetry classes of disordered systems can be found in Ref.~\cite{LudSchSto13}. 
In the context of disordered waveguides, fluctuations of the random matrix product is related to the question of the distribution of the transmission probability. 
The few analytical results are based on similar isotropic assumptions (see \cite{GerVas59,AntPasSly81} for the one-dimensional case and the review \cite{Bee97} for the multichannel case).
Other cases of product of random matrices belonging to invariant ensembles were analysed in Refs.~\cite{AkeBurKie14,For15} where both the Lyapunov spectra and related variances were derived (see Ref.~\cite{AkeIps15} for a review).

The main purpose of the present article is to derive explicit formulae for the variance of the logarithm of random matrix product when the matrices are  $2\times2$ and real, drawn from an arbitrary distribution.  
Consider a set of independent and identically distributed (i.i.d.) random matrices $M_1,\cdots,M_n$ belonging to a group $\mathrm{G}$, we denote their product 
\begin{equation}
  \label{eq:RMP}
  \Pi_n=M_nM_{n-1}\cdots M_1
  \:.
\end{equation}
Given a probability measure $\mu(\D M)$ over the group, one important question, which has been largely addressed both in the mathematical and the physical literature, is the determination of the (largest) Lyapunov exponent
\begin{equation}
  \lambda_1 = \lim_{n\to\infty} \frac{\mean{\ln ||\Pi_n\vec{x}_0||}}{n}
  \hspace{0.5cm}\mbox{where } ||\vec{x}_0||=1
  \:,
\end{equation}
controlling the growth rate of the matrix product. 
$\vec{x}_0$ is an initial vector on which the sequence of matrices acts and $||\vec{x}_0||$ its norm.
Strictly speaking, the averaging $\mean{\cdots}$ with respect to $\mu(\D M)$ is not needed as the logarithm is self averaging, however it is useful to keep it for comparison with the generalized Lyapunov exponent (below).
In earlier work, we have discussed several solvable cases involving matrices from the group $\mathrm{SL}(2,\mathbb{R})$ \cite{ComTexTou10,ComTexTou13}. 
Furthermore, when matrices are close to the identity matrix (i.e. in the \og continuum limit \fg{}), a general classification of solutions for the Lyapunov exponent was obtained in Ref.~\cite{ComLucTexTou13}.
The case of larger matrices remains a difficult problem in general, out of the scope of the methods developed in these papers.
Apart from the question of the average growth rate of the matrix product, one can also consider the fluctuations of random matrix products, which can be characterised by considering the generalized Lyapunov exponent (GLE)
\begin{equation}
  \label{eq:DefGLE}
  \widetilde{\Lambda}(q) \eqdef \lim_{n\to\infty} \frac{\ln \mean{||\Pi_n\vec{x}_0||^q}}{n}
  \:.
\end{equation}
It corresponds to the cumulant generating function of $\ln||\Pi_n\vec{x}_0||$. 
We will determine the GLE under the form of an expansion in powers of $q$, written for later convenience as  
\begin{equation}
  \label{eq:GLE-SmallLambda}
  \EXP{-\widetilde{\Lambda}(q)}
  =
  \lambda(q) = \sum_{n=0}^\infty \frac{\lambda_n}{n!} \, (-q)^n
  \hspace{1cm}
  \mbox{with}
  \hspace{0.25cm} 
  \lambda_0=1
  \:.
\end{equation}
It seems that the terminology \og \textit{generalized Lyapunov exponent} \fg{} has been introduced in the review of Paladin and Vulpiani \cite{PalVul87} (these authors refer to the two articles \cite{Fuj83,BenPalParVul85}, however the concept appeared earlier in the mathematical literature \cite{Tut65}, and is used in many works in order to prove generalized central limit theorems; see also \cite{BouLac85}).
Note that the study of \og \textit{finite time Lyapunov exponent} \fg{} corresponds to study the statistical properties of $(1/n)\ln||\Pi_n\vec{x}_0||$, i.e. to the same problem (this terminology is used in Refs.~\cite{SchTit02,GiaKurLecTai11} for example; see also references therein).

Let us mention few motivations from Physics for studying the GLE.
The concept of random matrix is relevant in dynamical systems, turbulence, statistical physics, propagation of waves in random media, etc (see the monographs \cite{Luc92,CriPalVul93}).
In dynamical systems, the study of fluctuations has been addressed by several authors \cite{Fuj83,BenPalParVul85,SchTit02,MalMar02,ZilPik03}, in connection with the multifractal analysis \cite{PalVul87} (multifractalilty was also investigated in the Anderson model in Ref.~\cite{PalVul87b}).
Spin chain models can be studied with transfer matrix method~: this provides a convenient framework to analyse the effect of disorder and lead to consider random matrix products. In this case $\ln||\Pi_n\vec{x}||$ is related to the free energy \cite{DerVanPom78,DerHil83,Luc93,CriPalVul93,FigMosKno98,ZilPik05}, which naturally raises the question of its fluctuations (cf. chapter 4 of \cite{CriPalVul93}).  
Products of $2\times2$ random matrices are also relevant for two-dimensional Ising spins with columnar disorder \cite{McCWu68,GenGiaGre17,ComGiaGre19}. 

The theory of random matrix products provides a suitable framework for quantum localisation in one-dimension \cite{Ish73,BouLac85,Luc92,CriPalVul93,Pen94} (see also the review article \cite{ComTexTou13}), which has raised many different questions.
Although the localisation properties were mostly discussed by studying the Lyapunov exponent, i.e. the mean value of the logarithm of the wave function, the question of the fluctuations was also considered in several works, mainly from the perspective of the reflection \cite{GerVas59,AntPasSly81,LifGrePas88} or transmission \cite{Mel80,Abr81,SteCheFabGog99} statistics.
The scaling approach has raised the question of the number of the relevant parameters (one or two) 
\cite{AndThoAbrFis80,CohRotSha88} (this question was rediscussed more recently \cite{DeyLisAlt00,DeyLisAlt01,SchTit03,TitSch03,DeyEreLis03,RamTex14}).
Analysis of the wave function statistics has allowed to characterize local density of states \cite{AltPri89} or Wigner time delay statistics \cite{TexCom99}.~\footnote{
  For a broader perspective on eigenfunction statistics, beyond 1D, cf. 
  \cite{Mir00,EveMir08}.
  }
As emphasized in Refs.~\cite{Tex00,TexHag10}, the characterization of the fluctuations becomes essential when the Lyapunov exponent vanishes, which can occur for the Dirac equation with random mass~; the question was later studied in \cite{RamTex14}.~\footnote{
  The vanishing of Lyapunov exponents can also occur in multichannel models \cite{BroMudSimAlt98,GraTex16}.
} 
Anomalous localisation (super- or sub-localisation) also requires some detailed study of fluctuations \cite{BooLuc07,BieTex08}.
More recently, the concept of GLE has allowed to unveil a connection between Anderson localisation and the problem of counting equilibria of a directed polymer in a random medium~\cite{FyoLeDRosTex18}.
The relation between random matrix products of $\mathrm{SL}(2,\mathbb{R})$ and the Schr\"odinger equation with generalized point-like scatterers was discussed in Refs.~\cite{ComTexTou10,ComTexTou13}. 
In the present paper, although this will not be the only case considered, some emphasis will be put on matrices of the form
\begin{equation}
  \label{eq:MatricesSchrodinger}
  M_n = 
  \begin{pmatrix}
    \cos\theta_n & -\sin\theta_n \\ 
    \sin\theta_n & \phantom{-}\cos\theta_n
  \end{pmatrix}
  \:
  \begin{pmatrix}
    1 & u_n \\ 0 & 1
  \end{pmatrix}
\end{equation}
and
\begin{equation}
  \label{eq:MatricesSchrodinger2}
  M_n = 
  \begin{pmatrix}
    \cosh\theta_n & \sinh\theta_n \\ 
    \sinh\theta_n & \cosh\theta_n
  \end{pmatrix}
  \:
  \begin{pmatrix}
    1 & u_n \\ 0 & 1
  \end{pmatrix}
  \:,
\end{equation}
known as transfer matrices for the Schr\"odinger equation with $\delta$-potentials
\begin{equation}
  \label{eq:Schrodinger}
  -\psi''(x) + V(x) \, \psi(x) = E\,\psi(x)
  \hspace{0.5cm}
  \mbox{where } 
  V(x)=\sum_n v_n\,\delta(x-x_n)
\end{equation}
The impurity positions are ordered, $x_1<x_2<\cdots<x_n<\cdots$.
The mapping between the quantum model and the problem of matrix product can be understood as follows~:
consider the vector $(\psi'(x)\,,\,k\,\psi(x))$, where $\psi(x)$ solves \eqref{eq:Schrodinger} for $E=+k^2$. 
In the free interval $]x_{n},x_{n+1}[$, the phase of the wave function performs a rotation of angle $\theta_n=k\ell_n>0$, where $\ell_n=x_{n+1}-x_n$. 
The matching of the wave function around the impurity $n$ reads 
$\psi'(x_n^+)-\psi'(x_n^-)=v_n\,\psi(x_n)$.
Gathering the two observations, we conclude that the evolution of the vector $\big(\psi'(x)\,,\,k\psi(x)\big)$ in the interval $[x_{n},x_{n+1}[$ is described by the transfer matrix~\eqref{eq:MatricesSchrodinger} for $\theta_n=k\ell_n$ and $u_n=v_n/k$~\cite{FriLlo60,Luc92,ComTexTou13}.
For negative energy, $E=-k^2$, 
the transfer matrices take the form \eqref{eq:MatricesSchrodinger2}.
Randomness can be introduced in different manners~: either in the weights $v_n$'s or in the positions $x_n$'s, or both \cite{LifGrePas88}.
The first case, considered in Subsection~\ref{subsec:Halperin}, corresponds to the generalized Kronig-Penney model (or "random alloy" model) describing a regular lattice of impurities with random weights \cite{Sch57,Ish73,DeyLisAlt01}. 
When impurity positions are random variables, a natural choice is to consider uncorrelated impurities with a uniform mean density $\rho$, corresponding to i.i.d. distances $\ell_n$ exponentially distributed $\mathrm{Proba}\{\ell_n>\ell\}=\EXP{-\rho\ell}$~: this is the Frisch-Lloyd model (or \og liquid metal \fg{} model) \cite{FriLlo60,BycDyk67,Ish73,Nie83}.
Finally, both sets of variables can be taken at random (\og liquid alloy \fg{} model), which allows to identify some solvable cases \cite{Nie83,ComTexTou10,ComTexTou13}~; this will be assumed in Sections~\ref{sec:Perturbation} and~\ref{sec:GLElocalisation}.

These remarks show that the GLE is also the generating function for the moments of the wave function, i.e. Eq.~\eqref{eq:DefGLE} can be rewritten as
\begin{equation}
  \label{eq:FluctPsiN}
  \widetilde{\Lambda}(q) = \lim_{n\to\infty} \frac{\ln\mean{|\psi(x_n^-)|^q}}{n}
  \:,
\end{equation}
where $x_n$ is the position of the $n$-th impurity.
On the other hand, in the context of localisation theory, it is more natural to define the GLE as 
\begin{equation}
  \label{eq:FluctPsiX}
  \Lambda(q) 
  \eqdef \lim_{x\to\infty} \frac{\ln\mean{|\psi(x)|^q}}{x}
  = \sum_{n=1}^\infty\frac{\gamma_n}{n!}\,q^n
  \:,
\end{equation}
where $\gamma_nx$ is the large $x$ behaviour of the cumulant of order $n$ of $\ln|\psi(x)|$.
When impurities form a regular lattice, $\ell_n=1/\rho$, the two definitions coincide, $\Lambda(q) = \rho\,\widetilde{\Lambda}(q)$,
however, within the Frisch-Lloyd model the fluctuations of the impurity positions make them different, $\Lambda(q) \neq \rho\,\widetilde{\Lambda}(q)$.
Both GLE will be studied here.

The analytical determination of the GLE is an extremely difficult problem in general.
Even numerically, the determination amounts to study the large deviations of random matrix products which grows exponentially. Thus the moments $\smean{||\Pi_N||^q}$ are dominated by rare configurations which are difficult to simulate and makes the numerical calculation rather tedious \cite{Van10,ChaLiPfiYai17,FyoLeDRosTex18} (this question was also addressed for dynamical systems in \cite{GiaKurLecTai11}); note however that the problem is much simpler when $q$ is an even integer with the use of the replica trick \cite{Pen82,BouGeoLeD86,BouGeoHanLeDMai86,Pen94,WeiMon96,FyoLeDRosTex18}.
The present article aims to elaborate on several ideas introduced by Tutubalin \cite{Tut65}~:
making use of the theory of group representation, Tutubalin has reformulated the determination of the GLE in terms of a spectral problem defined from an \textit{averaged representation}.
Here, 
we apply these general ideas to the case of the group $\mathrm{SL}(2,\mathbb{R})$. 
Previous analysis of the variance were limited to specific cases in \cite{SchTit02,RamTex14,ComGiaGre19}, corresponding all to the continuum limit of random matrix products. 
In contrast, we develop here a general method which can be applied to an arbitrary product of random matrices belonging to $\mathrm{SL}(2,\mathbb{R})$. 
As a result we obtain some explicit expressions for the variance $\widetilde{\Lambda}''(0)$ or $\Lambda''(0)$ in several cases.

In order to understand the origin of the spectral problem, let us underline at this step some analogy with the study of functionals of continuous stochastic processes, for which some general results are available in the mathematical literature~\cite{Bha82} (few general references from Physics literature are \cite{MajBra02,Maj05,ComDesTex05}).
Consider a \textit{stationary} stochastic process $z(x)$, with infinitesimal generator $\mathscr{G}_z$.
The study of the statistical properties of the functional $A[z(t)]=\int_0^x\D t\,\phi(z(t))$ can be achieved by introducing the moment generating function 
\begin{equation}
  \mathcal{Q}_x(z_0;q)
  =
  \big\langle
    \exp\big\{ q \int_0^x\D t\, \phi(z(t)) \big\} \, \big| \, z(0)=z_0
  \big\rangle
  \:,
\end{equation}
which obeys the backward equation
$\partial_x\mathcal{Q}_x(z;q)= \big[\mathscr{G}_{z}+q\,\phi(z)\big]\mathcal{Q}_x(z;q)$. 
Stationarity and short range correlations imply
$\mathcal{Q}_x(z_0;q)\sim\exp\big\{x\Lambda(q)\big\}$ for $x\to\infty$, where 
$\Lambda(q)$ is the largest eigenvalue of the operator $\mathscr{G}_{z}+q\,\phi(z)$.
$\Lambda(q)$ is the generating function for the cumulants of the functional, $\Lambda(q)=\gamma_1\,q+\gamma_2q^2/2+\mathcal{O}(q^3)$.
We can solve the eigenvalue problem $\big[\mathscr{G}_{z}+q\,\phi(z)\big]\Phi^\mathrm{L}_q(z)=\Lambda(q)\,\Phi^\mathrm{L}_q(z)$  by a perturbative method in $q$.
Expanding the eigenfunction as $\Phi^\mathrm{L}_q(z)=1+q\,L_1(z)+q^2\,L_2(z)+\cdots$, 
we obtain 
\begin{equation}
\gamma_1=\int\D z\,\phi(z)\,f(z)
\end{equation}
 at order $q^1$, 
where $f(z)$ is the stationary distribution, solving $\mathscr{G}_{z}^\dagger f(z)=0$. 
At order $q^2$, we get the following formula for the variance of the functional~\cite{Bha82}
\begin{equation}
  \gamma_2 = 2\int\D z\,L_1(z)\,\phi(z)\,f(z) 
  \hspace{0.5cm}\mbox{with}\hspace{0.5cm}
  \mathscr{G}_{z}L_1(z)=\gamma_1-\phi(z)
  \:,
\end{equation}
where $L_1(z)$ fulfills the condition $\int\D z\,L_1(z)f(z)=0$. 
For concreteness, let us give
a simple illustration~: we consider the functional $A[z(t)]=\int_0^x\D t\, z(t)^2$ where $z(t)$ is the Ornstein-Uhlenbeck process, $\partial_tz(t)=-2\kappa\,z(t)+\sqrt{2}\,\eta(t)$, and $\eta(t)$ a normalised Gaussian white noise. 
The generator is $\mathscr{G}_{z}=\partial_z^2-2\kappa\,z\,\partial_z$, thus $f(z)=\sqrt{\kappa/\pi}\exp\{-\kappa\,z^2\}$ and $\gamma_1=1/(2\kappa)$.  
It is easy to get $L_1(z)=z^2/(4\kappa)-1/(8\kappa^2)$ and thus 
$\gamma_2=1/(4\kappa^3)$.
  Here, 
  it is also possible to obtain the GLE $\Lambda(q)=\kappa-\sqrt{\kappa^2-q}$ for $q<\kappa^2$, and the large deviation function $\Phi$ controlling the distribution of the functional $\mathscr{P}_x(A)\sim\exp\big\{-x\,\Phi(A/x)\big\}$ for $x\to\infty$.
  We get 
  $
    \Phi(y)=\underset{q}{\mathrm{min}}\big\{q\,y-\Lambda(q)\big\}=(\kappa^2/y)\big[y-1/(2\kappa)\big]^2
  $.

The determination of the variance $\widetilde{\Lambda}''(0)$ of $\ln||\Pi_n\vec{x}_0||$ follows the same strategy. However, one has to identify the appropriate random process and the suitable functional, which requires to introduce more sophisticated tools.


\subsection{Main results}

In this article, most of the discussion is based on the Iwasawa decomposition of the group $\mathrm{SL}(2,\mathbb{R})$, corresponding to decompose each matrix as 
\begin{equation}
  \label{eq:Iwasawa}
  M = K(\theta)\,A(w)\,N(u) =
  \begin{pmatrix}
    \cos\theta & -\sin\theta \\ \sin\theta & \phantom{-}\cos\theta
  \end{pmatrix}
  \begin{pmatrix}
    \EXP{w} & 0 \\ 0 & \EXP{-w}
  \end{pmatrix}
  \begin{pmatrix}
    1 & u \\ 0 & 1
  \end{pmatrix}
\end{equation}
The calculation of the generalized Lyapunov exponent (GLE) is formulated in terms of a spectral problem based on a family of representations of the group acting on functions defined on the projective space and parametrised by a number $q$. 
In this representation, an element of the group takes the form
\begin{equation*}
  \mathscr{T}_{M}(q)   
  = \EXP{\theta\,\mathscr{D}_K(q)}
  \, \EXP{w\,\mathscr{D}_A(q)}
  \, \EXP{u\,\mathscr{D}_N(q)}
  \:,
\end{equation*}
where 
$\mathscr{D}_i(q)$'s are differential operators which form a representation of the Lie algebra. 
The GLE 
$
  \widetilde{\Lambda}(q)=-\ln \lambda(q)
$
is given by solving a spectral problem
\begin{equation}
  \label{eq:SpectralPbIntro}
  \big[\overline{\mathscr{T}}(q) \Phi^\mathrm{R}\big](z) = \frac{1}{\lambda(q)}\,\Phi^\mathrm{R}(z)
  \:,
  \hspace{0.5cm}\mbox{where}\hspace{0.25cm}
  \overline{\mathscr{T}}(q) \eqdef   \mean{   \mathscr{T}_{M}(q)  }_{M}  
\end{equation}
is the averaged transfer operator, $ \mean{ \cdots }_{M} $ denoting averaging over the group, i.e. over the set of parameters $(\theta,w,u)$. Here, $1/\lambda(q)$ is the largest eigenvalue of the operator $\overline{\mathscr{T}}(q)$ and $\Phi^\mathrm{R}(z)$ the corresponding right eigenvector. 
The spectral problem \eqref{eq:SpectralPbIntro} is precisely defined by some boundary conditions that will be discussed in \S~\ref{subsec:ARSP}.
%
The study of this spectral problem will be illustrated with several specific cases. 
We use a perturbative approach to solve the spectral problem and get the first coefficients of the expansion~\eqref{eq:GLE-SmallLambda}.

We first consider matrices of the form \eqref{eq:MatricesSchrodinger} or \eqref{eq:MatricesSchrodinger2} when angles are exponentially distributed,  $\mathrm{Proba}\{\theta_n>\theta\}=\EXP{-(\rho/k)\theta}$, and with arbitrary distribution $p(v)$ for the weights $v_n=k\,u_n$.
Introducing the L\'evy exponent $\levy(s)=\rho\,\big[1-\hat{p}(s)\big]$, where $\hat{p}(s)$ is the Fourier transform of the weight distribution,  we show that the coefficients $\lambda_n$'s of the expansion \eqref{eq:GLE-SmallLambda} can be expressed in terms of  the recessive solution of the differential equation (for $s\geq0$) 
\begin{equation*}
  -\hat{f}''(s) + \left[ E-\frac{\levy(s)}{\I s} \right]\hat{f}(s) = 0
  \:,
  \hspace{1cm}\mbox{with }
  \hat{f}(0)=1 
  \hspace{0.25cm}\mbox{and}\hspace{0.25cm}
  \hat{f}(s) \underset{s\to+\infty}{\longrightarrow} 0
  \:,
\end{equation*}
We set $E=+k^2$ for matrices \eqref{eq:MatricesSchrodinger} and $E=-k^2$ for matrices \eqref{eq:MatricesSchrodinger2}. 
We recover a known expression of the  Lyapunov exponent
$
  \lambda_1 = -\lambda'(0) =-\rho^{-1}\,\im[\hat{f}'(0^+)]
$
and derive an integral representation for 
\begin{equation*}
   \lambda_2 
    = - \int_0^\infty\frac{ \D s}{s}
      \, 
    \re\left[
      \left(  
         2\lambda_1\hat{p}(s)\,  - \frac{\I}{\rho}\, \deriv{}{s}
      \right)
      \hat{f}(s)^2
    \right]
   \:.
\end{equation*} 
$\lambda_2$ characterizes the fluctuations of the wave function at the $n$-th impurity
\begin{equation*}
     \widetilde{\Lambda}''(0) 
      = \lim_{n\to\infty} \frac{\mathrm{Var}\big(\ln ||\Pi_n\vec{x}_0||\big)}{n}
      = \lim_{n\to\infty} \frac{\mathrm{Var}\big(\ln |\psi(x_n^-)|\big)}{n}
  =\lambda_1^2-\lambda_2
     \:.
\end{equation*} 
We also study the GLE \eqref{eq:FluctPsiX} for the same model. 
The variance of $\ln|\psi(x)|$,
\begin{equation*}
    \Lambda''(0) 
      = \lim_{x\to\infty} \frac{\mathrm{Var}\big(\ln |\psi(x)|\big)}{x}
    =\gamma_2
     \:,
\end{equation*} 
is given by the integral 
\begin{equation*}
   \gamma_2  
    =  \int_0^\infty\frac{ \D s}{s}
      \, 
    \re\left[
      \left(  
         2\gamma_1  -  \I \, \deriv{}{s}
      \right)
      \hat{f}(s)^2
    \right]
   \:.
\end{equation*} 
Although $\Lambda(q) \neq \rho\,\widetilde{\Lambda}(q)$, the first derivatives coincide, $\gamma_1=\Lambda'(0) = \rho\,\widetilde{\Lambda}'(0)=\rho\,\lambda_1$.
L\'evy exponent of the form  $\levy(s)=\rho\,\big[1-\hat{p}(s)\big]$ corresponds to the case where $\int_0^x\D t\, V(t)$ is a compound Poisson process. 
Because any L\'evy process can be obtained by taking some appropriate limit of compound Poisson processes, these formulae apply to the case where $\int_0^x\D t\, V(t)$ is \textit{any} L\'evy process. 


A by-product of this compact formula allows us to rediscuss the criterion for \og single parameter scaling \fg{} (SPS) proposed by Deych, Lisyansky and Altshuler \cite{DeyLisAlt00,DeyLisAlt01}. We show that the Frisch-Lloyd model offers a counter-example where Deych et al.'s criterion fails to identify the boundary of the SPS regime.

Finally we analyze the continuum limit of random matrix products, 
i.e. when the matrices are close to the identity matrix $M\to\mathbf{1}_2$.
This corresponds to study the random walk over the group on large scales so that it can be described as a Brownian motion in the group.
The study of the continuum limit of matrix products has attracted the interest both in Physics \cite{ComLucTexTou13} and Mathematics \cite{ValVir14,GenGiaGre17,ComGiaGre19} literature.
For the Iwasawa decomposition, matrices are labelled by three parameters $(\theta,\,w,\,u)$.
Denoting by $\vec{\mu}=(\btheta,\,\bw,\,\bu)$ the vector of mean values and by $\covma^2$ the covariance matrix of the three parameters, the continuum limit is achieved by taking the limit where the nine parameters go to zero, all scaling in the same way (as well as the GLE).
The spectral problem \eqref{eq:SpectralPbIntro} then involves a differential operator
\begin{equation*}
  \label{eq:SpectralPbContinuumIntro}
  \left\{
   \frac{1}{2} \underline{\mathscr{D}}(q)\cdot\covma^2\cdot\underline{\mathscr{D}}(q)
   + \frac{1}{2} \vec{c}\cdot
     \Big(\underline{\mathscr{D}}(q)\times\underline{\mathscr{D}}(q)\Big)
   +    \vec{\mu} \cdot \underline{\mathscr{D}}(q)
  \right\}
  \Phi^\mathrm{R}(z)
  = \widetilde{\Lambda}(q)\, \Phi^\mathrm{R}(z)
  \:.
\end{equation*}
The vector $\underline{\mathscr{D}}(q)=\big(\mathscr{D}_K(q)\,,\,\mathscr{D}_A(q)\,,\,\mathscr{D}_N(q)\big)$ collects the three infinitesimal generators introduced above and the vector $\vec{c}$ is built from the correlations of the three parameters.
This will be illustrated on few examples. One of these cases corresponds to matrices in a two-parameter subgroup of $\mathrm{SL}(2,\mathbb{R})$, for which the spectral problem 
can be solved exactly.


\subsection{Outline}

In Section~\ref{sec:RepresentationsSL2R}, 
we introduce several useful representations of the group $\mathrm{SL}(2,\mathbb{R})$.
Using these ideas, we show in Section~\ref{sec:SpectralPb} that the determination of the generalized Lyapunov exponent (GLE) amounts to solving a spectral problem. 
In Section~\ref{sec:Transformations}, we show how integral transforms can simplify the resolution of the spectral problem.
Section~\ref{sec:Perturbation} presents a perturbative approach for the first terms of the GLE's Taylor expansion in powers of $q$. 
In Section~\ref{sec:GLElocalisation}, we describe a modified version of the problem, when the number of matrices is also a fluctuating quantity, which is more appropriate for quantum localisation problems.
Finally, we discuss the continuum limit of random matrix products in Section~\ref{sec:ContinuumLimit}.
The paper is closed by concluding remarks in Section~\ref{sec:Conclusion}.

\section{Several representations of the group $\mathrm{SL}(2,\mathbb{R})$}
\label{sec:RepresentationsSL2R}


\subsection{Matrix representation}

The elements of the group $\mathrm{SL}(2,\mathbb{R})$ are represented by $2\times2$ real matrices \begin{equation}
  \label{eq:MatrixOfSL2R}
  M = 
  \begin{pmatrix}
    a & b \\ c & d
  \end{pmatrix}
  \hspace{1cm}\mbox{with }
  ad-bc=1
  \:.
\end{equation}
Each element can be parametrised by three real parameters, depending on a choice of decomposition.
In \cite{ComTexTou10,ComLucTexTou13,ComTexTou13}, we have emphasized the role of the Iwasawa decomposition \eqref{eq:Iwasawa}, where the three matrices $K(\theta)$, $A(w)$ and $N(u)$ 
belong to the compact, Abelian and nilpotent subgroups, denoted $\mathrm{K}$, $\mathrm{A}$ and $\mathrm{N}$, respectively.
For each subgroup, we introduce an infinitesimal generator.
For the rotations, we have 
\begin{equation}
  \Gamma_K \eqdef \lim_{\theta\to0} \frac{K(\theta) - K(0)}{\theta}
  \hspace{0.5cm}\mbox{i.e.}\hspace{0.5cm}
  K(\theta) = \EXP{\theta\Gamma_K}
  \:,
\end{equation}
and similar relations for the two other subgroups.
Hence, the three generators related to the Iwasawa decomposition are
\begin{equation}
  \Gamma_K  
  =
  \begin{pmatrix}
    0  & -1 \\ 
    1  & \phantom{-}0
  \end{pmatrix}
  \:,
  \hspace{0.5cm}
  \Gamma_A
  =
  \begin{pmatrix}
    1 & \phantom{-}0 \\ 
    0 & -1
  \end{pmatrix}
  \:,
  \hspace{0.5cm}
  \Gamma_N
  =
  \begin{pmatrix}
    0 & 1 \\ 
    0 & 0
  \end{pmatrix}
\end{equation} 
and the Lie algebra 
takes the form
\begin{equation}
  \label{eq:LieAlgebra}
  [\Gamma_K , \Gamma_A]=2\Gamma_K +4\Gamma_N
  \:,
  \hspace{0.5cm}
  [\Gamma_A , \Gamma_N]=2\Gamma_N
  \:,
  \hspace{0.25cm}\mbox{and}\hspace{0.25cm}
  [\Gamma_N , \Gamma_K]=\Gamma_A
  \:.
\end{equation}
It follows that the only two-parameter subgroup is generated by $\Gamma_A$ and~$\Gamma_N$.

For concreteness, we mostly focus on the Iwasawa decomposition in the article, although the ideas are general and apply to any other choice.
In general, we write the Lie algebra
\begin{equation}
 \label{eq:DefStructureCstes}
  \left[  \Gamma_i \, , \, \Gamma_j \right]
  = \sum_{k=1}^3 c_{ijk}\,\Gamma_k
\end{equation}
in terms of the structure constants  $c_{ijk}=-c_{jik}$ of the group.
For the Iwasawa decomposition, we see from \eqref{eq:LieAlgebra} that $c_{121}=-c_{211}=2$, $c_{123}=-c_{213}=4$, 
$c_{233}=-c_{323}=2$ and $c_{312}=-c_{132}=1$ and all other coefficients equal to zero.

\paragraph{Another decomposition used in the paper.---}

As we have seen in Section~\ref{sec:Intro}, the subgroup $\widetilde{\mathrm{K}}$ of matrices 
\begin{equation}
  \widetilde{K}(\theta)
  =
  \begin{pmatrix}
    \cosh\theta & \sinh\theta \\ \sinh\theta & \cosh\theta
  \end{pmatrix}
\end{equation}
plays a role for the quantum models. 
The subgroup is characterised by the infinitesimal generator
\begin{equation}
  \Gamma_{\widetilde{K}}
  =
  \begin{pmatrix}
    0 & 1 \\ 1 & 0
  \end{pmatrix}
  \:.
\end{equation}
An alternative useful decomposition replacing \eqref{eq:Iwasawa} is  
\begin{equation}
  \label{eq:IwasawaTilde}
  M = \widetilde{K}(\theta)\,A(w)\, N(u)
  \:,
\end{equation}
where the Lie algebra of $\mathrm{SL}(2,\mathbb{R})$ now takes a different form~:
\begin{equation}
  [\Gamma_{\widetilde{K}} , \Gamma_A]=2\Gamma_{\widetilde{K}} - 4\Gamma_N
  \:,
  \hspace{0.5cm}
  [\Gamma_A , \Gamma_N]=2\Gamma_N
  \:,
  \hspace{0.25cm}\mbox{and}\hspace{0.25cm}
  [\Gamma_N , \Gamma_{\widetilde{K}}]=\Gamma_A
  \:.
\end{equation}

\subsection{Group action on $\overline{\mathbb{R}}$ : M\"obius transformations}

Our analysis of the random matrix products involves several representations, or more precisely \textit{group actions}, introduced in the remaining part of the section.
Let us first recall the definition of a group action~:
given a group $\mathrm{G}$ and a space $\mathfrak{X}$, a group action over $\mathfrak{X}$ is a map
$(\mathrm{G},\mathfrak{X})\overset{\mathcal{F}}{\longrightarrow}\mathfrak{X}$ such that, for $x\in\mathfrak{X}$ and $M\in\mathrm{G}$, the map $(M,x)\mapsto\mathcal{F}_M(x)$ satisfies
\begin{align}
  \mathcal{F}_{\mathbf{1}}(x) = x 
  \hspace{1cm}\mbox{and}\hspace{1cm}
  \mathcal{F}_{M_1}\left(\mathcal{F}_{M_2}(x)\right) = \mathcal{F}_{M_1M_2}(x) 
  \:,
\end{align}
where $\mathbf{1}$ is the identity and $M_1,\:M_2$ are two elements of $\mathrm{G}$.

The first useful group action involves M\"obius transformations, and arises as follows. 
The matrix product  $\Pi_n=M_n\cdots M_2M_1$ can be thought of as a random walk in the group, or a random walk in the vector space on which the matrices act, if one considers the sequence of vectors
\begin{equation}
   \{\vec{x}_0,\,\vec{x}_1,\cdots,\,\vec{x}_n,\cdots\}
   \hspace{1cm}\mbox{with}\hspace{1cm}
   \vec{x}_n=M_n\vec{x}_{n-1}
   \:.
\end{equation}
We can also map this sequence onto a random walk in the projective space.
For $2\times2$ matrices acting on $\vec{x}=(x,y)\in\mathbb{R}^2$, the projective space is the space of directions, $z=x/y\in\overline{\mathbb{R}}=\mathbb{R}\cup\{\infty\}$.
The projective line $\overline{\mathbb{R}}$ plays the role of the space $\mathfrak{X}$ and each matrix \eqref{eq:MatrixOfSL2R} is represented by a M\"obius transformation~:
\begin{equation}
  \label{eq:HomographicTransf}
  z \ \mapsto\ z' = \mathcal{M}(z) = \frac{a\,z+b}{c\,z+d}
  \:.
\end{equation}
The analysis of the random sequence of vectors is replaced by that of the Markov chain 
\begin{equation}
  \label{eq:RiccatiRecursion}
  \{z_0,\,z_1,\cdots,\,z_n,\cdots\}
   \hspace{1cm}\mbox{with}\hspace{1cm}
  z_n=\mathcal{M}_n(z_{n-1})
  \:.
\end{equation}

\paragraph{The infinitesimal generators and the Lie algebra.---}

Accordingly, we can introduce new expressions of the infinitesimal generators.
Consider a subgroup $\{M(\alpha)\}$ (for example $\mathrm{K}$, $\mathrm{A}$ or $\mathrm{N}$), related to the set of M\"obius transformations $\{\mathcal{M}_\alpha\}$.
We define the generator as 
\begin{equation}
  g(z) \eqdef - \lim_{\alpha\to0} \frac{\mathcal{M}_\alpha(z) - \mathcal{M}_0(z)}{\alpha}
  \:,
  \hspace{1cm}\mbox{i.e.}\hspace{0.5cm}
  \mathcal{M}_\alpha(z)  \underset{\alpha\to0}{\simeq} z-\alpha\,g(z)
  \:.
\end{equation}
For example, consider $K(\theta)\in\mathrm{K}$~:
\begin{equation}
  \mathcal{K}_\theta(z) = \frac{z\cos\theta-\sin\theta}{z\sin\theta+\cos\theta}
  \underset{\theta\to0}{\simeq} 
  z - \theta\, (1+z^2)
\end{equation}
hence $g_K(z) = 1+z^2$.
Similarly, for the subgroups $\mathrm{A}$ and $\mathrm{N}$, we deduce
\begin{equation}
  g_K(z) = 1+z^2\:,
  \hspace{0.5cm}
  g_A(z) = -2z\:,
  \hspace{0.5cm}
  g_N(z)   = -1
\end{equation}
(cf. Table~\ref{tab:InfGenerators}).

Combining two M\"obius transformations close to the identity, here $\theta\to0$ and $w\to0$, we have 
$
\mathcal{K}_\theta\big(\mathcal{A}_w(z)\big) - \mathcal{A}_w\big(\mathcal{K}_\theta(z)\big)
\simeq -  \theta\,w\,\mathcal{W}[g_K,g_A]
$\:, 
where $\mathcal{W}[f,g]\eqdef fg'-f'g$ denotes the Wronskian of two functions.
This shows that the Lie algebra \eqref{eq:LieAlgebra} is realized as follows 
\begin{equation}
  \label{eq:LieAlgebra2}
  \mathcal{W}[g_K,g_A]
  =2g_K(z)+4g_N(z)  \:,
  \hspace{0.25cm}
  \mathcal{W}[g_A,g_N]=2g_N(z)
  \mbox{ and }
  \mathcal{W}[g_N,g_K]=g_A(z)
  \:.
\end{equation}
More generally, the relations can be written in terms of the structure constants $c_{ijk}$, defined in Eq.~\eqref{eq:DefStructureCstes}, as
\begin{equation}
  \label{eq:giLieAlgebra}
  \mathcal{W}[g_i(z),g_j(z)]=\sum_{k=1}^3 c_{ijk}\, g_k(z)
  \:.
\end{equation}

\subsection{Group action on functions defined on the projective line}
\label{subsec:ActionOnFunctions}

The study of the Lyapunov exponent~in Ref.~\cite{ComLucTexTou13} has led us to introduce a second representation (group action). 
In this case the space $\mathfrak{X}$ is identified with the space of functions defined on the projective line $\overline{\mathbb{R}}$.
For each $M\in\mathrm{SL}(2,\mathbb{R})$ and for a function $f(z)$, we define
\begin{equation}
  \label{eq:Repres3}
  \left[
     \mathscr{T}_M f
  \right](z) 
  \eqdef 
  \deriv{\mathcal{M}^{-1}(z)}{z}\,
  f\!\left( \mathcal{M}^{-1}(z) \right)
\end{equation}
where 
\begin{equation}
  \mathcal{M}^{-1}(z)=\frac{d\,z-b}{-c\,z+a}
\end{equation}
is the inverse transformation. 
Strictly speaking, according to the discussion of the next subsection, this representation involves the \textit{multiplier} 
\begin{equation}
  \deriv{\mathcal{M}^{-1}(z)}{z} = \frac{1}{(-c\,z+a)^2}
  \:.
\end{equation}

We can associate to the random sequence \eqref{eq:RiccatiRecursion} a sequence of probability distribution functions $f_{n}(z)=\mean{ \delta(z-z_{n})}$.
The operator \eqref{eq:Repres3} allows to characterize the evolution of the density due to the action of the random M\"obius transformation at each step~: 
\begin{equation}
  \label{eq:DistribRecursion}
   f_{n}(z) = \mean{ \left[
     \mathscr{T}_M f_{n-1}
  \right](z)  }_M
   \:,
\end{equation}
where $\mean{\cdots}_M$ is an average over the group.
In the limit $n\to\infty$, the distribution is expected to reach a limit law $f(z)$~\cite{Fur63,BouLac85}, in the distributional sense, solution of the Furstenberg integral equation
\begin{equation}
  \label{eq:FurstenbergEquation}
   f(z) = \mean{ \left[
     \mathscr{T}_M f
  \right](z)  }_M
  \:.
\end{equation}
In Physics literature, \eqref{eq:FurstenbergEquation} is known as the Dyson-Schmidt equation~\cite{LifGrePas88,Luc92}.

Let us now derive the infinitesimal generators associated with the representation~\eqref{eq:Repres3}.
For one of the three one-parameter subgroups ($\mathrm{K}$, $\mathrm{A}$ or $\mathrm{N}$ for instance), 
we write an element as $M(\alpha)=\EXP{\alpha\Gamma}=1+\alpha\,\Gamma+\mathcal{O}(\alpha^2)$, 
associated with the M\"obius transform $\mathcal{M}_\alpha(z)=1-\alpha\,g(z)+\mathcal{O}(\alpha^2)$ for $\alpha\to0$. 
We expand
\begin{align}
  \big[
     \mathscr{T}_{M(\alpha)} f
  \big](z) 
  &= \deriv{}{z} \int^{\mathcal{M}_\alpha^{-1}(z)}\D t\, f(t)
  = \deriv{}{z} \left( 1 + \alpha\,g(z)\,\deriv{}{z} +\mathcal{O}(\alpha^2)\right)\int^{z}\D t\, f(t)
  \nonumber\\
  &= f(z) + \alpha\,\deriv{}{z}\left[ g(z)\, f(z)\right] +\mathcal{O}(\alpha^2)
\end{align}
Identification with the form
\begin{equation}
  \left[
     \mathscr{T}_M f
  \right](z) 
  = f(z)
  +\left( \theta \, \mathscr{D}_K + w\, \mathscr{D}_A + u\, \mathscr{D}_N \right) f(z)
  +\cdots
\end{equation}
finally gives
\begin{equation}
  \label{eq:InfGenCLTT13}
  \mathscr{D}_i = \deriv{}{z}\,g_i(z)
  \hspace{0.5cm}
  \mbox{for } 
  i\in\{ K, \,  A, \, N  \}
  \:.
\end{equation}
The notation must be understood as 
$\mathscr{D}_if(z)=\deriv{}{z}\left[g_i(z)f(z)\right]$.

%
%
%
%
%

\subsubsection{The invariant measure for a one-parameter subgroup}

Consider a subgroup $\{M(\alpha)\}$ ($\mathrm{K}$, $\mathrm{A}$ or $\mathrm{N}$, for instance), characterised  by the generator $g(z)$ (in the projective space $\overline{\mathbb{R}}$).
The measure invariant under the action of the elements of the subgroup is 
\begin{equation}
  \rho(z) \, \D z = \frac{\D z }{g(z)}
  \:.
\end{equation}
%
%
Using the representation $\mathscr{T}_{M(\alpha)}=\exp\big\{\alpha\deriv{}{z}g(z)\big\}$, it is pretty obvious that the density $\rho(z)=1/g(z)$ is invariant under such transformations~:
\begin{equation}
  \left[ \mathscr{T}_{M(\alpha)}\rho \right](z) = \rho(z)
  \hspace{0.5cm}\forall\alpha
\:.
\end{equation}


Apart this simple case of one-parameter subgroup, the determination of the invariant measure, solution of the Furstenberg equation \eqref{eq:FurstenbergEquation}, in the general case  is an extremely difficult problem.



\subsection{Jacobians}


Jacobians turn out to be central in order to introduce a group action useful for the determination of the GLE.
Given a certain measure $\rho(z)\,\D z$ on the projective space, for each element $M\in\mathrm{SL}(2,\mathbb{R})$ and its associated M\"obius transform $\mathcal{M}$, we define the Jacobian by
\begin{equation}
  \label{eq:DefJ}
   J(M,z) = \frac{\rho(z')\,\D z'}{\rho(z)\,\D z}
   \hspace{0.5cm}\mbox{for }
   z' = \mathcal{M}(z)
   \:,
\end{equation}
characterizing the transformation of the measure $\rho(z)\,\D z$ under the M\"obius transform (a geometrical interpretation is discussed in \cite{ComTexTou19}).
It satisfies the two following properties~:
\begin{enumerate} 

\item[(i)]
  $J(e,z)=1$ where $e$ is the identity of the group.

\item[(ii)]
  The chain property~:
  $J(M_2M_1,z)=J(M_2,\mathcal{M}_1(z))\,J(M_1,z)$.  
\end{enumerate}
We also introduce the \textit{additive cocycle} \cite{FurKes60,Tut65,BouLac85}
\begin{equation}
  \label{eq:DefCocycle}
  \sigma(M,z) = \ln J(M,z)
  \:.
\end{equation}

\subsubsection{Three specific cases}

Let us introduce three natural choices of Jacobian to which we will refer later.
\begin{enumerate}
\item
  For the flat measure $\rho_N(z)\,\D z=\D z/|g_N(z)|=\D z$, i.e. the measure invariant under the action of the subgroup~$\mathrm{N}$.
  The Jacobian related to the matrix \eqref{eq:MatrixOfSL2R} takes the form
  \begin{equation}
    \label{eq:JN}
    J_N(M,z) = \deriv{\mathcal{M}(z)}{z} = \frac{1}{(c\,z+d)^2}  
    \:.
  \end{equation}
  where $\mathcal{M}(z)$ is the M\"obius transformation \eqref{eq:HomographicTransf} related to \eqref{eq:MatrixOfSL2R}.

\item
  We now consider the measure $\rho_A(z)\,\D z=\D z/|g_A(z)|=\D z/|2z|$ invariant under the Abelian subgroup $\mathrm{A}$. We have 
  \begin{equation}
    \label{eq:JA}
     J_A(M,z) 
     = \frac{\rho_A(\mathcal{M}(z))}{\rho_A(z)} \, J_N(M,z)
     = \frac{z}{(a\,z+b)(c\,z+d)}
     \:.
  \end{equation}

\item
  Finally we consider the measure  $\rho_K(z)\,\D z=\D z/g_K(z)=\D z/(1+z^2)$ invariant under the action of the compact subgroup $\mathrm{K}$. Thus 
  \begin{equation}
    \label{eq:JK}
     J_K(M,z) 
     = \frac{\rho_K(\mathcal{M}(z))}{\rho_K(z)} \, J_N(M,z)
     = \frac{1+z^2}{(a\,z+b)^2 + (c\,z+d)^2 }
     \:.
  \end{equation}
\end{enumerate}

Because $\rho_N(z)\,\D z$ is the measure invariant under the action of $N(u)\in\mathrm{N}$, we have 
\begin{equation}
  \label{eq:PorpertyJ}
  J_N(N(u),z)=1
  \hspace{0.25cm}
  \mbox{ and similarly} 
  \hspace{0.25cm}
  J_A(A(w),z)=1
  \:, 
  \hspace{0.25cm}
  J_K(K(\theta),z)=1
\end{equation}
for $A(w)\in\mathrm{A}$ and $K(\theta)\in\mathrm{K}$. 
This will be used below in order to simplify the calculations.

A key observation for the following is that, given the vector 
\begin{equation}
  \vec{x} = \frac{1}{\sqrt{1+z^2}}
  \begin{pmatrix}
    z \\ 1
  \end{pmatrix}
  \:,
\end{equation}
we have 
\begin{equation}
  \label{eq:KeyObservation}
  ||M\vec{x}|| = \frac{ \sqrt{(a\,z+b)^2 + (c\,z+d)^2} }{ \sqrt{1+z^2} } = J_K^{-1/2}(M,z)
  \:.
\end{equation}
This remark will be central in order to express the norm $||\Pi_N\vec{x}||$ involved in the definition of the GLE, Eq.~\eqref{eq:DefGLE}.

\subsubsection{A simple application~: a formula for the Lyapunov exponent}

As a simple application of the concept of Jacobian and additive cocycle, we deduce a formula for the Lyapunov exponent $\lambda_1$ of the RMP. We start from the Furstenberg formula~\cite{Fur63,BouLac85} 
\begin{align}
  \lambda_1 = \int\D z\,f(z) \,
  \bigg\langle
    \ln\frac{\left\lVert M\begin{pmatrix}
       z \\ 1
    \end{pmatrix}\right\rVert}{\left\lVert \begin{pmatrix}
       z \\ 1
    \end{pmatrix}\right\rVert}  
    \bigg\rangle_{\!\!M}
  &=-\frac12 \int\D z\,f(z) \,\mean{ \ln J_K(M,z) }_M
\end{align}
where the average is taken over the matrix $M$.
As a result we can rewrite the Lyapunov exponent as an average of the cocycle~\eqref{eq:DefCocycle}
\begin{equation}
  \label{eq:LyapunovFromCocycle}
  \lambda_1 = -\frac12 \int\D z\,f(z) \,\mean{ \sigma(M,z) }_M
  \:.
\end{equation}
We stress that this formula is not restricted to choosing $J_K$, but is valid \textit{for any choice of Jacobian}, as 
\begin{align}
  \int\D z\,f(z) \,\mean{ \ln J(M,z) }_M
  &= \int\D z\,f(z) \,\mean{ \ln\left[\frac{\rho(\mathcal{M}(z))}{\rho(z)} J_N(M,z)\right] }_M
  \nonumber\\
  &= \int\D z\,f(z) \,\mean{ \ln J_N(M,z) }_M
  \:,
\end{align}
where the last equality follows from the Furstenberg equation \eqref{eq:FurstenbergEquation}. This leads to
$\int\D z\,f(z) \,\mean{\ln\rho(\mathcal{M}(z))}_M=\int\D z\,f(z) \,\ln\rho(z)$. 

In particular for $J=J_N$, the expression of the cocycle is simple, cf.~\eqref{eq:JN}~:
\begin{equation}
  \label{eq:GeneralFormulaLyapunov}
  \lambda_1  =  \int\D z\,f(z) \,\mean{ \ln|c\,z+d| }_M
  = \int\D z\,f(z) \,\mean{ \ln\left|(z+u)\EXP{w}\sin\theta+\EXP{-w}\cos\theta)\right| }_M
  \:.
\end{equation}


\subsection{A family of representations with multipliers}
\label{subsec:Family}

Our aim is to study the generalized Lyapunov exponent \eqref{eq:DefGLE}. 
For this purpose, the remark \eqref{eq:KeyObservation} suggests to weight the transfer operator \eqref{eq:Repres3} by the Jacobian to the power $q$.
This leads to introduce a family of representations (or more precisely of ``group actions'') parametrised by~$q$~:
\begin{equation}
  \label{eq:Repres4}
    \boxed{
  \left[
     \mathscr{T}_M^{(\rho)}(q) f
  \right](z) 
  \eqdef
  J^{q/2}(M^{-1},z) 
  \,
  \deriv{\mathcal{M}^{-1}(z)}{z}\,
  f\!\left( \mathcal{M}^{-1}(z) \right)
  }
\end{equation}
The operator depends on the Jacobian \eqref{eq:DefJ}, i.e. on the related density $\rho(z)$, what we have made explicit in the notation [in the following sections, we will simplify the notation as $\mathscr{T}_M^{(\rho)}(q)\longrightarrow\mathscr{T}_M(q)$].
In the paper, we will choose $q\in\mathbb{R}$.
The operator \eqref{eq:Repres4} for $q=0$ coincides with \eqref{eq:Repres3}.
We remark that this new representation involves two different types of Jacobians~:
\begin{equation}
  \left[
     \mathscr{T}_M^{(\rho)}(q) f
  \right](z) 
  =
  J^{q/2}(M^{-1},z)  \,  J_N(M^{-1},z) \,
  f\!\left( \mathcal{M}^{-1}(z) \right)
  \:.
\end{equation}
In Appendix~\ref{app:RepresMatheux}, the relation with another choice of representation is discussed~: Eq.~\eqref{eq:Repres4Yves} involves only one type of Jacobian, however we have found it less convenient for the subsequent analysis.  
We will show in the next section why \eqref{eq:Repres4} is useful in order to determine the GLE.
Let us first discuss in detail some properties of this family of representations indexed by $q$.

\subsubsection{Infinitesimal generators}
\label{subsubsec:InfGenDq}

The new representation defines a set of $q$-dependent infinitesimal generators $\mathscr{D}_i^{(\rho)}(q)$.
We proceed in a similar way as above~:
for $M\to\boldsymbol{1}_2$, i.e. $\theta\to0$, $w\to0$ and $u\to0$ in the Iwasawa representation, we define three functions $h_K^{(\rho)}$, $h_A^{(\rho)}$ and $h_N^{(\rho)}$ by
\begin{equation}
  \label{eq:ExpansionJ}
  J(M,z) = 1 -2 \theta\, h_K^{(\rho)}(z) -2 w\, h_A^{(\rho)}(z) -2 u\, h_N^{(\rho)}(z) + \cdots
  \:.
\end{equation}
The functions $h_i^{(\rho)}(z)$'s carry the dependence in $\rho$ [i.e. on the choice of Jacobian $J$ in Eq.~\eqref{eq:Repres4}] of the infinitesimal generator. 
With these functions, we deduce the expressions of the infinitesimal generators for the representation \eqref{eq:Repres4}, defined by 
\begin{equation}
  \label{eq:Representation4Elements}
  \mathscr{T}_M^{(\rho)}(q) 
  = \EXP{\theta\,\mathscr{D}_K^{(\rho)}(q)}
  \, \EXP{w\,\mathscr{D}_A^{(\rho)}(q)}
  \, \EXP{u\,\mathscr{D}_N^{(\rho)}(q)}
  \:.
\end{equation}
Writing
\begin{equation}
   \left[
     \mathscr{T}_M^{(\rho)}(q) f
  \right](z) 
  = f(z) 
  + \theta\, \mathscr{D}_K^{(\rho)}(q) f(z) 
  + w\,      \mathscr{D}_A^{(\rho)}(q) f(z) 
  + u\,      \mathscr{D}_N^{(\rho)}(q) f(z) 
  +\cdots
  \:,
\end{equation}
expansion of \eqref{eq:Repres4} for $\theta\to0$, $w\to0$ and $u\to0$ gives
\begin{equation}
  \label{eq:GeneratorsDdeQ}
  \boxed{
   \mathscr{D}_i^{(\rho)}(q) = \deriv{}{z} \, g_i(z) + q\, h_i^{(\rho)}(z)
   }
  \hspace{0.5cm}
  \mbox{for } 
  i\in\{ K, \, A, \, N  \}
  \:.
\end{equation}

\paragraph{Choosing the Jacobian $J_N$~:}

A specific choice which will play an important role is $J=J_N$. 
We denote by $\{M(\alpha)\}$ the elements of a one parameter subgroup with 
$\mathcal{M}_\alpha(z)\simeq z - \alpha\,g(z)$ for $\alpha\to0$, where $g(z)$ represents the associated generator.
We have
\begin{equation}
  J_N(M(\alpha),z) =  \deriv{\mathcal{M}_\alpha(z)}{z}  
  \underset{\alpha\to0}{\simeq}  1 - \alpha\,g'(z)
  \:.
\end{equation}
Thus, comparison with \eqref{eq:ExpansionJ} gives
\begin{equation}
    h_i^{(\rho_N)}(z) = \frac12\, g_i'(z)   
    \hspace{0.5cm}
  \mbox{for } 
  i\in\{ K, \,  A, \, N  \}
\end{equation}
(cf. Table~\ref{tab:InfGenerators}).
The corresponding infinitesimal generators can be written as 
\begin{equation}
   \mathscr{D}_i^{(\rho_N)}(q) = \deriv{}{z}\, g_i(z) + \frac{q}{2}\, g_i'(z)
   \:.
\end{equation}


\paragraph{Choosing an arbitrary Jacobian $J$~:}

Instead of considering the flat measure $\rho_N(z)\,\D z=\D z$ associated with $J_N$, consider the measure $\rho(z)\,\D z$ and
\begin{equation}
   J(M,z)=J_N(M,z)\,\frac{ \rho\left(\mathcal{M}(z)\right) }{ \rho(z) }
   \:.
\end{equation}
Inserting $\mathcal{M}_\alpha^{-1}(z)\simeq z + \alpha\,g(z)$ for $\alpha\to0$,  we find 
\begin{equation}
  J(M(\alpha)^{-1},z) 
  \simeq 1 
  + \alpha \, 
  \left( 
     g'(z) + g(z)\,\frac{\rho'(z)}{\rho(z)} 
  \right)
  \:.
\end{equation}
Thus, going from the flat measure $\rho_N(z)\,\D z=\D z$ to an arbitrary measure $\rho(z)\,\D z$, the functions $h_i^{(\rho)}(z)$'s controlling the infinitesimal generators are transformed as
\begin{equation}
  \label{eq:FromHnToH}
  h_i^{(\rho)}(z) = h_i^{(\rho_N)}(z) + \frac12 \, g_i(z)\, \frac{\rho'(z)}{\rho(z)} 
\end{equation}
or more explicitly
\begin{equation}
  \label{eq:ExpressionOfHi}
   \boxed{
      h_i^{(\rho)}(z) 
      = \frac{1}{2}
      \left(
          g_i'(z) + g_i(z)\, \deriv{\ln\rho(z)}{z} 
      \right)
      = \frac{1}{2\rho(z)}\,\big( g_i(z)\,\rho(z)\big)'
   }
\end{equation} 
The corresponding transformation of the generators can be written as 
\begin{equation}
   \mathscr{D}_i^{(\rho)}(q)
  =\rho(z)^{-q/2}\, \mathscr{D}_i^{(\rho_N)}(q) \, \rho(z)^{q/2}
  \:.
\end{equation}
and accordingly, the transfer operators as 
\begin{equation}
  \label{eq:TransformationJNtoJ}
  \boxed{
   \mathscr{T}_M^{(\rho)}(q)
  =\rho(z)^{-q/2}\, \mathscr{T}_M^{(\rho_N)}(q) \, \rho(z)^{q/2}
  }
\end{equation}
This is an equivalence relation connecting different \textit{realizations} of the same \textit{representation}.

The transformation \eqref{eq:FromHnToH} is convenient to construct the different sets of functions involved in the infinitesimal generators.
For example, we have $h_K^{(\rho_N)}(z)=z$ and $h_A^{(\rho_N)}(z)=-1$. 
Hence 
$
    h_K^{(\rho_A)} = z - \frac12\, (1+z^2)/z = (z^2-1)/(2z)
$
and 
$
    h_A^{(\rho_A)} = -1 + \frac12\, 2z/z = 0
$, etc.  
Table~\ref{tab:InfGenerators} summarizes all possible functions in relation with the cases discussed in the paper.

One can see on the table that we have the property
\begin{equation}
  h_i^{(\rho_i)}(z)=0
  \hspace{0.5cm}\mbox{for} \hspace{0.5cm}
  i\in\{K,\,A,\,N\}
  \:.
\end{equation}
This obviously originates from Eq.~\eqref{eq:PorpertyJ}.
We will make use of this remark in order to simplify the analysis below.

\begin{table}[!ht]
\centering
\begin{tabular}{|c|l|cccccc|}
\hline
  & & & \multicolumn{5}{c|}{choice of measure $\rho(z)\,\D z$ \& Jacobian $J$ } \\
  subgroup & generator &      & $J_N$ & $J_A$ & $J_K$ & $J_{\widetilde{K}}$ & $J_{N_-}$\\
  \hline
  $\widetilde{\mathrm{K}}$ &  $g_{\widetilde{K}}(z) = -1+z^2$  
                           & $h_{\widetilde{K}}(z)$: 
                           & $z$ 
                           & $\frac{z^2+1}{2z}$ 
                           & $\frac{2z}{z^2+1}$
                           & $0$
                           & $\frac{1}{z}$
  \\[0.125cm]
  \hline
  $\mathrm{K}$     & $g_K(z) = 1+z^2$ 
                   & $h_K(z)$: 
                   & $z$ 
                   & $\frac{z^2-1}{2z}$ 
                   & $0$ 
                   & $\frac{2z}{1-z^2}$
                   & $-\frac{1}{z}$
  \\[0.125cm]
  $\mathrm{A}$     &$g_A(z) = -2z$ 
                             & $h_A(z)$: 
                             & $-1$ 
                             & $0$ 
                             & $\frac{z^2-1}{z^2+1}$
                             & $\frac{z^2+1}{z^2-1}$
                             & $1$
  \\[0.125cm]
  $\mathrm{N}$     &$g_N(z) = -1$ 
                            & $h_N(z)$: 
                            & $0$ 
                            & $\frac{1}{2z}$ 
                            & $\frac{z}{z^2+1}$
                            & $\frac{z}{z^2-1}$
                             & $\frac{1}{z}$
  \\[0.125cm]
  \hline
  $\mathrm{N}_-$     &$g_{N_-}(z) = z^2$ 
                            & $h_{N_-}(z)$: 
                            & $z$ 
                            & $\frac{z}{2}$ 
                            & $\frac{z}{z^2+1}$
                            & $-\frac{z}{z^2-1}$
                            & $0$
  \\
  \hline
\end{tabular}
\caption{\it 
The functions $g_i(z)$ and $h_i(z)$ involved in the infinitesimal generators \eqref{eq:GeneratorsDdeQ} of the family of representations introduced in \S~\ref{subsec:Family}. 
The last line corresponds to the subgroup of matrices of the form $N_-(u)=N(u)^\mathrm{T}$ also considered in Refs.~\cite{ComTexTou10,ComTexTou19}.
}
\label{tab:InfGenerators}
\end{table}

\paragraph{Representation of the Lie algebra~:}

As an interesting exercice we can check that the operators \eqref{eq:GeneratorsDdeQ} 
provide a representation of the Lie algebra (\ref{eq:LieAlgebra},\ref{eq:DefStructureCstes}). 
For this purpose we remark that the commutator can be expressed as 
\begin{equation}
   \left[ \mathscr{D}_i^{(\rho)}(q) \, , \, \mathscr{D}_j^{(\rho)}(q) \right]
   = \deriv{}{z} \, \mathcal{W}[g_i(z),g_j(z)]
   + q \left(  g_i(z) h_j'(z) -  h_i'(z) g_j(z) \right)
   \:,
\end{equation}
where we have dropped the label $^{(\rho)}$ in functions $h_i$ for clarity. 
%
From \eqref{eq:ExpressionOfHi}, we get
\begin{equation}
  g_i(z) h_j'(z) -  h_i'(z) g_j(z) 
  =\frac{1}{2}\left(
   \deriv{ \mathcal{W}[g_i,g_j]}{z} + \mathcal{W}[g_i,g_j]\,\deriv{\ln\rho(z)}{z}
  \right)
  = \sum_k c_{ijk}\, h_k(z)
\end{equation}
where we have used \eqref{eq:giLieAlgebra} (the structure constants $c_{ijk}$ were introduced in \eqref{eq:DefStructureCstes} above).
We have thus recovered the Lie algebra
\begin{equation}
   \left[ \mathscr{D}_i^{(\rho)}(q) \, , \, \mathscr{D}_j^{(\rho)}(q) \right] 
   = \sum_k c_{ijk}\,  \mathscr{D}_k^{(\rho)}(q) 
   \:,
\end{equation}
which is equivalent to say that \eqref{eq:Repres4} is a representation of the group.

In the rest of the paper, we will drop the label $^{(\rho)}$ in the operators, in order to lighten the notations.

\subsubsection{Adjoint operator}

For $q\in\mathbb{R}$, the adjoint of \eqref{eq:Repres4}
\begin{equation}
  \label{eq:Repres4adjoint}
  \boxed{
  \big[
     \mathscr{T}_M^\dagger(q) f
  \big](z) 
  =
  J^{-q/2}(M,z) 
  \,
  f\!\left( \mathcal{M}(z) \right)
  }
\end{equation}
is defined with respect to the following inner product
\begin{equation}
  \label{eq:DefinitionScalarProduct}
  \braket{\psi}{f} \eqdef \int\D z \, \psi(z)^*\, f(z)
  \:.
\end{equation}

Let us verify the relation between the two operators (\ref{eq:Repres4},\ref{eq:Repres4adjoint})~:
\begin{align}
  \braket{\psi}{\mathscr{T}_M(q)f}
  &= 
  \int \D z \, \psi(z)^*\, J^{q/2}(M^{-1},z) \,\deriv{\mathcal{M}^{-1}(z)}{z} 
  \:
  f\!\left( \mathcal{M}^{-1}(z) \right)
\end{align}
Performing the change of variable $z=\mathcal{M}(y)$ we obtain
\begin{align}
  \int \D y\, \, \psi\left( \mathcal{M}(y) \right)^*\, 
  J^{q/2}(M^{-1},\mathcal{M}(y)) \, f(y) 
  &=\int \D z\, J^{-q/2}(M,z)\, \psi\left( \mathcal{M}(z) \right)^*\, f(z)
  \nonumber\\
  &= \braket{\mathscr{T}_M^\dagger(q)\psi}{f}
\end{align}
The last step makes use of $J(M^{-1},\mathcal{M}(z))\,J(M,z)=J(M^{-1}M,z)=1$, i.e.
\begin{equation}
  \label{eq:AUsefulPropertyWithTheJacobian}
  J(M^{-1},\mathcal{M}(z))=1/J(M,z)
  \:.
\end{equation}

We point out a relation between the transfer operator \eqref{eq:Repres4} and its adjoint \eqref{eq:Repres4adjoint}, that will be useful below.
Discussion is more simple for $J=J_N$, then
\begin{align}
  \label{eq:Eq2.65}
    \big[
     \mathscr{T}_M(q) f
  \big](z) 
  &=
  J_N^{1+q/2}(M^{-1},z) 
  \,
  f\!\left( \mathcal{M}^{-1}(z) \right)
  \\
    \big[
     \mathscr{T}_M^\dagger(q) f
  \big](z) 
  &=
  J_N^{-q/2}(M,z) 
  \,
  f\!\left( \mathcal{M}(z) \right)
  \:.
\end{align}
Thus  
\begin{equation}
  \label{eq:SymTdaggerT}
   \mathscr{T}_{M^{-1}}(-q-2) = \mathscr{T}_M^\dagger(q)
   \:.
\end{equation}

The adjoint operator can be expressed in terms of adjoints of the infinitesimal generators
\begin{equation}
  \label{eq:Representation4AdjointElements}
  \mathscr{T}_M^\dagger(q) 
  = 
  \EXP{u\,\mathscr{D}_N^\dagger(q)}  
  \, \EXP{w\,\mathscr{D}_A^\dagger(q)}
   \, \EXP{\theta\,\mathscr{D}_K^\dagger(q)}
  \:,
\end{equation}
by expanding \eqref{eq:Repres4adjoint} as $\mathscr{T}_M(q) $, we get the expressions of the adjoints of the generators \eqref{eq:GeneratorsDdeQ}
\begin{equation}
  \label{eq:AdjointInfinGene}
   \mathscr{D}_i^\dagger(q) = -g_i(z)\,\deriv{}{z} + q\, h_i(z)
  \hspace{0.5cm}
  \mbox{for } 
  i\in\{ K, \, A, \, N  \}
  \:.
\end{equation}
It is also straightforward to check that
$
  \braket{\psi}{\mathscr{D}_i(q)f} = \braket{\mathscr{D}_i^\dagger(q)\psi}{f} 
$.

We stress that the operators \eqref{eq:AdjointInfinGene} do \textit{not} provide a representation of the group since
$\mathscr{T}_{M_1}^\dagger(q)\mathscr{T}_{M_2}^\dagger(q)=\mathscr{T}_{M_2M_1}^\dagger(q)$.


\section{The generating function and the spectral problem}
\label{sec:SpectralPb}

Having all the tools in place, we can tackle the main problem. 
Our aim is to analyse the moment generating function for the product of random matrices~\eqref{eq:RMP}
\begin{equation}
  \boxed{
  Q_n(z_0;q) \eqdef \mean{||\Pi_n\vec{x}_0||^q}_{M_1,\cdots,M_n}
  }
  \hspace{0.5cm}\mbox{where}\hspace{0.5cm}
  \vec{x}_0 = \frac{1}{\sqrt{1+z_0^2}}
  \begin{pmatrix}
    z_0 \\ 1
  \end{pmatrix}
  \:,
\end{equation}
$\mean{\cdots}_M$ denoting the average over the group.
We now show that its large $n$ behaviour is controlled by the largest eigenvalue of a certain linear operator.

\subsection{The moment generating function}

Using the relation \eqref{eq:KeyObservation} and the chain rule, we can rewrite the norm of the product as 
\begin{equation}
   ||\Pi_n\vec{x}_0|| = J_K^{-1/2}(\Pi_n,z_0)
   = \prod_{i=1}^n J_K^{-1/2}(M_i,z_{i-1})
   \hspace{1cm}\mbox{where }
   z_i=\mathcal{M}_i(z_{i-1})
  \:.
\end{equation}
Thus
\begin{equation}
  Q_n(z_0;q) 
  = \mean{ \prod_{i=1}^n J_K^{-q/2}(M_i,z_{i-1}) }_{M_1,\cdots,M_n}
  \:.
\end{equation}
For convenience, we prefer to express the product in terms of the Jacobian $J_N$. We use
\begin{equation}
  J_K(M,z) = \frac{\rho_K(\mathcal{M}(z))}{\rho_K(z)}\, J_N(M,z)
\end{equation}
where $\rho_K(z)\,\D z=\D z/(1+z^2)$ is the measure introduced above.
We end with the convenient form
\begin{equation}
  \label{eq:68}
  Q_n(z_0;q) 
  = \rho_K(z_0)^{q/2} \mean{ 
      \rho_K(z_{n})^{-q/2} \prod_{i=1}^n J_N^{-q/2}(M_i,z_{i-1}) 
     }_{M_1,\cdots,M_n}
     \:.
\end{equation}


\subsection{Forward evolution}

We introduce the \og \textit{propagator} \fg{} 
\begin{align}
  \label{eq:PropagatorFEvol}
  P_n(z|z_0;q) 
  &\eqdef
  \mean{ \delta(z-z_n) \, \prod_{i=1}^n J_N^{-q/2}(M_i,z_{i-1}) }_{M_1,\cdots,M_n}
\end{align}
which is related to the generating function \eqref{eq:68} by 
\begin{equation}
  \label{eq:CharacterisitcFunctionFromPropagator}
  Q_n(z_0;q) 
  =\rho_K(z_0)^{q/2} \int\D z\,  \rho_K(z)^{-q/2} \: P_n(z|z_0;q) 
  \:.
\end{equation}
We now derive a recursion for the propagator.
We split the product as 
\begin{equation}
  \prod_{i=1}^n J^{-q/2}(M_i,z_{i-1})
  = J^{-q/2}(M_n,z_{n-1})\prod_{i=1}^{n-1} J^{-q/2}(M_i,z_{i-1})
\end{equation}
and use 
\begin{equation}
  \delta(z-z_n) = \delta\left(z-\mathcal{M}_n(z_{n-1})\right)
  =\deriv{\mathcal{M}_n^{-1}(z)}{z}\,\delta\left(z_{n-1}-\mathcal{M}_n^{-1}(z)\right)
  \:.
\end{equation}
Thus 
\begin{align}
  \label{eq:5.14}
  &\mean{ \delta(z-z_n) \, \prod_{i=1}^n J^{-q/2}(M_i,z_{i-1}) }_{M_1,\cdots,M_n}
  \nonumber
  \\
  = 
  &\mean{ 
  \deriv{\mathcal{M}_n^{-1}(z)}{z}\,\delta\left(z_{n-1}-\mathcal{M}_n^{-1}(z)\right)
  J^{-q/2}(M_n,\mathcal{M}_n^{-1}(z)) 
  \prod_{i=1}^{n-1} J^{-q/2}(M_i,z_{i-1})
  }
  \:.
\end{align}
Now 
inserting \eqref{eq:AUsefulPropertyWithTheJacobian}
in \eqref{eq:5.14}, we obtain 
\begin{align}
  \label{eq:ForwardEvolution}
  P_n(z|z_0;q)
   &= 
   \mean{
     J_N^{q/2}(M^{-1},z) 
     \,
     \deriv{\mathcal{M}^{-1}(z)}{z}\,
     P_{n-1}\left( \mathcal{M}^{-1}(z)|z_0;q\right)
   }_M
   \\
   &=
   \mean{
     J_N^{1+q/2}(M^{-1},z) 
     \,
     P_{n-1}\left( \mathcal{M}^{-1}(z)|z_0;q\right)
   }_M
\end{align}
We recognize the action of the operator $\mathscr{T}_M(q)$ introduced in Eq.~\eqref{eq:Repres4} or \eqref{eq:Eq2.65}, for $J=J_N$.

For $q=0$, the propagator
\begin{equation}
    P_n(z|z_0;0) = \mean{\delta(z-z_n)} = f_n(z)
\end{equation}
coincides with the distribution of $z_n$, and thus we recover 
\eqref{eq:DistribRecursion}, as it should.



\subsection{Backward evolution}

In the study of functionals of stochastic processes, the use of backward equations is often more convenient, see for instance Ref.~\cite{Maj05} (the study of Brownian functionals is usually performed with the backward equations in the mathematical literature \cite{Zus10,ComMey15}). Let us follow here also the same strategy and define a backward evolution for $Q_n(z_0;q) $.

The backward evolution can be used to generate the product of Jacobians.
We apply the adjoint operator \eqref{eq:Repres4adjoint} on the flat function $1(z)\equiv1$~:
\begin{align}
  &\big[
     \mathscr{T}_{M_n}^\dagger(q) 1
  \big](z_{n-1}) 
  =
  J^{-q/2}(M_n,z_{n-1}) 
  \\
  &\big[
     \mathscr{T}_{M_{n-1}}^\dagger(q)\mathscr{T}_{M_n}^\dagger(q) 1
  \big](z_{n-2}) 
  =
  J^{-q/2}(M_{n-1},z_{n-2}) \,  J^{-q/2}(M_n,\underbrace{  \mathcal{M}_{n-1}(z_{n-2}) }_{\equiv z_{n-1}} ) 
  \nonumber
  \\[-0.5cm]
  \nonumber
  &\hspace{1cm} \vdots \hspace{1cm}  \vdots \hspace{1cm} \vdots 
  \\
  &\big[
     \mathscr{T}_{M_{1}}^\dagger(q)\cdots\mathscr{T}_{M_{n-1}}^\dagger(q)\mathscr{T}_{M_n}^\dagger(q) 1
  \big](z_{0}) 
  = 
  \prod_{i=1}^n J^{-q/2}(M_i,z_{i-1}) 
  \:.
\end{align}
with  $z_i=\mathcal{M}_i\big(z_{i-1}\big)=\mathcal{M}_i\big(\cdots\mathcal{M}_{2}\big(\mathcal{M}_1\big(z_{0}\big)\big)\cdots\big)$.

This allows to write the generating function in the convenient form
\begin{equation}
  \label{eq:QnBackward}
    Q_n(z_0;q) 
  = \mean{ \big[
     \mathscr{T}_{M_{1}}^\dagger(q)\cdots\mathscr{T}_{M_{n-1}}^\dagger(q)\mathscr{T}_{M_n}^\dagger(q) 1
  \big](z_{0})  }
  \hspace{1cm}\mbox{for }
  J=J_K
  \:,
\end{equation}
This avoids the introduction of the propagator.



\subsection{The averaged representation and the spectral problem}
\label{subsec:ARSP}

The forward evolution \eqref{eq:ForwardEvolution} can be rewritten more formally and iterated   
\begin{align}
  P_n(z|z_0;q)
  &= \mean{
    \left[\mathscr{T}_{M_n}(q)P_{n-1}\right](z|z_0;q)
  }_{M_n}
  \\
  &= \mean{
    \left[\mathscr{T}_{M_n}(q)\cdots\mathscr{T}_{M_1}(q)P_{0}\right](z|z_0;q)
  }_{M_1,\cdots,M_n}
\end{align}
for $J=J_N$.
Making use of the statistical independence of the matrices we have
\begin{equation}
     \mean{
     \mathscr{T}_{M_n}(q) \cdots \mathscr{T}_{M_1}(q) 
      }_{M_1,\cdots,M_n}
  =
  \left[ \overline{\mathscr{T}}(q) \right]^n
  \hspace{0.5cm}\mbox{where } 
  \overline{\mathscr{T}}(q) \eqdef
  \mean{
   \mathscr{T}_{M}(q)
  }_{M}  
  \:.
\end{equation}
At this stage it is useful to rewrite the propagator with notations similar to those of quantum mechanics~:
\begin{equation}
  \label{eq:RightEvolutionPropagator}
   P_n(z|z_0;q)
   =\bra{z} 
      \overline{\mathscr{T}}(q) ^n
   \ket{z_0}  
   \hspace{1cm}\mbox{for }
   J=J_N
   \:,
\end{equation}
which makes clear that the large $n$ behaviour of the propagator is controlled by \textit{largest} eigenvalue of the average operator $\overline{\mathscr{T}}(q)$, denoted for later convenience $1/\lambda(q)$.

On the other hand, \eqref{eq:QnBackward} leads us to introduce the average of the adjoint operator
\begin{equation}
  \overline{\mathscr{T}}^\dagger(q) \eqdef
  \big\langle 
     \mathscr{T}_{M}^\dagger(q)
  \big\rangle_{M}  
  \:,
\end{equation}
hence
\begin{equation}
    Q_n(z_0;q) 
    = \Big[ \overline{\mathscr{T}}^\dagger(q)^n\, 1 \Big](z_0)
   \hspace{1cm}\mbox{for }
   J=J_K
  \:.
\end{equation}

The fact that the generating function involves different realizations for two choices of Jacobian $J_N$ or $J_K$ can be understood from the relation \eqref{eq:TransformationJNtoJ}.
Averaging of this latter equation gives, reintroducing the label $^{(\rho)}$ for a moment,
\begin{equation}
  \label{eq:TransformationTbar}
  \overline{\mathscr{T}}^{(\rho)}(q)
   = \rho(z)^{-q/2} \, \overline{\mathscr{T}}^{(\rho_N)}(q) \, \rho(z)^{q/2} 
\end{equation}
and shows that the spectrum of eigenvalues of the tranfer operator is independent of the choice of Jacobian $J$.
This is a crucial observation, which will be used below in order to simplify the spectral problem.
We consider the spectral problem for arbitrary choice of Jacobian $J$~:
\begin{equation}
  \label{eq:TheSpectralPb}
  \boxed{
  \left[\overline{\mathscr{T}}(q) \Phi^\mathrm{R}_q\right](z) 
  = \frac{1}{\lambda(q)} \Phi^\mathrm{R}_q(z)
  \hspace{0.5cm}\mbox{and}\hspace{0.5cm}
  \left[\overline{\mathscr{T}}^\dagger(q) \Phi^\mathrm{L}_q\right](z) 
  = \frac{1}{\lambda(q)} \Phi^\mathrm{L}_q(z)
  }
\end{equation}
where $\Phi^\mathrm{R}_q(z)$ and $\Phi^\mathrm{L}_q(z)$ are the two related right and left eigenvectors satisfying $\int\D z\,\Phi^\mathrm{L}_q(z)\Phi^\mathrm{R}_q(z)=1$. 
In \eqref{eq:TheSpectralPb}, $1/\lambda(q)$ is the \textit{largest eigenvalue} of the operator $\overline{\mathscr{T}}(q)$.
As shown in Refs.~\cite{Tut65,BouLac85} this eigenvalue is \textit{isolated}, which is a crucial remark here~:
provided $q$ is sufficiently close to zero, the transfer operator takes the form 
$\overline{\mathscr{T}}(q)=\lambda(q)^{-1}\,\mathscr{N}(q) + \mathscr{Q}(q)$ where $\mathscr{N}(q)$ is a rank-one projection such that $\mathscr{N}(q)\mathscr{Q}(q)=\mathscr{Q}(q)\mathscr{N}(q)=0$ and the spectral radius of $\mathscr{Q}(q)$ is smaller than $\lambda(q)^{-1}$
(see chapter 5 of \cite{BouLac85} for a detailed discussion of this question for the linear group $\mathrm{GL}(n,\mathbb{R})$, or \cite{BurMen16} for a summary of the required conditions). 
The spectral problem \eqref{eq:TheSpectralPb} is precisely defined by some \textit{boundary conditions}.
Using basic properties of representation theory \cite{GelGraVil66}, it is possible to show (Appendix~\ref{app:RepresentationTheory}) that the invariant subspaces of functions defined on the projective line consist of functions such that the two limits
\begin{equation}
  \label{eq:AsymptoticBehaviourPhiR}
  \boxed{
  \lim_{z\to-\infty} (-z)^{\eta}\, \Phi^\mathrm{R}_q(z)
  \hspace{0.25cm}\mbox{and}\hspace{0.25cm}
  \lim_{z\to+\infty} z^{\eta}\, \Phi^\mathrm{R}_q(z)
  \hspace{0.25cm}\mbox{exist and are equal}
}
\end{equation}
The exponent $\eta$, called the ``degree'' of the representation, depends on $q$ and on the choice of Jacobian $J$.
In particular, we show in Appendix~\ref{app:RepresentationTheory} that $\eta=2+q$ for $J=J_N$.
The statement \eqref{eq:AsymptoticBehaviourPhiR} is equivalent to say that the eigenvector behaves as 
$\Phi^\mathrm{R}_q(z)  \simeq  \mathcal{A}_q\, |z|^{-\eta}$ at infinity, 
the crucial point being that the power law decays at $z\to+\infty$ and $z\to-\infty$ are controlled by the \textit{same} coefficient $\mathcal{A}_q$ (which may vanish).
For $J\neq J_N$, this observation further relies on the asumption that the density $\rho(z)$ entering the definition of the Jacobian $J$ is a symmetric function. 
Note that the left eigenvector presents the asymptotic behaviour $\Phi^\mathrm{L}_q(z)  \sim  |z|^{\eta-2}$.
For more details, cf. the discussion in Appendix~\ref{app:RepresentationTheory} where the exponent $\eta$ is related to the properties of the representation.

%

Because the largest eigenvalue of the transfer operator is isolated, it dominates the large $n$ limit of the propagator, hence we can write
\begin{equation}
  P_n(z|z_0;q) \underset{n\to\infty}{\simeq}
  \lambda(q)^{-n}\:\Phi^\mathrm{R}_q(z)\Phi^\mathrm{L}_q(z_0)
\end{equation}
where $\Phi^\mathrm{R}_q(z)$ and $\Phi^\mathrm{L}_q(z)$ solve the spectral problem for $J=J_N$.
Thus, the $n$-dependence of the generating function is
 \begin{equation}
   Q_n(z_0;q) \underset{n\to\infty}{\sim} \lambda(q)^{-n}
 \end{equation}
and the GLE 
\eqref{eq:DefGLE} is deduced from the solution of the spectral problem \eqref{eq:TheSpectralPb}~:
\begin{equation}
  \boxed{
  \widetilde\Lambda(q) = -\ln\lambda(q)
  }
\end{equation}
One has thus to determine the largest eigenvalue $1/\lambda(q)$ of the operator $\overline{\mathscr{T}}(q)$.


\subsection{Illustration 1 : products of matrices $KN$ (Frisch-Lloyd model)}
\label{subsec:FL1}

We make explicit the spectral problem \eqref{eq:TheSpectralPb} by considering the case of matrices of the form \eqref{eq:MatricesSchrodinger}. 
The connection with the quantum localisation problem requires to consider either fixed or exponentially distributed angles $\theta_n$'s. 
We make the second choice~: $\mathrm{Proba}\{\theta_n>\theta\}=\EXP{-\rho\theta}$. 
We consider an arbitrary distribution $p(u)$ for the $u_n$'s.
For simplicity we first set $k=1$, and reintroduce it afterwards.
The averaged operator is
\begin{equation}
  \label{eq:AveragedTFrischLloy}
  \overline{\mathscr{T}}(q) 
  = \mean{ \EXP{\theta\,\mathscr{D}_K(q)} \, \EXP{u\,\mathscr{D}_N(q)} }
  = \mean{\EXP{\theta\,\mathscr{D}_K(q)} }_\theta \,  \mean{\EXP{u\,\mathscr{D}_N(q)}}_u
\end{equation}
At this point, we can use the freedom on the choice of measure (i.e. Jacobian) to simplify the calculations. 
The distribution of the angles being exponential, i.e. $\mean{\EXP{-r\,\theta} }_\theta=1/(1+r/\rho)$, the part
 $\mean{\EXP{\theta\,\mathscr{D}_K(q)} }_\theta$ takes a simple form.
Thus it is more advantageous to simplify $\mathscr{D}_N(q)$ by choosing $J\equiv J_N$ with $h_N=0$ (cf. Table~\ref{tab:InfGenerators}). 
Then 
\begin{equation}
  \label{eq:InfGenFrischLloyd}
 \mathscr{D}_K(q) = \deriv{}{z}(1+z^2) + q\, z
 \hspace{0.5cm}\mbox{and}\hspace{0.5cm}
 \mathscr{D}_N(q)=\mathscr{D}_N = -\deriv{}{z}
 \end{equation} 
and we have
\begin{equation}
   \overline{\mathscr{T}}(q) 
  = \frac{1}{1-\rho^{-1}\mathscr{D}_K(q)}\,\mean{\EXP{u\,\mathscr{D}_N}}_u
\end{equation}
Note that $\mathscr{D}_N$ generates translations on the infinite line and has a purely imaginary spectrum, hence $\mean{\EXP{u\,\mathscr{D}_N}}_u$ is well defined.
Similarly, we show in Appendix~\ref{app:DiagDK} that $\mathscr{D}_K$ and $\mathscr{D}_K(q)$ are of the same nature and have the same purely imaginary (discrete) spectrum of eigenvalues. Thus $1-\rho^{-1}\mathscr{D}_K(q)$ is invertible.
With \eqref{eq:AveragedTFrischLloy}, we can rewrite Eq.~\eqref{eq:TheSpectralPb} under the form
\begin{equation}
  \label{eq:SPFrischLloyd0}
     \left[ 1 - \rho^{-1}\mathscr{D}_K(q)  \right] \Phi^\mathrm{R}_q(z)
  = \lambda(q) \,\mean{\EXP{u\,\mathscr{D}_N}}_u  \Phi^\mathrm{R}_q(z)
  \:.
\end{equation}
Using that $\EXP{u\,\mathscr{D}_N}=\EXP{-u\deriv{}{z}}$ is the translation operator, elementary manipulations lead to 
\begin{equation}
  \label{eq:SPFrischLloyd}
  \left[ \deriv{}{z}(k^2+z^2) + q\, z \right]\Phi^\mathrm{R}_q(z)
 + \rho\left[ \lambda(q)\mean{\Phi^\mathrm{R}_q(z-v)}_v  - \Phi^\mathrm{R}_q(z)\right]
  =0
  \:.
\end{equation}
For convenience for the following discussions, we have reintroduced~\footnote{
  The parameter $k$ can be reintroduced by performing the substitutions
  $z\to z/k$, $\rho\to\rho/k$ and $u\to v/k$.
} 
the parameter $k$, defined in Section~\ref{sec:Intro} after Eq.~\eqref{eq:Schrodinger} (so that $\mean{\cdots}_u\to\mean{\cdots}_v$, as $u_n=v_n/k$). 
We postpone to a later section the resolution of this equation.

We discuss the asymptotic behaviour of the solution.
The $z\to\infty$ limit of Eq.~\eqref{eq:SPFrischLloyd} gives 
\begin{equation}
  \left[ \deriv{}{z}z^2 + q\, z \right]\Phi^\mathrm{R}_q(z) \simeq 0
  \:.
\end{equation}
Combined with the general property \eqref{eq:AsymptoticBehaviourPhiR} we conclude that
\begin{equation}
  \label{eq:AsymptoticBehaviourForPhiR}
  \Phi^\mathrm{R}_q(z) \simeq \frac{\mathcal{A}_q}{|z|^{2+q}}
  \hspace{1cm}\mbox{for }
  z \to \pm\infty
  \:.
\end{equation}
This condition was conjectured and verified numerically in Ref.~\cite{FyoLeDRosTex18}.
In particular, for $q=0$, the eigenvector coincides with the invariant density $\Phi^\mathrm{R}_0(z)= f(z)$ and the relation corresponds to the well-known Rice formula \cite{Kot76}
\begin{equation}
  \label{eq:RiceFormula}
  \lim_{z\to-\infty}z^2f(z) = \lim_{z\to+\infty}z^2f(z) = \IDoS 
\end{equation}
expressing that $f(z)$ is a stationary distribution with constant current $-\IDoS$, where $\IDoS$~coincides with the integrated density of states (IDoS) per unit length of the quantum model~\eqref{eq:Schrodinger} \cite{ComTexTou10}.

\subsection{Illustration 2~: products of matrices $\widetilde{K}A$ (Dirac or supersymmetric case)}
\label{subsec:SuSy1}

Another type of random matrices considered in the literature are 
\begin{equation}
  \label{eq:MatricesSusy2}
  M_n = 
  \begin{pmatrix}
    \cosh\theta_n & \sinh\theta_n \\ 
    \sinh\theta_n & \cosh\theta_n
  \end{pmatrix}
  \:
  \begin{pmatrix}
    \EXP{w_n} & 0 \\ 0 & \EXP{-w_n}
  \end{pmatrix}
  \:.
\end{equation}
Setting $\theta_n=k\ell_n$, one can show that they are tranfer matrices for the Dirac equation with a mass made of $\delta$-peaks \cite{RamTex14}
\begin{equation}
  \label{eq:Dirac}
  \big[\I\sigma_2\partial_x+\sigma_1\mass(x)\big]\Psi(x)=\varepsilon\,\Psi(x)
  \hspace{0.5cm}
  \mbox{where } 
  \mass(x)=\sum_nw_n\delta(x-x_n)
  \:,
\end{equation}
for imaginary energy~\footnote{
the case of matrices $KA$, i.e. when the first matrix in \eqref{eq:MatricesSusy2} is replaced by a rotation, corresponds to a real energy~\cite{RamTex14}.} 
$\varepsilon=\I k\in\I\mathbb{R}$.
Here, $\Psi(x)$ is a bi-spinor and $\sigma_i$ the Pauli matrices.
Equivalently these matrices are transfer matrices for the so-called supersymmetric Schr\"odinger equation~\cite{ComTexTou10} (for a review on supersymmetric quantum mechanics with disorder, see Ref.~\cite{ComTex98}). 
The matrices \eqref{eq:MatricesSusy2} are also known as transfer matrices for Ising spin chain in random field \cite{DerHil83}, a problem which has attracted a lot of attention (we will come back to this point in \S~\ref{sec:ContinuumSusy}).

We now discuss the form taken by the spectral problem \eqref{eq:TheSpectralPb} in case of matrices \eqref{eq:MatricesSusy2}, with exponentially distributed angles $\theta_n$'s and with arbitrary weight distribution $p(w)$.
The choice $J=J_A$ simplifies the operator 
$\mean{\EXP{w\,\mathscr{D}_A(q)}}_w=\mean{\EXP{w\,\mathscr{D}_A}}_w$ (cf. Table~\ref{tab:InfGenerators}), 
and thus makes the calculations more simple.
Infinitesimal generators are then 
\begin{equation}
 \mathscr{D}_{\widetilde{K}}(q) = \deriv{}{z}(-1+z^2) + \frac{q}{2}\,\left(z+\frac{1}{z}\right)
 \hspace{0.5cm}\mbox{and}\hspace{0.5cm}
 \mathscr{D}_A(q)=\mathscr{D}_A = -2\deriv{}{z}z
\end{equation} 
and the spectral problem now takes the form
\begin{equation}
  \label{eq:SPsusy0}
     \left[ 1 - \rho^{-1}\mathscr{D}_{\widetilde{K}}(q)  \right] \Phi^\mathrm{R}_q(z)
  = \lambda(q) \,\mean{\EXP{w\,\mathscr{D}_A}}_w  \Phi^\mathrm{R}_q(z)
  \:.
\end{equation}
The operator
~\footnote{
   Eq.~\eqref{eq:AnOperator} is proven by using Mellin transform, defined by \eqref{eq:DefMellin}.
   First we remark that $-\deriv{}{z}\big[z\varphi(z)\big]$ is transformed into
   $s\,\check{\varphi}(s)$, thus $\EXP{-2w\,\deriv{}{z}z} \varphi(z)\to\EXP{2ws}\check{\varphi}(s)$.
   Second we use that the dilatation $\lambda\,\varphi(\lambda z)$ takes the form $\lambda^{-s}\check{\varphi}(s)$ and choose $\lambda=\EXP{-2w}$.
   \textsc{Qed.}
}
\begin{equation}
  \label{eq:AnOperator}
  \EXP{w\,\mathscr{D}_A} \varphi(z)
  =  \EXP{-2w\,\deriv{}{z}z} \varphi(z)
  = \EXP{-2w}\, \varphi(z\EXP{-2w})
\end{equation}
is a dilatation.
Thus, we can rewrite \eqref{eq:SPsusy0} as 
\begin{align}
   \label{eq:SPsusy}
   &\left[
     \deriv{}{z}(-k^2+z^2) + \frac{q}{2}\,\left(z+\frac{k^2}{z}\right)
   \right]\Phi^\mathrm{R}_q(z)
   \nonumber\\
   &\hspace{1.5cm}
   + \rho\left[ \lambda(q)\mean{\EXP{-2w}\Phi^\mathrm{R}_q(z\EXP{-2w})}_w  - \Phi^\mathrm{R}_q(z)\right]
   =0 
\end{align}
where $k$ was reintroduced.~\footnote{
  the parameter $k$ is reintroduced thanks to $z\to z/k$ and $\rho\to\rho/k$.}
For $q=0$, we check that we recover the equation for the invariant measure given in \cite{BieTex08}.  
We will see below how to simplify the analysis by making use of an appropriate integral transform.

The spectral problem \eqref{eq:SPsusy} is supplemented by the boundary condition 
\begin{equation}
  \Phi^\mathrm{R}_q(z) \simeq \frac{\mathcal{A}_q}{|z|^{2+q/2}}
  \hspace{1cm}\mbox{for }
  z \to \pm\infty
  \:.
\end{equation}
The exponent is obtained by large $z$ analysis of \eqref{eq:SPsusy}. 
The coefficient $\mathcal{A}_q$ might vanish.


\section{Integral transforms}
\label{sec:Transformations}

The role of integral transforms (Fourier, Mellin or Hilbert) for the calculation of the Lyapunov exponent was emphasized in Refs.~\cite{Luc04,ComTexTou11,ComLucTexTou13,GraTexTou14}.  
Group theoretical methods provide a unified approach to the study of special functions and integral transforms \cite{Vil78}.
Fourier, Laplace and Mellin transforms are connected with certain group representations.
The representation space is constructed from a basis of vectors which diagonalizes a generator of a subgroup.
We will use a similar idea in our study of the transfer operator.



\subsection{Matrices $KN$ or $\widetilde{K}N$ (Frisch \& Lloyd case) : Fourier transform}
\label{subsec:IntTransfSchrod}

We now simplify the spectral problem \eqref{eq:SPFrischLloyd}.
Because the disorder is on the weights $v_n=k\,u_n$, we have seen that the choice of Jacobian $J=J_N$ simplifies the action of the operator $\mathscr{D}_N(q)$ as $h_N=0$.
Furthermore, the operator $\mathscr{D}_N(q)=\mathscr{D}_N$ can be diagonalised by performing a Fourier transform 
\begin{equation}
  \hat{\varphi}(s) = \int_{-\infty}^{+\infty}\D z\, \varphi(z)\, \EXP{-\I sz}
  \:.
\end{equation}

\subsubsection{Preliminary~: L\'evy exponent}

Following \S~\ref{subsec:FL1}, we consider the case where angles $\theta_n$ have an exponential distribution.
This is related to the disordered model \eqref{eq:Schrodinger} for impurities at random and independent positions for a uniform mean density $\rho$.
It is convenient to introduce the process $Y(x)=\int_0^x\D t\, V(t)$ related to the random potential $V(x)=\sum_n v_n\,\delta(x-x_n)$, where the impurity positions are now ordered, $x_1<x_2<\cdots$, with distance $\ell_n=x_{n+1}-x_n$ exponentially distributed. 
The process thus exhibits random jumps $Y(x_n^+)=Y(x_n^-)+v_n$ at random \og times \fg{}, occuring with the \og rate \fg{} $\rho$, and is thus Markovian. It can be characterized by the L\'evy exponent $\levy(s)$, defined by 
\begin{equation}
  \mean{\EXP{-\I s Y(x)}} 
  = \EXP{-x\,\levy(s)}
  \:.
\end{equation}
Denoting by $p(v)$ the distribution of the weights $v_n$'s, it is expressed as
\begin{equation}
  \label{eq:LECPP}
  \levy(s) = \rho \left[ 1 - \hat{p}(s) \right] 
  \:,
  \hspace{1cm}\mbox{where }
  \hat{p}(s)=\int\D v\,p(v)\,\EXP{-\I sv}
\end{equation}
is the Fourier transform of the weight distribution $p(v)$.
The process $Y(x)$ is known as a \og compound Poisson process \fg{}  (see \cite{ComTexTou11,GraTexTou14}). 
Such process corresponds to a subclass of L\'evy processes \cite{App04}.
However, because more general L\'evy processes can be obtained by considering appropriate limits of compound Poisson processes, we conclude that 
the results of the sections~\ref{subsec:IntTransfSchrod}, \ref{sec:PeturbFL} and \ref{sec:GLElocalisation} apply to the case where  $Y(x)=\int_0^x\D t\, V(t)$ is \textit{any} L\'evy process (see \cite{ComTexTou11,GraTexTou14} for other examples of L\'evy processes, or the monograph \cite{App04}). 
Another simple case is the Gaussian white noise
$
  \label{eq:LEGN}
  \levy(s)  = \frac{\sigma}{2}s^2 
$,
discussed below as a certain continuum limit of the compound Poisson process.

\subsubsection{New form of the spectral problem}
 
Action of the infinitesimal generators on the Fourier transform is 
\begin{equation}
  \widehat{\mathscr{D}}_j(q)=\I s\,g_j\big(\I\deriv{}{s}\big) + q\,h_j\big(\I\deriv{}{s}\big)
  \hspace{0.5cm}
  \mbox{for }
   j\in\{ K, \,  A, \, N  \}
  \:,
\end{equation}
 or explicitly
\begin{align}
  \widehat{\mathscr{D}}_K(q) 
  = \I s \left( 1 -\deriv{^2}{s^2} \right) + \I\, q\, \deriv{}{s}
  \:,
  \hspace{0.5cm}
  \widehat{\mathscr{D}}_A(q) = 2s\,\deriv{}{s} - q
  \:,
  \hspace{0.5cm}
  \widehat{\mathscr{D}}_N(q) 
  = -\I\,s
  \:.
\end{align}
The Fourier transformation has allowed to diagonalize the operator $\mean{\EXP{u\,\mathscr{D}_N}}_u$ involved in \eqref{eq:SPFrischLloyd0}, whose action simply becomes the multiplication by the Fourier transform of the weight distribution $\hat{p}(s)=\mean{\EXP{-\I us}}_u$.
Reintroducing the parameter $k$ in Eq.~\eqref{eq:SPFrischLloyd0}, or Fourier transforming Eq.~\eqref{eq:SPFrischLloyd}, the spectral problem takes the form
\begin{equation}
  \label{eq:SPFrischLloydFourier}
  \boxed{
  \I s \left[ -\deriv{^2}{s^2} + k^2 - \frac{\levy(s)}{\I s} \right] 
  \widehat{\Phi}^\mathrm{R}_q(s)
  =
  \left[ \rho\,\hat{p}(s)\left[1-\lambda(q)\right] -\I q\deriv{}{s}\right] 
  \widehat{\Phi}^\mathrm{R}_q(s)
  }
\end{equation}
where the L\'evy exponent \eqref{eq:LECPP} characterizes the disorder potential.
The boundary conditions \eqref{eq:AsymptoticBehaviourForPhiR} imply the behaviour
\begin{equation}
\label{eq:BCForPhiRFourier}
  \widehat{\Phi}^\mathrm{R}_q(s)
   \simeq \widehat{\Phi}^\mathrm{R}_q(0) + \omega_q\, |s|^{q+1} +\mbox{ regular (analytic) terms}
   \hspace{0.5cm}\mbox{for }
   s\to0
\end{equation}
 (see the Subsection~\ref{subsec:PhiRSmallSbehaviour} of Appendix~\ref{app:SmallSbehaviours}).
The coefficient $\omega_q$ is proportional to $\mathcal{A}_q$, i.e. must be \textit{real}~; in Ref.~\cite{Tex19b}, the spectral problem is studied in the high energy limit.
This study shows that imposing that $\omega_q$ is real provides the secular equation allowing to determine the eigenvalue $1/\lambda(q)$.

Equation \eqref{eq:SPFrischLloydFourier} will be analysed below by a perturbative approach by expanding $\lambda(q)$ in powers of~$q$.


\subsection{Matrices $\widetilde{K}A$ (supersymmetric case) : Mellin transform}
\label{subsec:MellinTransfForSusy}

We now reformulate the spectral problem \eqref{eq:SPsusy}.
Following the same logic as in the previous subsection, we now aim to diagonalize the operator $\mean{\EXP{w\,\mathscr{D}_A}}_w$. 
The appropriate transform is now the Mellin transform
\begin{equation}
  \label{eq:DefMellin}
  \check{\varphi}(s) = \int_0^\infty\D z\, z^s \, \varphi(z)
  \:,
\end{equation}
as already pointed out in \cite{ComTexTou11}. 
In this case the stationary distribution $f(z)=\Phi^\mathrm{R}_0(z)$ has its support on $\mathbb{R}_+$ \cite{ComTexTou11},~\footnote{
  The fact that $f(z)$ has support in $\mathbb{R}_+$ can be understood as follows~:
the action of matrices \eqref{eq:MatricesSusy2} in the projective line can be analysed by considering the (Riccati) stochastic process $z(x)$ corresponding to the diffusion operator 
$\smean{\EXP{\theta\mathscr{D}_{\widetilde{K}}}}_\theta\smean{\EXP{w\mathscr{D}_{A}}}_w$.
  The action of the matrix $A(w_n)$ corresponds to $z(x_{n}^+)=z(x_{n}^-)\,\EXP{2w_n}$ while the action of $\widetilde{K}(\theta_n)$ to the evolution $\deriv{}{x}z(x)=k^2-z(x)^2$ for $x\in]x_{n},x_{n+1}[$, with $\theta_n=k(x_{n+1}-x_{n})$.
  $z=k$ is a fixed point of the free evolution, hence $z(x)\in[k,\infty[$ when $w_n>0$, and $z(x)\in[0,k]$ when $w_n<0$ (see \S~4.3.2 of \cite{ComTexTou10}). If $w_n$'s have random signs, the random process belongs to $\mathbb{R}_+$, which is thus the support of the distribution~$f$.
}
Hence the definition \eqref{eq:DefMellin} 
(for matrices $KA$, the support of $f$ is $\mathbb{R}$ and one has to introduce a pair of Mellin transforms \cite{ComTexTou19}). 
The action of the infinitesimal generators are now
\begin{align}
  &\widecheck{\mathscr{D}}_{\widetilde{K}}(q) \check{\varphi}(s)
  = \left( s + \frac{q}{2} \right) \check{\varphi}(s-1)
  + \left(-s + \frac{q}{2} \right) \check{\varphi}(s+1)
  \:,
  \\
  &\widecheck{\mathscr{D}}_{A} \check{\varphi}(s) = 2s\,\check{\varphi}(s)
  \hspace{0.5cm}\mbox{and}\hspace{0.25cm}
  \widecheck{\mathscr{D}}_{N}(q) \check{\varphi}(s) 
  = \left( s + \frac{q}{2} \right) \check{\varphi}(s-1)
  \:.
\end{align}
The operator $\mean{\EXP{w\,\mathscr{D}_A}}_w$ is thus replaced by the multiplication by $\tilde{p}(-2s)$, the Laplace transform of the weight distribution 
\begin{equation}
  \tilde{p}(r) = \int\D w\, p(w) \, \EXP{-rw} = \hat{p}(-\I r)
  \:,
\end{equation}
where $\hat{p}(s)$ is the Fourier transform defined above.
 
Hence  \eqref{eq:SPsusy} is now replaced by the finite difference equation
\begin{align}
  &s\left[
    -\widecheck{\Phi}^\mathrm{R}_q(s+1) + k^2\, \widecheck{\Phi}^\mathrm{R}_q(s-1)
    -\frac{\levy(2\I s)}{s}\widecheck{\Phi}^\mathrm{R}_q(s)
  \right]
  \nonumber
  \\
  &
  \hspace{1cm}
  = \rho\, \hat{p}(2\I s)\,\left[ 1-\lambda(q) \right]\widecheck{\Phi}^\mathrm{R}_q(s)
  -\frac{q}{2}\left[
    \widecheck{\Phi}^\mathrm{R}_q(s+1) + k^2\, \widecheck{\Phi}^\mathrm{R}_q(s-1)
  \right]
  \:,
  \label{eq:SPsusyMellin}
\end{align}
where $k$ was reintroduced.  
The distribution of the weights $w_n$ must be chosen in such a way that $\mean{\EXP{2sw_n}}$ exists 
(if not, the phenomenon of \og  superlocalisation  \fg{} occurs, cf. \cite{BieTex08} and references therein).
The equation is written in a form appropriate for perturbative expansion in $q$, similarly to Eq.~\eqref{eq:SPFrischLloydFourier}.

The equation will be further discussed in the next section.

\section{Perturbative approach and a formula for the variance $\widetilde{\Lambda}''(0)$ }
\label{sec:Perturbation}

Making use of the simplifications introduced in Section~\ref{sec:Transformations}, our aim is now to solve the central spectral problem \eqref{eq:TheSpectralPb}.  
We will use a perturbative approach, providing a systematic method for the calculation of the moments of the logarithm of the matrix product.
The method is similar to the perturbative method used in Ref.~\cite{SchTit02}, which has considered a stochastic problem related to the continuum limit of random matrix products of type $KN$.
Here, the perturbative approach is not restricted to the continuum limit. 
We will mostly focus on the case of matrices of type $KN$ or $\widetilde{K}N$, i.e. analyse Eq.~\eqref{eq:SPFrischLloydFourier}, although the method can be applied to other types of matrices in principle (we will also briefly discuss the case of matrices of type $\widetilde{K}A$, i.e. Eq.~\eqref{eq:SPsusyMellin}).

\subsection{Products of matrices $KN$ and $\widetilde{K}N$ (Frisch-Lloyd model)}
\label{sec:PeturbFL}

Our starting point is the equation \eqref{eq:SPFrischLloydFourier}, which we solve recursively by a perturbative method in the parameter $q$.
We expand the eigenvalue (moment generating function) as 
\begin{equation}
  \label{eq:ExpansionSmallLambda}
  \lambda(q) 
  = \sum_{n=0}^\infty\frac{(-1)^n}{n!}\,\lambda_n\,q^n
  = 1 - \lambda_1\, q + \frac{1}{2}\lambda_2\,q^2 + \cdots
  \:,
\end{equation}
 where $\lambda_1$ is the Lyapunov exponent,
and the corresponding eigenvector
\begin{equation}
  \label{eq:ExpansionPhiR}
  \widehat{\Phi}^\mathrm{R}_q(s) = \widehat{R}_0(s) + q\,\widehat{R}_1(s) + q^2\,\widehat{R}_2(s) + \cdots
\end{equation}
We now discuss the equations corresponding to each order.
Perturbative expansions for the spectrum of Hermitian and non Hermitian operators have been discussed in Ref.~\cite{VisLyu60}.
Here we assume that such an expansion exists for sufficiently small~$q$.

\subsubsection{Order $q^0$~: Green's function and invariant density}

Order $q^0$ term of \eqref{eq:SPFrischLloydFourier} is 
\begin{equation}
  \label{eq:7.3}
  \I s\left[ -\deriv{^2}{s^2} + k^2 - \frac{\levy(s)}{\I s}  \right] \widehat{R}_0(s)=0
  \hspace{0.25cm}\Rightarrow\hspace{0.25cm}
  \left[ -\deriv{^2}{s^2} + k^2 - \frac{\levy(s)}{\I s}  \right] \widehat{R}_0(s)
  = c_0\,\delta(s)
\end{equation}
where $c_0$ is some constant to be determined.
Let us introduce the recessive solution $\varphi(s)$ of the homogeneous equation 
(i.e. $\varphi(s)\to0$ for $s\to+\infty$)
\begin{equation}
  \label{eq:HomogeneousDE}
  \boxed{
  \left[ -\deriv{^2}{s^2} + E - \frac{\levy(s)}{\I s}  \right]  \varphi(s) = 0
  }
\end{equation}
where $E=k^2$.
Because $\levy(-s)^*=\levy(s)$, the function $\varphi(-s)^*$ is also solution of the differential equation. 
If it differs from $\varphi(s)$, it is the independent solution vanishing at $-\infty$.

Two examples discussed in the paper are~:
\begin{itemize}
\item
  The compound Poisson process \eqref{eq:LECPP} with exponentially distributed weights,
  $\levy(s)=\I\rho \bv s/(1+\I s\bv)$ where $\mean{v_n}=\bv$. 
  For $E=+k^2$, the solution of Eq.~\eqref{eq:HomogeneousDE} is~\cite{ComTexTou10,GraTexTou14} 
  \begin{equation}
  \label{eq:PhiFrischLloyd}
  \varphi(s) = W_{-\frac{\I\rho}{2k},\frac12}(2k(s-\I/\bv))
  \:,
  \end{equation}
  where $W_{\lambda,\mu}(z)$ is the Whittaker function \cite{gragra}.
  Here $\varphi(-s)^*$ is an independent solution of the differential equation.
  
  For $E=-k^2$, we have
  \begin{equation}
  \label{eq:PhiFrischLloydNegEnergy}
  \varphi(s) = W_{-\frac{\rho}{2k},\frac12}(2k(\I s+1/\bv))
  \:.
  \end{equation}
  In this case we have $\varphi(-s)^*=\varphi(s)$. The second independent solution involves the Whittaker function $M_{\lambda,\mu}(z)$.
  
\item  
  The Gaussian white noise $\levy(s)  = \frac{\sigma}{2}s^2$.
  The solution of Eq.~\eqref{eq:HomogeneousDE} is then
  \begin{equation}
  \label{eq:PhiHalperin}
    \varphi(s) 
    = \mathrm{Ai}\left(-E-\I\sigma s/2\right) - \I\, \mathrm{Bi}\left(-E-\I\sigma s/2\right)
    \:.
\end{equation}
\end{itemize}


The Gaussian white noise is rather special as it can be achieved for $\rho\to\infty$ with vanishing weights, leading to a L\'evy exponent growing at infinity.
In general, for finite $\rho$, we have $\levy(s)\simeq\rho$ for $s\to\pm\infty$. 
As a result, considering positive energy $E=+k^2$, the solutions of \eqref{eq:HomogeneousDE} display the asymptotic behaviours $s^{\pm\I\rho/(2k)}\EXP{\pm ks}$.
The recessive solution thus behaves as $\varphi(s)\sim s^{-\I\rho/(2k)}\EXP{-ks}$, and differs from $\varphi(-s)^*$ which vanishes for $s\to-\infty$. We conclude that $\varphi(s)$ blows up for $s\to-\infty$.
The Wronskian of the two solutions 
\begin{equation}
  \label{eq:Wronskian}
  \mathcal{W} \equiv \mathcal{W}\left[ \varphi(s) \, , \, \varphi(-s)^* \right]
  = -2 \re\left[\varphi(0)\varphi'(0)^*\right]
\end{equation}
is constant due to the absence of a first derivative in \eqref{eq:HomogeneousDE}.
We will see below that the situation is different for $E=-k^2$.
There are in fact two regimes of solution for Eq.~\eqref{eq:HomogeneousDE}.

\paragraph{First regime: the Wronskian is $\mathcal{W}\neq0$.--}

With the two independent solutions, we can construct the Green's function, solution of
\begin{equation}
  \label{eq:DiffEqGF}
  \left[ -\deriv{^2}{s^2} + E - \frac{\levy(s)}{\I s}  \right] 
  G(s,s') = 
  \delta(s-s')  
  \:.
\end{equation}
We have
\begin{equation}
  \label{eq:GreenFct}
  G(s,s') = \frac{ 1 }{ \mathcal{W} }\, \varphi(s_>)\varphi(-s_<)^*
\end{equation}
where 
$s_>=\max{s}{s'}$ and  $s_<=\min{s}{s'}$.

Note the useful properties~: 
\begin{enumerate}
\item [(i)]
  $G(s,s')=G(s',s)$.
\item [(ii)]
  $G(-s,-s')=G(s,s')^*$.  
\end{enumerate}



The invariant density $f(z)$ 
is the stationary distribution of the Riccati variable.
It solves \eqref{eq:SPFrischLloyd} for $q=0$, hence we have $\widehat{R}_0(s) = \hat{f}(s)$, where
\begin{equation}
  \hat{f}(s) = 2\pi \IDoS \, G(s,0)
   = 
     \begin{cases}
     \displaystyle
    \frac{\varphi(s)}{\varphi(0)} & \mbox{for } s>0 
    \\[0.25cm]
     \displaystyle
    \frac{\varphi(-s)^*}{\varphi(0)^*} & \mbox{for } s<0 
  \end{cases}
\end{equation}
The prefactor corresponds to the constant introduced in \eqref{eq:7.3}~: it is determined from the Rice formula \eqref{eq:RiceFormula}, leading to $c_0=-\hat{f}'(0^+)+\hat{f}'(0^-)=2\pi \IDoS $, where  
\begin{equation}
  \label{eq:IDoS}
  \IDoS  
  = -\frac{1}{\pi}\frac{\re\left[\varphi(0)\varphi'(0)^*\right]}{|\varphi(0)|^2}
  = \frac{ \mathcal{W} }{ 2\pi|\varphi(0)|^2 }
\end{equation}
has a physical interpretation~: it corresponds with the IDoS per unit length of the disordered model (cf. \cite{GraTexTou14} for instance or \cite{Kot76}).

The Green's function is used in paragraphs~\ref{subsubsec:FLorder1} and \ref{subsubsec:FLorder2}.


\paragraph{Second regime~: the Wronskian is $\mathcal{W}=0$.--}

This hence corresponds to the case where the IDoS $\IDoS$ of the disordered model vanishes.
$\varphi(-s)^*$ is not the second independent solution of the differential equation \eqref{eq:HomogeneousDE}, which has still to be identify.
The paragraph~\ref{subsubsec:CaseNzero} at the end of the section will be devoted to the case~$\mathcal{W}=0$ (i.e. $\IDoS =0$).

\subsubsection{Order $q^1$ -- The Lyapunov exponent}
\label{subsubsec:FLorder1}

Order $q^1$ terms of equation \eqref{eq:SPFrischLloydFourier} read
\begin{equation}
  \label{eq:Order1}
  \I s\, 
  \left\{
    - \deriv{^2}{s^2} + E - \frac{\levy(s)}{\I s} 
  \right\}
  \widehat{R}_1(s) = 
  \left( \rho\lambda_1\hat{p}(s) - \I \deriv{}{s} \right) \hat{f}(s) 
  \:.
\end{equation}

A first step is to obtain a formula for the Lyapunov exponent $\lambda_1$.
In Appendix~\ref{app:SmallSbehaviours} we show that $\re\big[\I s\,\widehat{R}''_1(s)\big]\to0$ as $s\to0^+$ (while $\im\big[\I s\,\widehat{R}''_1(s)\big]$ reaches a finite value).
Thus, keeping the real part of Eq.~\eqref{eq:Order1} and taking the limit $s\to0$ leads to 
\begin{equation}
  \label{eq:Lyapunov}
  \rho\lambda_1 = -\im\left[\hat{f}'(0^\pm)\right]
  = \frac{\im\left[\varphi(0)\varphi'(0)^*\right]}{|\varphi(0)|^2}
  \:.
\end{equation}
This is nothing but the counterpart of the well known formula 
$\rho\lambda_1=\dashint\D z\,z\,f(z)$ \cite{LifGrePas88} (see also \cite{ComTexTou10,ComLucTexTou13} for a discussion in relation with random $2\times2$ matrix products).
We also recall the useful relation \cite{ComTexTou10,ComTexTou11,GraTexTou14} 
\begin{equation}
  \label{eq:DerivativeFhat}
   \hat{f}'(0^\pm) = \mp \pi \, \IDoS  - \I\,\rho\lambda_1
   \:.
\end{equation}

We now solve \eqref{eq:Order1}, i.e.
\begin{equation}
\label{eq:111}
  \left\{
    - \deriv{^2}{s^2} + E - \frac{\levy(s)}{\I s} 
  \right\}
  \widehat{R}_1(s) = 
  \tilde{c}_1 \,\delta(s) 
  + \frac{\rho\lambda_1 \hat{p}(s)\, \hat{f}(s) - \I\, \hat{f}'(s)}{\I s} 
  \:.
\end{equation}
where the constant $\tilde{c}_1$ is related to the normalisation condition $\int\D z\,R_1(z)$. 
A difficulty comes from the singular behaviour of the right hand side 
\begin{equation}
  \frac{\rho\lambda_1\hat{p}(s)\, \hat{f}(s) - \I\, \hat{f}'(s)}{\I s} 
  \simeq \frac{\pi \IDoS }{|s|}
  \hspace{0.5cm}\mbox{for }
  s\to0
  \:.
\end{equation}
This shows that $-\widehat{R}_1''(s)\simeq{\pi \IDoS }/{|s|}$ for $s\to0$, leading to the behaviour $\widehat{R}_1(s)\simeq\widehat{R}_1(0)-\pi \IDoS |s|(\ln|s|-1)$ (see Appendix~\ref{app:SmallSbehaviours}).
The solution of \eqref{eq:111} is
\begin{equation}
  \widehat{R}_1(s) 
  = \tilde{c}_1 \hat{f}(s)
  + 
   \int_{-\infty}^{+\infty}\D t\, 
  G(s,t) \, \mathrm{Pf}\left(\frac{\rho\lambda_1\hat{p}(t)\, \hat{f}(t) - \I\, \hat{f}'(t)}{\I t} \right)  
  \:,
\end{equation}
where $\mathrm{Pf}$ denotes the finite part (cf. Appendix~\ref{app:Pf}). The finite part is required in order to make the integral meaningful. 
More explicitly, the integral rewrites 
\begin{align}
  \label{eq:ExpressionR1}
  &\widehat{R}_1(s) 
  = \tilde{c}_1 \hat{f}(s)
  \\\nonumber
  &+ 
  \lim_{\epsilon\to0^+}
  \left[ 
  \left( \int_{-\infty}^{-\epsilon}+\int_{\epsilon}^\infty\right)\D t\, 
  G(s,t) \, \frac{\rho\lambda_1\hat{p}(t)\, \hat{f}(t) - \I\, \hat{f}'(t)}{\I t}   
  +2\pi \IDoS \, G(s,0)\, \ln\epsilon
  \right]
\end{align}
where $\epsilon$ is a regulator.
We see that the last term is simply $\hat{f}(s)\ln\epsilon$, hence the $\ln\epsilon$ could be absorbed in $\tilde{c}_1$.

\subsubsection{Order $q^2$ -- Fluctuations}
\label{subsubsec:FLorder2}

We identify the order $q^2$ terms of Equation \eqref{eq:SPFrischLloydFourier}~: 
\begin{equation}
  \label{eq:Order2}
  \I s\, 
  \left\{
    - \deriv{^2}{s^2} + E - \frac{\levy(s)}{\I s} 
  \right\}
  \widehat{R}_2(s) = 
  \left( \rho\lambda_1\hat{p}(s) - \I \deriv{}{s} \right) \widehat{R}_1(s) - \frac12\rho\lambda_2\,\hat{p}(s)\,\hat{f}(s)
  \:.
\end{equation}
In Appendix~\ref{app:SmallSbehaviours} we show that $\re\big[\I s\,\widehat{R}''_2(s)\big]\to0$ as $s\to0$ (while the imaginary part diverges logarithmically). 
Keeping the real part of Eq.~\eqref{eq:Order2} and taking the limit $s\to0$, we get
\begin{equation}
  \label{eq:Lambda2}
   \rho\lambda_2 = 2 
   \left(
     \rho\lambda_1\,\widehat{R}_1(0) + \im\left[\widehat{R}_1'(s)\right]_{s=0}
     \right)
\end{equation}
(in Ref.~\cite{ComTexTou19}, the choice of normalisation $\widehat{R}_n(0)=\delta_{n,0}$ was made, which simplifies the relation).
We have now to introduce \eqref{eq:ExpressionR1} in that expression.
An important remark is that the term $\tilde{c}_1 \hat{f}(s)$ does not contribute to $\lambda_2$ by virtue of \eqref{eq:Lyapunov}, hence we will simply drop it below for simplicity.
Similarly, the term $\sim\ln\epsilon$ in \eqref{eq:ExpressionR1} will not contribute to $\lambda_2$.

Let us first consider $\widehat{R}_1(0)$.
Making use of the symmetries of $\hat{f}(s)$ and $G(s,t)$, we get  
\begin{equation}
  \label{eq:116}
  \widehat{R}_1(0) = 2 \frac{|\varphi(0)|^2}{\mathcal{W}}
  \int_\epsilon^\infty\D s\,
  \re\left[
    \hat{f}(s)\,\frac{\rho\lambda_1\hat{p}(s)\, \hat{f}(s) - \I\, \hat{f}'(s)}{\I s} 
  \right]
  +\ln\epsilon
  \:,
\end{equation}
where we have omitted $\lim_{\epsilon\to0^+}$ for clarity.
Similarly, we consider the imaginary part of the derivative~: we get 
\begin{align}
  \label{eq:117}
  &\im\left[\widehat{R}_1'(s)\right]_{s=0}
  \\\nonumber
  &= -2 \im\left[
    \frac{\varphi(0)\varphi'(0)^*}{\mathcal{W}}
    \int_\epsilon^\infty\D s\,
    \hat{f}(s)\,\frac{\rho\lambda_1\hat{p}(s)\, \hat{f}(s) - \I\, \hat{f}'(s)}{\I s}     
  \right]
  -\rho\lambda_1\ln\epsilon
  \:.
\end{align}
We easily check that the two expressions (\ref{eq:116},\ref{eq:117}) are finite in the limit $\epsilon\to0^+$ (note however that the real part of $\widehat{R}_1'(s)$ is logarithmically divergent for $s\to0$).

Combining the two equations and using (\ref{eq:Wronskian},\ref{eq:Lyapunov}), we can now send the regulator $\epsilon$ to zero. We finally obtain 
\begin{equation}
  \label{eq:FinalLambda2}
  \boxed{
    \rho\lambda_2
    = -   \int_0^\infty \frac{\D s}{s}\,       
    \re\left[
      \left(  
         2\rho\lambda_1\hat{p}(s)\,  - \I\, \deriv{}{s}
      \right)
      \hat{f}(s)^2
    \right]
  }
\end{equation}
which is a central result.
Be aware that real part and integral cannot be reversed.
It applies to any random Schr\"odinger operator \eqref{eq:Schrodinger} provided that $\int_0^x\D t\,V(t)$ is a L\'evy process, i.e. is the disorder $V(x)$ has purely local correlations (cf. Appendix 1 of Ref.~\cite{GraTexTou14}).
In terms of the localisation model \eqref{eq:Schrodinger}, it characterizes the fluctuations of the wave function at the $n$-th impurity
\begin{equation}
  \widetilde{\Lambda}''(0)=
  \lambda_1^2-\lambda_2
  = \lim_{n\to\infty} \frac{\mathrm{Var}\big(\ln |\psi(x_n^-)|\big)}{n}
  \:.
\end{equation}
The integral representation can be used to analyse the limit $k\to\infty$ of physical interest~:
in this weak disorder limit, one gets $\lambda_2\simeq-\lambda_1$, i.e.
\begin{equation}
  \widetilde{\Lambda}''(0) \simeq \widetilde{\Lambda}'(0)
  \hspace{0.5cm}\mbox{for }
  k\to\infty
  \:.
\end{equation}
This limit will be rediscussed in more detail in Section~\ref{sec:GLElocalisation} (we will explain its relation with the so-called \og single scaling parameter \fg{} property).
The variance is plotted in Fig.~\ref{fig:FL0}.

\subsubsection{The case $\IDoS =0$ : case $E=-k^2$ (matrices $\widetilde{K}N$) with $v_n>0$}
\label{subsubsec:CaseNzero}

The recessive solution $\varphi(s)$ of the homogeneous equation \eqref{eq:HomogeneousDE} 
plays a central role. The Green's function has been constructed from the two independent solutions $\varphi(s)$ and $\varphi(-s)^*$. 
However, when the Wronskian $\mathcal{W}$ (i.e. the IDoS $\IDoS$)
vanishes, the two solutions are not independent, as it is clear from \eqref{eq:IDoS}. Hence the solution of \eqref{eq:111} cannot be constructed in this way.
This is for instance the case for a potential with positive weights $v_n\geq0$ in the region $E=-k^2<0$.

We first analyse the asymptotic behaviour of $\varphi(s)=\varphi(-s)^*$. 
For $s\to\pm\infty$ we have $\levy(s)\to\rho$, hence the differential equation takes the form
\begin{equation}
  \left[ - \deriv{^2}{s^2} - k^2 - \frac{\rho}{\I s} \right] \varphi(s) \simeq0
  \hspace{0.5cm}\mbox{for } 
  s\to\pm\infty
  \:.
\end{equation}
The recessive solution displays the behaviour
\begin{equation}
  \label{eq:AsymptVarphi}
  \varphi(s) \sim (\I s)^{-\rho/(2k)}\EXP{-\I ks}
  \hspace{0.5cm}\mbox{for } 
  s\to\pm\infty
\end{equation}
and thus vanishes for both $s\to+\infty$ and $s\to-\infty$ (this is not the case for $E=k^2$ for which the recessive solution $\varphi(s)$ blows up for $s\to-\infty$).
There exists a second solution which thus grows at infinity
\begin{equation}
  \label{eq:AsymptChi}
  \chi(s) \sim (\I s)^{+\rho/(2k)}\EXP{\I ks}
  \hspace{0.5cm}\mbox{for } 
  s\to\pm\infty
  \:.
\end{equation}
We introduce the Wronskian 
\begin{equation}
  \mathcal{W}_- = \mathscr{W}\left[ \varphi \, , \, \chi \right] 
  \:.
\end{equation}

We can now express the solution of \eqref{eq:111} with the help of these two independent solutions
\begin{equation}
  \label{eq:BareExpressionR1}
  \widehat{R}_1(s) 
  = \varphi(s)\int_0^s \frac{\D t}{\mathcal{W}_-}\,\sigma(t)\,\chi(t)
  - \chi(s)\int_0^s \frac{\D t}{\mathcal{W}_-}\,\sigma(t)\,\varphi(t)
  +\tilde{c}_1\,\varphi(s) + \tilde{d}_1\,\chi(s)
  \:,
\end{equation}
where we have introduced the notation
\begin{equation}
  \sigma(s) = \frac{\rho\lambda_1 \hat{p}(s)\, \hat{f}(s) - \I\, \hat{f}'(s)}{\I s} 
\end{equation}
for the source term of \eqref{eq:111}.
From the previous remarks, we deduce the asymptotic behaviour
\begin{equation}
  \sigma(s)\,\chi(s) 
  \sim 1/s
  \hspace{0.5cm}\mbox{for } s\to\pm\infty
  \:,
\end{equation}
so that the first term of \eqref{eq:BareExpressionR1} behaves as 
$s^{-\rho/(2k)}\EXP{-\I ks}\ln|s|$ at large $s$, i.e. vanishes for $s\to\pm\infty$.
The combination $\sigma(s)\,\varphi(s)\sim s^{-1-\rho/k}\EXP{-2\I ks}$ decays faster, hence the second term of \eqref{eq:BareExpressionR1} diverges in the same way as $\chi(s)$ for $s\to\pm\infty$.
Being the Fourier transform of a normalisable function, $\widehat{\Phi}_q^\mathrm{R}(s)$ vanishes at infinity and so does $\widehat{R}_1(s)$. The condition that \eqref{eq:BareExpressionR1} vanishes for $s\to+\infty$ gives
\begin{equation}
  \tilde{d}_1 = \int_0^\infty \frac{\D t}{\mathcal{W}_-}\,\sigma(t)\,\varphi(t)
  \:,
\end{equation}
i.e.
\begin{equation}
  \label{eq:ExpressionR1neg}
  \widehat{R}_1(s) 
  = \varphi(s)\int_0^s \frac{\D t}{\mathcal{W}_-}\,\sigma(t)\,\chi(t)
  + \chi(s)\int_s^{+\infty} \frac{\D t}{\mathcal{W}_-}\,\sigma(t)\,\varphi(t)
  +\tilde{c}_1\,\varphi(s) 
  \:.
\end{equation}
The value of $\tilde{c}_1$ is not important as the last term does not contribute to $\lambda_2$.

We get the expression of $\lambda_2$ by sending $s\to0$ in \eqref{eq:Order2} (when $\IDoS =0$, $s\,\widehat{R}_2''(s)$ vanishes for $s\to0$). 
Finally, we obtain
\begin{align}
   \rho\lambda_2 &= 2 
     \rho\lambda_1\,\widehat{R}_1(0) -\I \, \widehat{R}_1'(0)
   = 2
   \left(
     \rho\lambda_1\,\chi(0) -\I \, \chi'(0)
     \right)\,\tilde{d}_1
\\
  \label{eq:FinalLambda2bis}
    &=  2
    \left(
     \frac{\I\varphi'(0)}{\varphi(0)}\,\chi(0) -\I \, \chi'(0)
     \right)\,
     \int_0^\infty \frac{\D s}{\mathcal{W}_-}\,\sigma(s)\,\varphi(s)
     = -2
     \I\int_0^\infty\D s \,\sigma(s)\,\hat{f}(s)
      \:.
\end{align}
By using Laplace transforms instead of Fourier transform, we  prove below that the expression is real.
With this remark in mind, we conclude that \eqref{eq:FinalLambda2bis} coincides with \eqref{eq:FinalLambda2}.
We have thus extended the validity of Eq.~\eqref{eq:FinalLambda2} to the case $\IDoS=0$, and have provided an alternative derivation of this formula.


For the case of positive weights $v_n$, it can be advantageous to deal with the Laplace transforms of the invariant density 
\begin{equation}
  \label{eq:FLaplace}
  \tilde{f}(r) = \hat{f}(-\I r) = \int\D z\, f(z) \, \EXP{-rz}
\end{equation}
and the weight distribution $\tilde{p}(r) = \hat{p}(-\I r)$, which are real functions.
The differential equation for the Laplace transform,
\begin{equation}
  \Big[
    \deriv{^2}{r^2} - k^2 - \frac{1}{r}\,\mathcal{L}(-\I r)
  \Big]
  \tilde{f}(r) =0
  \:,
\end{equation}
for $r>0$, 
now involves \textit{real} functions since $\mathcal{L}(-\I r)=\rho\,\big[1-\tilde{p}(r)\big]$.
The solution presents the asymptotic behaviour $\tilde{f}(r)\sim r^{-\rho/(2k)}\EXP{-kr}$.
Eventually, we obtain
\begin{equation}
  \label{eq:FinalLambda2Convenient}
   \rho\lambda_2   
     = 2
     \int_0^\infty\frac{\D r}{r} \,
     \left[ \tilde{p}(r)\,\tilde{f}'(0) \tilde{f}(r) - \tilde{f}'(r) \right]
     \,\tilde{f}(r) 
     \:,
\end{equation}
where we have used $\rho\lambda_1=-\tilde{f}'(0)$. 
The form \eqref{eq:FinalLambda2Convenient} now involves an integral which converges much better than \eqref{eq:FinalLambda2}.

\paragraph{Illustration~1 : the subgroup $\widetilde{\mathrm{K}}$.--- } 
 
We first consider a trivial case, although instructive for the following~:
when the L\'evy exponent vanishes, $\levy(s)=0$, which corresponds to consider products of matrices of the one-parameter subgroup of matrices $\{\widetilde{K}(\theta)\}$.
We have obviously $\tilde{f}(r)=\EXP{-kr}$ with $r>0$. 
We find $\rho\lambda_1=-\tilde{f}'(0)=k$ and \eqref{eq:FinalLambda2Convenient} gives $\lambda_2=0$, thus
\begin{equation}
  \lambda(q)=1-qk/\rho
\end{equation}
for $q<\rho/k$.
As a result, the GLE is 
\begin{equation}
  \label{eq:TrivialGLESubgroupKtilde}
  \widetilde{\Lambda}(q)=-\ln(1-qk/\rho)
  \hspace{0.5cm}\mbox{for }q<\rho/k
  \:.
\end{equation}
We recognize the generating function for the Poisson distribution.

The result \eqref{eq:TrivialGLESubgroupKtilde} can be recovered easily by noticing that
$\Pi_n=\Pi_{i=1}^n\widetilde{K}(\theta_i)=\widetilde{K}(\sum_i\theta_i)$, hence $\ln||\Pi_n\vec{x}_0||\simeq\sum_{i=1}^n\theta_i$, which is indeed characterised by the mean 
$\widetilde{\Lambda}'(0)=k/\rho$ 
and the variance
$\widetilde{\Lambda}''(0)=(k/\rho)^2$.

\begin{figure}[!ht]
\centering
\includegraphics[scale=0.6]{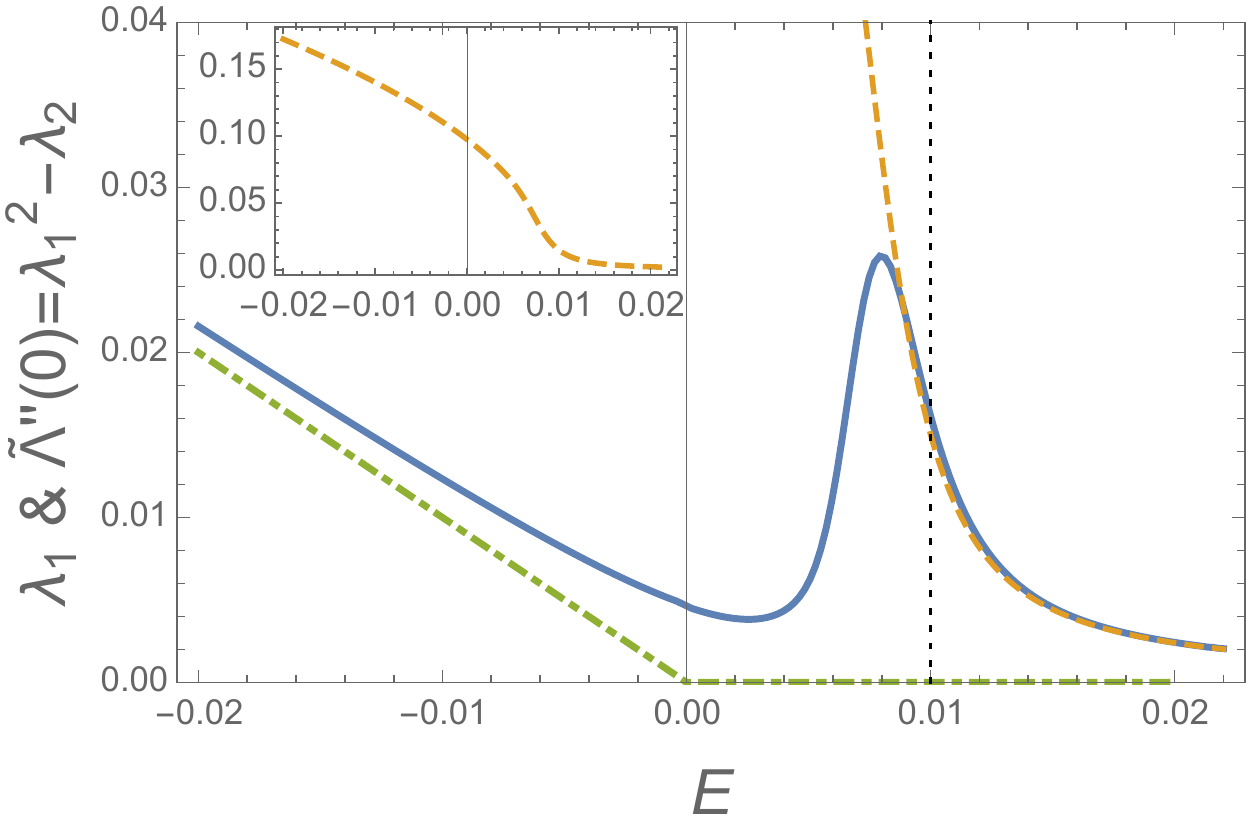}
\caption{\it  The Lyapunov exponent $\lambda_1$ (dashed orange line) and the variance $\lambda_1^2-\lambda_2$  (blue line) of the random matrix product (\ref{eq:MatricesSchrodinger},\ref{eq:MatricesSchrodinger2}) matrices) for exponentially distributed angles 
$\mathrm{Proba}\{\theta_n>k\ell\}=\EXP{-\rho\ell}$ 
and exponentially distributed weights 
$\mathrm{Proba}\{u_n>y/k\}=\EXP{-y/\bv}$. 
We choose $\rho=1$ with exponentially distributed weights with $\bv=0.01$.
For $E=+k^2$, we use Eqs.~(\ref{eq:PhiFrischLloyd},\ref{eq:Lyapunov},\ref{eq:FinalLambda2}) and, for $E=-k^2$, Eqs.~(\ref{eq:PhiFrischLloydNegEnergy},\ref{eq:FinalLambda2}) or equivalently Eqs.~(\ref{eq:FinalLambda2Convenient},\ref{eq:FtildeFL}).
The green dot-dashed line is $\widetilde{\Lambda}''(0)$ for $v_n=0$.
The dotted vertical line indicates the value of $\rho\bv$.
The inset shows $\lambda_1$ on a larger scale.
}
\label{fig:FL0}
\end{figure}

\paragraph{Illustration 2~: } 
 
We consider the case of exponentially distributed weights, where $\varphi(s)$ is given by \eqref{eq:PhiFrischLloydNegEnergy}.
We first check the asymptotic behaviour of $\varphi(s)$~: \eqref{eq:PhiFrischLloydNegEnergy} behaves as 
$\varphi(s)\simeq(2\I ks)^{-\rho/(2k)}\EXP{-\I ks-k/\bv}$ for $s\to\pm\infty$. 
The second independent solution introduced above is obviously the second Whittaker function
\begin{align}
  \label{eq:ChiFrischLloydNegEnergy}
  \chi(s) &= M_{-\frac{\rho}{2k},\frac12}(2k(\I s+1/\bv))
  \simeq \frac{ (2\I ks)^{\rho/(2k)} }{ \Gamma(1+\rho/(2k)) } \, \EXP{\I ks+k/\bv}
  \hspace{0.5cm}\mbox{for } 
  s\to\pm\infty
  \:.
\end{align}
The Wronskian of the two solutions is explicitly $\mathcal{W}_-=2\I k/\Gamma(1+\frac{\rho}{2k})$.

Finally, we remark that the integral representation \eqref{eq:FinalLambda2Convenient} is convenient as the Laplace transform of the invariant density
\begin{equation}
  \label{eq:FtildeFL}
  \tilde{f}(r)  = \frac{ W_{-\frac{\rho}{2k},\frac{1}{2}}\left( 2k (r+1/\bv) \right) }
                 { W_{-\frac{\rho}{2k},\frac{1}{2}}\left( 2k/\bv \right) } 
\end{equation}
decays exponentially, which makes the integral absolutely convergent~;
we use $\tilde{p}(r)=\mean{\EXP{-rv_n}}=1/(1+\bv r)$.
The result of the integration is plotted in Fig.~\ref{fig:FL0}. 
Note that the variance presents the same growth 
$\widetilde{\Lambda}''(0)\simeq(k/\rho)^2$ for large $k$ as for the subgroup $\widetilde{\mathrm{K}}$, which is explained by the fact that matrices $\widetilde{K}(\theta)$ become prominent in this limit.


\subsection{Products of matrices $\widetilde{K}A$ (Dirac or supersymmetric case)}
\label{sec:PerturbativeSusy}


We now show that \eqref{eq:SPsusyMellin} can be analysed with the same type of perturbative approach as in the previous section. 
We expand the Mellin transform in powers of $q$~:
\begin{equation}
  \widecheck{\Phi}^\mathrm{R}_q(s) = \widecheck{R}_0(s) + q\,\widecheck{R}_1(s) + q^2\,\widecheck{R}_2(s) + \cdots
\end{equation}

At order $q^0$ we recover the equation for the Mellin transform of the invariant density $\widecheck{R}_0(s)=\check{f}(s)$, obtained in Ref.~\cite{ComTexTou11}~:
\begin{equation}
  \label{eq:EqMellinTransff}
  - \check{f}(s+1) - \frac{\levy(2\I s)}{s}\,\check{f}(s) + k^2\,\check{f}(s-1) = 0
  \:.
\end{equation} 
We now identify the order $q^1$ terms of \eqref{eq:SPsusyMellin}~:
\begin{align}
  \label{eq:EqCheckR1}
  &s\left[
    -\widecheck{R}_1(s+1) -\frac{\levy(2\I s)}{s}\widecheck{R}_1(s) + k^2\, \widecheck{R}_1(s-1)
  \right]
  \nonumber\\
  &\hspace{0.5cm}
  = \rho\lambda_1\hat{p}(2\I s)\,\check{f}(s)
  -\frac{1}{2}\left[    \check{f}(s+1) + k^2\, \check{f}(s-1)  \right]
\end{align}
Considering the limit $s\to0$ we get a formula for the Lyapunov exponent~\cite{ComTexTou11}
\begin{equation}
  \label{eq:LyapMellin}
  \rho\lambda_1 = \frac{1}{2}\left[    \check{f}(1) + k^2\, \check{f}(-1)  \right]
  \:.
\end{equation}
Identifying the $q^2$ terms of \eqref{eq:SPsusyMellin} gives
\begin{align}
  &s\left[
    -\widecheck{R}_2(s+1) -\frac{\levy(2\I s)}{s}\widecheck{R}_2(s) + k^2\, \widecheck{R}_2(s-1)
  \right]
  \\\nonumber
  &\hspace{0.5cm}
  = -\frac{1}{2}\rho\lambda_2\hat{p}(2\I s)\,\check{f}(s)
   + \rho\lambda_1\hat{p}(2\I s)\,\widecheck{R}_1(s)
  -\frac{1}{2}\left[  \widecheck{R}_1(s+1) + k^2\, \widecheck{R}_1(s-1)  \right]
  \:,
\end{align}  
whose $s\to0$ limit provides the expression
\begin{equation}
  \rho\lambda_2 = 2\,  \rho\lambda_1\,\widecheck{R}_1(0)
  - \widecheck{R}_1(1) - k^2\, \widecheck{R}_1(-1)
  \:.
\end{equation}
The next step requires to find $\widecheck{R}_1(s)$, i.e. solve a difference equation with a source term.
Note that we only need the solution for $s\in\mathbb{Z}$.
One can introduce a Green's function, solving 
\begin{equation}
    -\widecheck{G}_{s+1,s'} -\frac{\levy(2\I s)}{s}\widecheck{G}_{s,s'} + k^2\, \widecheck{G}_{s-1,s'}
    = \delta_{s,s'}
\end{equation}
and write the solution of \eqref{eq:EqCheckR1} as 
\begin{equation}
  \widecheck{R}_1(s)
   = 
   \tilde{c}_1\,   \check{f}(s)
   +
   \sum_{s'\ (\neq0)}
   \widecheck{G}_{s,s'} 
   \frac{2\rho\lambda_1\hat{p}(2\I s')\,\check{f}(s') -  \check{f}(s'+1) - k^2\, \check{f}(s'-1) }{2s'}
  \:.
\end{equation}
We will not pursue further this analysis here.



\section{Application for localisation theory~: the variance $\Lambda''(0)=\gamma_2$}
\label{sec:GLElocalisation}

The correspondence between the random matrix problem and the quantum localisation model \eqref{eq:Schrodinger} relies on the fact that the matrices $M_n=K(\theta_n)N(u_n)$ are transfer matrices for the Schr\"odinger equation.
As discussed in Section~\ref{sec:Intro}, the GLE \eqref{eq:DefGLE} characterizes the fluctuations of the wave function at the $n$-th impurity, cf. Eq.~\eqref{eq:FluctPsiN}.
On the other hand, in the context of localisation theory, it is more natural to define the GLE as \eqref{eq:FluctPsiX}.
Hence, we have to distinguish two cases~:
\begin{enumerate}[label=({\it\roman*}),leftmargin=*,align=left,itemsep=+0.1cm]

\item
  \textit{Fixed angles (lattice of impurities)~:}
  The case of fixed angles $\theta_n=\btheta=k/\rho$ $\forall\:n$ describes a regular lattice of impurities with lattice spacing $\ell_n=1/\rho$.
Thus we have 
\begin{equation}
   \Lambda(q) = \rho\widetilde{\Lambda}(q) 
   \:.
\end{equation}
Inserting the expansions \eqref{eq:ExpansionSmallLambda} and \eqref{eq:FluctPsiX} in that relation, 
we can express the cumulants of $\ln |\psi(x)|$ in terms of the coefficients $\lambda_n$'s studied in the previous Section~:
\begin{equation}
  \gamma_1=\rho\,\lambda_1
  \:,
  \hspace{1cm}
  \gamma_2=\rho\,(\lambda_1^2-\lambda_2)
  \:,
  \hspace{1cm}
  \mbox{etc.}
\end{equation}
$\lambda_n$ can be determined by solving \eqref{eq:TheSpectralPb} for 
$
\overline{\mathscr{T}}(q)
=\EXP{\btheta\mathscr{D}_K(q)}\,\smean{\EXP{w\mathscr{D}_A(q)}\EXP{u\mathscr{D}_N(q)}}_{w,u}
$.
The continuum limit of this problem will be discussed in Section~\ref{sec:ContinuumLimit}. 

\item
    \textit{Exponentially distributed angles (impurities with random positions)~:}
The case discussed in the previous sections corresponds to exponentially distributed angles.
This case is interesting as it describes uncorrelated impurities with random positions for a uniform mean density $\rho$. 
In the definition \eqref{eq:FluctPsiN}, the fact that the position of the $n$-th impurity presents fluctuations of order $\sim\sqrt{n}\sim\sqrt{\rho x}$ contributes to the fluctuations of 
$\ln|\psi(x_n^-)|$. 
As we mentioned, for the localisation problem, it is more natural to quantify the fluctuations at a fixed distance $x$ and define the GLE according to \eqref{eq:FluctPsiX}, with this time 
\begin{equation}
\Lambda(q) \neq \rho\widetilde{\Lambda}(q)
\:.
\end{equation}
In order to stress the difference between the two definitions, we rewrite \eqref{eq:FluctPsiX} as 
\begin{equation}
  \label{eq:DefLambdaLocalisation}
  \Lambda(q) 
  = \lim_{x\to\infty} \frac{\ln \mean{||\Pi_{\mathscr{N}(x)}\vec{x}_0||^q}}{x}
  = \lim_{x\to\infty} \frac{\ln\smean{|\psi(x_{\mathscr{N}(x)}^-)|^q}}{x}
\end{equation}
where averaging is taken both over the matrices and the Poisson process $\mathscr{N}(x)$.
We recall that a Poisson process of intensity $\rho$, the number of impurities in the interval $[0,x]$, has probability distribution
\begin{equation}
\mathrm{Proba}\{  \mathscr{N}(x) = n \} =  \frac{(\rho x)^n}{n!}\,\EXP{-\rho x}
\:.
\end{equation}
\end{enumerate}
In the rest of the Section, we focus on the second situation.


\subsection{Another spectral problem}

The formulation of the spectral problem takes a different form than in Section~\ref{sec:SpectralPb}, although it involves the same tools.
Random matrices have the form $M_n=K(\theta_n)A(w_n)N(u_n)\equiv K(\theta_n)B_n$.
When the lengths $\ell_n=\theta_n/k$ are exponentially distributed, we can introduce a Markovian process~\footnote{
  The Riccati variable is related to the wave function solution of \eqref{eq:Schrodinger} by 
  $z(x)=\psi'(x)/\psi(x)$ \cite{ComTexTou10}.
} $z(x)$. 
Action of the rotation corresponds to the evolution
\begin{equation}
  \label{eq:RiccatiFreeEvol}
  z'(x) = -k^2 - z(x)^2
\end{equation}
while the ``impurity'' corresponds to the jump
\begin{equation}
  \label{eq:RiccatiJump}
  z(x_n^+) = \mathcal{B}_n\left( z(x_n^-) \right)
\end{equation}
occuring with probability rate $\rho$ \cite{ComTexTou10}.
For matrices $M_n=\widetilde{K}(\theta_n)B_n$, the evolution \eqref{eq:RiccatiFreeEvol} is replaced by $z'(x) = +k^2 - z(x)^2$.
Instead of \eqref{eq:PropagatorFEvol}, we introduce 
\begin{align}
  \mathcal{P}_x(z|z_0;q) 
  \eqdef
  \mean{ \delta(z-z(x)) \, \prod_{i=1}^{\mathscr{N}(x)} J_N^{-q/2}(M_i,z(x_{i-1}^-)) } 
\:.
\end{align}
controlling the moment generating function
\begin{align}
  \mathcal{Q}_x(z_0;q) 
  &=\mean{|\psi(x)|^q} = \mean{||\Pi_{\mathscr{N}(x)}\vec{x}_0|| ^q}
  = \mean{ \prod_{i=1}^{\mathscr{N}(x)} J_K^{-q/2}(M_i,z(x_{i-1}^-)) }
  \nonumber  \\
  &= \rho_K(z_0)^{q/2} \int\D z\,  \rho_K(z)^{-q/2} \: \mathcal{P}_x(z|z_0;q) 
  \:.
\end{align}
If we consider the evolution of the propagator in the interval $[x,x+\delta x]$, we have to take into account two types of events~:
(\textit{i})
with probability $1-\rho\delta x$, the interval is free of impurity and the process is governed by the free evolution \eqref{eq:RiccatiFreeEvol}, i.e. $z(x+\delta x)\simeq z(x) - g_K(z(x))\,\delta x$, corresponding to a rotation of angle $\delta\theta=\delta x$ (we set $k=1$ for simplicity).
(\textit{ii})
The interval contains an impurity with probability $\rho\delta x$, inducing the jump \eqref{eq:RiccatiJump}.
Correspondingly, 
\begin{align}
  &\mathcal{P}_{x+\delta x}(z|z_0;q)
  \nonumber\\
  &\hspace{0.5cm}
  \simeq 
  (1-\rho\delta x) \,\left(1 + q\, h_K(z)\,\delta x\right)
  \,
  \mathcal{P}_{x}(z + g_K(z)\,\delta x|z_0;q)\,\left(1 + g_K'(z)\,\delta x\right)
  \nonumber\\
  &\hspace{1.5cm}
  + \rho\delta x\, 
  \mean{ 
    J_N^{q/2}(B^{-1},z)\,
    \mathcal{P}_{x}(\mathcal{B}^{-1}(z)|z_0;q) \deriv{\mathcal{B}^{-1}(z)}{z} 
    }_B
      \:.
\end{align}
Using the notation introduced above $\mathscr{T}_B(q)=\EXP{w\,\mathscr{D}_A(q)}\EXP{u\,\mathscr{D}_N(q)}$ and sending $\delta x\to0$, we get 
\begin{equation}
  \label{eq:coucou}
  \derivp{}{x}\mathcal{P}_{x}(z|z_0;q)
  =     \left[ 
       \mathscr{D}_K(q)   
       + \rho \left( \mean{\EXP{w\,\mathscr{D}_A(q)}\EXP{u\,\mathscr{D}_N(q)}}  -1 \right) 
      \right] \mathcal{P}_{x}(z|z_0;q)
      \:.
\end{equation}
As a check, for $q=0$, we recover the equations controlling the disribution of the process $z(x)$ given either for $w=0$ or for $u=0$ in Ref.~\cite{BieTex08}.
Writing 
\begin{equation}
  \mathcal{Q}_x(z_0;q) 
  = \rho_K(z_0)^{q/2} \int\D z\,  \rho_K(z)^{-q/2} \: \mathcal{P}_{x}(z|z_0;q) 
  \underset{x\to\infty}{\sim} \EXP{x\,\Lambda(q)}
\end{equation}
we deduce that the GLE is the largest eigenvalue of the operator involved in \eqref{eq:coucou}, i.e. we now have to solve the spectral problem
\begin{equation}
  \label{eq:SpectralPbExpRL}
  \boxed{
     \left[ 
       \mathscr{D}_K(q)   
       + \rho \left( \mean{\EXP{w\,\mathscr{D}_A(q)}\EXP{u\,\mathscr{D}_N(q)}}  -1 \right) 
      \right] \Phi^\mathrm{R}_q(z)
  = \Lambda(q) \,  \Phi^\mathrm{R}_q(z)
  }
\end{equation}
This equation replaces \eqref{eq:TheSpectralPb}.
Boundary conditions are also given by \eqref{eq:AsymptoticBehaviourPhiR}.
As in Section~\ref{sec:SpectralPb}, we stress that making different choices of Jacobian $J$ (in the infinitesimal generators) corresponds to different realizations of the same spectral problem. 
Equation \eqref{eq:SpectralPbExpRL} can be rewritten as
\begin{equation}
     \left[ 1 - \rho^{-1}\mathscr{D}_K(q)  \right] \Phi^\mathrm{R}_q(z)
  =  \left[ \mean{\EXP{w\,\mathscr{D}_A(q)}\EXP{u\,\mathscr{D}_N(q)}}  - \rho^{-1}\Lambda(q) \right] \Phi^\mathrm{R}_q(z)
  \:,
\end{equation}
which is a form closer to \eqref{eq:SPFrischLloyd0}.
We can give more explicit forms~:
\begin{itemize}
\item 
\textit{Matrices $KN$~:}
  setting $w_n=0$ and choosing the measure $J=J_N$, Eq.~\eqref{eq:SpectralPbExpRL} reads
  \begin{equation}
    \label{eq:SPFrischLloydLoc}
    \left[ \deriv{}{z}(k^2+z^2) + q\, z \right]\Phi^\mathrm{R}_q(z)
   + \rho\left[ \mean{\Phi^\mathrm{R}_q(z-v)}_v  - \Phi^\mathrm{R}_q(z)\right]  
    = \Lambda(q)\, \Phi^\mathrm{R}_q(z)
    \:, 
  \end{equation}
  which replaces \eqref{eq:SPFrischLloyd}, with boundary conditions \eqref{eq:AsymptoticBehaviourForPhiR}. 

\item
\textit{Matrices $\widetilde{K}A$~:} 
Setting $u_n=0$ and choosing $J=J_A$, we get 
\begin{align}
   \label{eq:SPsusyLoc}
   &\left[
     \deriv{}{z}(-k^2+z^2) + \frac{q}{2}\,\left(z+\frac{k^2}{z}\right)
   \right]\Phi^\mathrm{R}_q(z)
   \nonumber\\
   &\hspace{1.5cm}
   + \rho\left[ \mean{\EXP{-2w}\Phi^\mathrm{R}_q(z\EXP{-2w})}_w  - \Phi^\mathrm{R}_q(z)\right]
   =\Lambda(q)\,\Phi^\mathrm{R}_q(z)
   \:,
\end{align}
replacing \eqref{eq:SPsusy}.
\end{itemize}

\subsection{Application~: the Schr\"odinger operator with a random potential (matrices $KN$ or $\widetilde{K}N$)}
\label{subsec:RSO}

As we have discussed at length, the resolution of the spectral problem can be simplified by using appropriate transforms. 
In this paragraph, let us illustrate the problem by considering the case of random matrices \eqref{eq:MatricesSchrodinger} or \eqref{eq:MatricesSchrodinger2}. 
After Fourier transform of \eqref{eq:SPFrischLloydLoc} we get 
\begin{equation}
  \label{eq:LocalisationEqForPhiR}
  \boxed{
  \I s \left[ -\deriv{^2}{s^2} + E - \frac{\levy(s)}{\I s} \right] 
  \widehat{\Phi}^\mathrm{R}_q(s)
  =
  \left[ \Lambda(q)  -\I q\deriv{}{s}\right] 
  \widehat{\Phi}^\mathrm{R}_q(s)
  }
\end{equation}
which presents a similar structure than \eqref{eq:SPFrischLloydFourier}.
Boundary conditions for this spectral problem are also given by \eqref{eq:BCForPhiRFourier} with $\omega_q\in\mathbb{R}$~;
see \S~\ref{subsec:PhiRSmallSbehaviour} for a detailed discussion.
The resolution follows the same steps as in \S~\ref{sec:PeturbFL}, simply taking into account of the change in the right hand side of \eqref{eq:SPFrischLloydFourier}.
We inject \eqref{eq:ExpansionPhiR} and 
\begin{equation}
  \Lambda(q) = \sum_{n=1}^\infty \frac{\gamma_n}{n!}\,q^n
\end{equation}
in \eqref{eq:LocalisationEqForPhiR}.
Eventually, we get the Lyapunov exponent 
\begin{equation}
  \label{eq:Gamma1}
  \gamma_1  = -\im\left[\hat{f}'(0^\pm)\right]  \equiv \rho\lambda_1
\end{equation}
and also 
\begin{equation}
  \label{eq:DerivativeFhat2}
  \hat{f}'(0^\pm) = \pm\,\pi \, \IDoS  -\I\,\gamma_1
  \:,
\end{equation}
where $\IDoS $ is the IDoS of the quantum model. 
The variance is given by 
\begin{equation}
  \label{eq:Gamma2}
   \gamma_2 = -2 
   \left(
     \gamma_1\,\widehat{R}_1(0) + \im\left[\widehat{R}_1'(s)\right]_{s=0}
     \right)
\end{equation}
where 
\begin{equation}
  \widehat{R}_1(s) 
  = \tilde{c}_1 \hat{f}(s)
  + 
   \int_{-\infty}^{+\infty}\D t\, 
  G(s,t) \, \mathrm{Pf}\left(\frac{\gamma_1\, \hat{f}(t) - \I\, \hat{f}'(t)}{\I t} \right)  
  \:,
\end{equation}
where $\mathrm{Pf}$ is the finite part (cf. Appendix~\ref{app:Pf}).
Some algebra similar to that of \S~\ref{sec:PeturbFL} eventually leads to 
\begin{equation}
  \label{eq:FinalGamma2}
  \boxed{
     \gamma_2
    =    \int_0^\infty \frac{\D s}{s}\,       
    \re\left[
      \left(  
         2\gamma_1  - \I\, \deriv{}{s}
      \right)
      \hat{f}(s)^2
    \right]
  }
\end{equation}
This is another important result of the paper.~\footnote{
  A similar form was obtained in chapter 9 of \cite{ItzDro89} by using the replica trick, for the specific case of the Halperin model (continuum limit of matrix products of type $KN$).
The relation with the variance $\gamma_2$ was however not established and the divergence of the integral in the absence of the $\re[\cdots]$ was not discussed.
}
Note that the result is valid for both $E=+k^2$ and $E=-k^2$.
The expression \eqref{eq:FinalGamma2} can be compared to \eqref{eq:FinalLambda2}, which shows that there is no simple relation between $\gamma_2=\Lambda''(0)$ and $\lambda_1^2-\lambda_2=\widetilde{\Lambda}''(0)$, i.e. between the fluctuations of $\ln|\psi(x)|$ and $\ln|\psi(x_n^-)|$. 

\paragraph{The case $\IDoS=0$.--}

In order to compute $\gamma_2$ for $E=-k^2$, it is convenient to use
\begin{equation}
     \gamma_2  = -2
     \int_0^\infty\frac{\D r}{r} \,
     \left[ \tilde{f}'(0) \tilde{f}(r) - \tilde{f}'(r) \right]
     \,\tilde{f}(r) 
     \:,
\end{equation}
where $\tilde{f}(r)$ is the Laplace transform \eqref{eq:FLaplace} of the invariant density.
This formula is obtained from \eqref{eq:FinalGamma2} by a rotation of the contour of integration and is analogous to \eqref{eq:FinalLambda2Convenient} (hence it can be derived along the lines of \S~\ref{subsubsec:CaseNzero}).

\paragraph{High energy limit and single parameter scaling.---}

The high energy limit $E\to+\infty$ can be analysed by performing a perturbative analysis of the solution $\hat{f}(s)$ of equation \eqref{eq:HomogeneousDE} in the disorder (i.e. in the L\'evy exponent). Some algebra gives 
\begin{equation}
  \gamma_1 \simeq \frac{1}{4k} \int\D s \, \EXP{-2k|s|}\, \deriv{}{s}\left[\frac{\levy(s)}{s}\right]
  \hspace{0.5cm}\mbox{for }
  E=k^2\to\infty
  \:.
\end{equation}
Assuming a regular expansion for the L\'evy exponent,
$\levy(s)\simeq\rho(\I s\mean{v_n}+(s^2/2)\mean{v_n^2})$ for $s\to0$, 
we recover the well-known behaviour \cite{AntPasSly81,LifGrePas88} $\gamma_1 \simeq \rho\mean{v_n^2}/(8E)$.
A similar, but more cumbersome analysis of \eqref{eq:FinalGamma2} leads to 
\begin{equation}
  \label{eq:SPSnormal}
  \gamma_1\simeq\gamma_2
  \hspace{0.5cm}\mbox{for } E\to+\infty
  \:.
\end{equation}
This relation is known as \og \textit{single parameter scaling} \fg{}, and has played an important role in localisation theory \cite{AbrAndLicRam79,AndThoAbrFis80,CohRotSha88}.  
Details on the high energy analysis of $\gamma_2$ and a broader perspective can be found in Ref.~\cite{Tex19b}. This paper also provides a generalization of \eqref{eq:SPSnormal}, to more general disorder models.

\subsection{Illustration~: exponentially distributed weights}

\begin{figure}[!ht]
\centering
\includegraphics[scale=0.6]{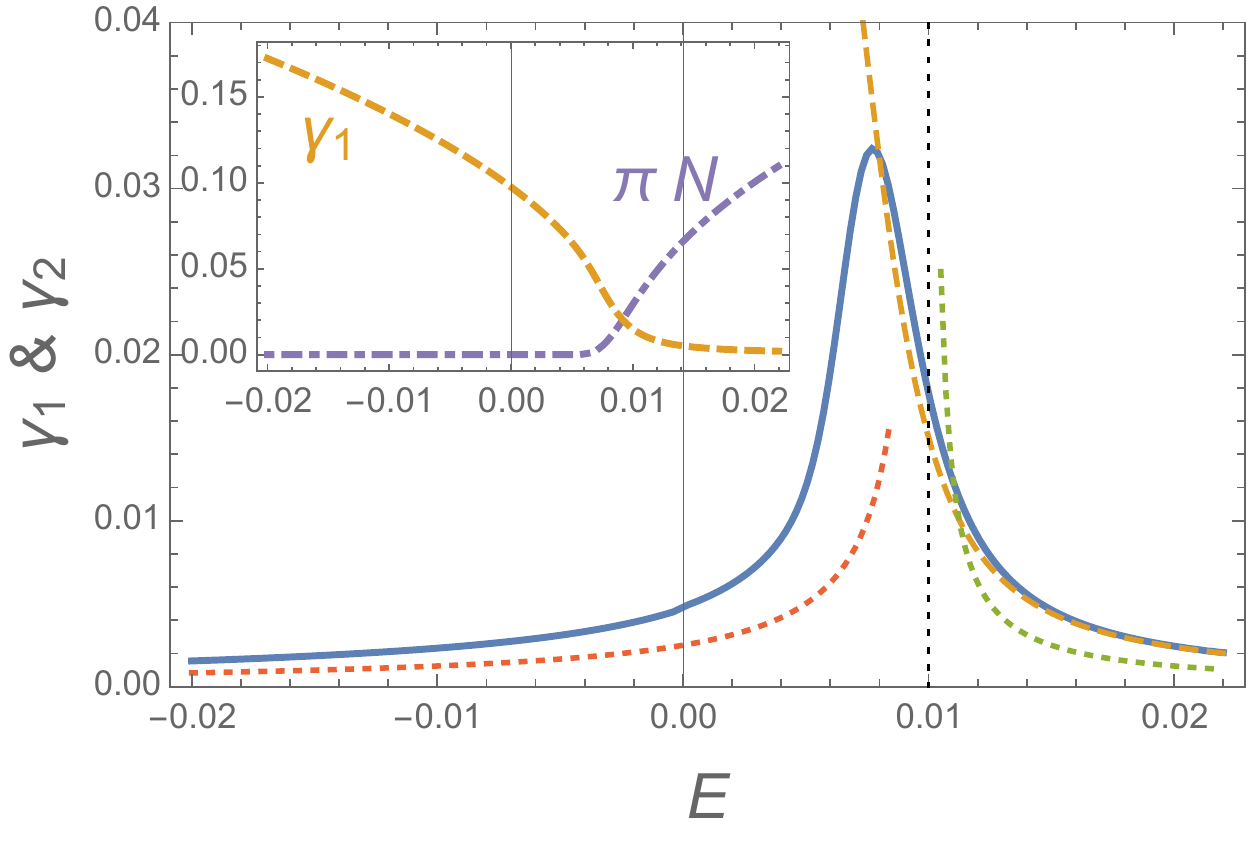}
\caption{\it  The Lyapunov exponent $\gamma_1$ (dashed orange line) and the variance $\gamma_2$ (blue line) for the Frisch-Lloyd model for $\rho=1$ with exponentially distributed weights with $\mean{v_n}=0.01$, from Eqs.~(\ref{eq:PhiFrischLloyd},\ref{eq:Lyapunov},\ref{eq:FinalGamma2}).
Dotted lines are discussed in the text.
Inset show $\gamma_1$ and the IDoS $\IDoS$ on a larger scale.
}
\label{fig:FLhd}
\end{figure}

We now inject the solution \eqref{eq:PhiFrischLloyd} for $E=+k^2$ (or \eqref{eq:PhiFrischLloydNegEnergy} for $E=-k^2$) in Eq.~\eqref{eq:FinalGamma2} in order to make an explicit calculation of $\gamma_2$. 
The integral is computed numerically.
We set $\rho=1$. 
Comparing Fig.~\ref{fig:FL0} and Fig.~\ref{fig:FLhd} shows the difference, at the level of the variance, of the two problems encoded in the different definitions (\ref{eq:DefGLE},\ref{eq:FluctPsiN}) and \eqref{eq:FluctPsiX}.


\paragraph{High density of impurity.---}

First, we consider $\mean{v_n}=0.01$ and plot $\gamma_1$ and $\gamma_2$ in Fig.~\ref{fig:FLhd}. 
We see that, increasing the energy above $\rho\mean{v_n}$, the curves for $\gamma_1$ and $\gamma_2$ become rapidly very close, corresponding to \og single parameter scaling \fg{}.

The high density regime corresponds to the continuum limit~: 
in the limit $\rho\to\infty$, the Frisch-Lloyd model coincides with the Halperin model discussed in Subsec.~\ref{subsec:Halperin} and in Appendix~\ref{app:Halperin}.
Indeed, we check that Fig.~\ref{fig:FLhd} shows the same behaviours as Fig.~\ref{fig:Halperin}.
This remark allows to make comparison with simple limiting behaviours.
A more accurate comparison with Eqs.~(\ref{eq:HalperinHighEnergy},\ref{eq:HalperinLowEnergy}) can be made by shifting the energy by $\rho\mean{v_n}$ and writing the perturbative result $\gamma_1\simeq\gamma_2\simeq\rho\mean{v_n^2}/\big[8(E-\rho\mean{v_n})\big]$ (dotted green line). 
For a large negative energy, we observe a convergence towards the behaviour 
$\gamma_2\simeq\rho\mean{v_n^2}/\big[4(\rho\mean{v_n}-E)\big]$ (dotted red line) found in \cite{RamTex14} (cf. Eq.~\eqref{eq:HalperinLowEnergy} below and also Appendix~\ref{app:PhForm}).

\begin{figure}[!ht]
\centering
\includegraphics[scale=0.6]{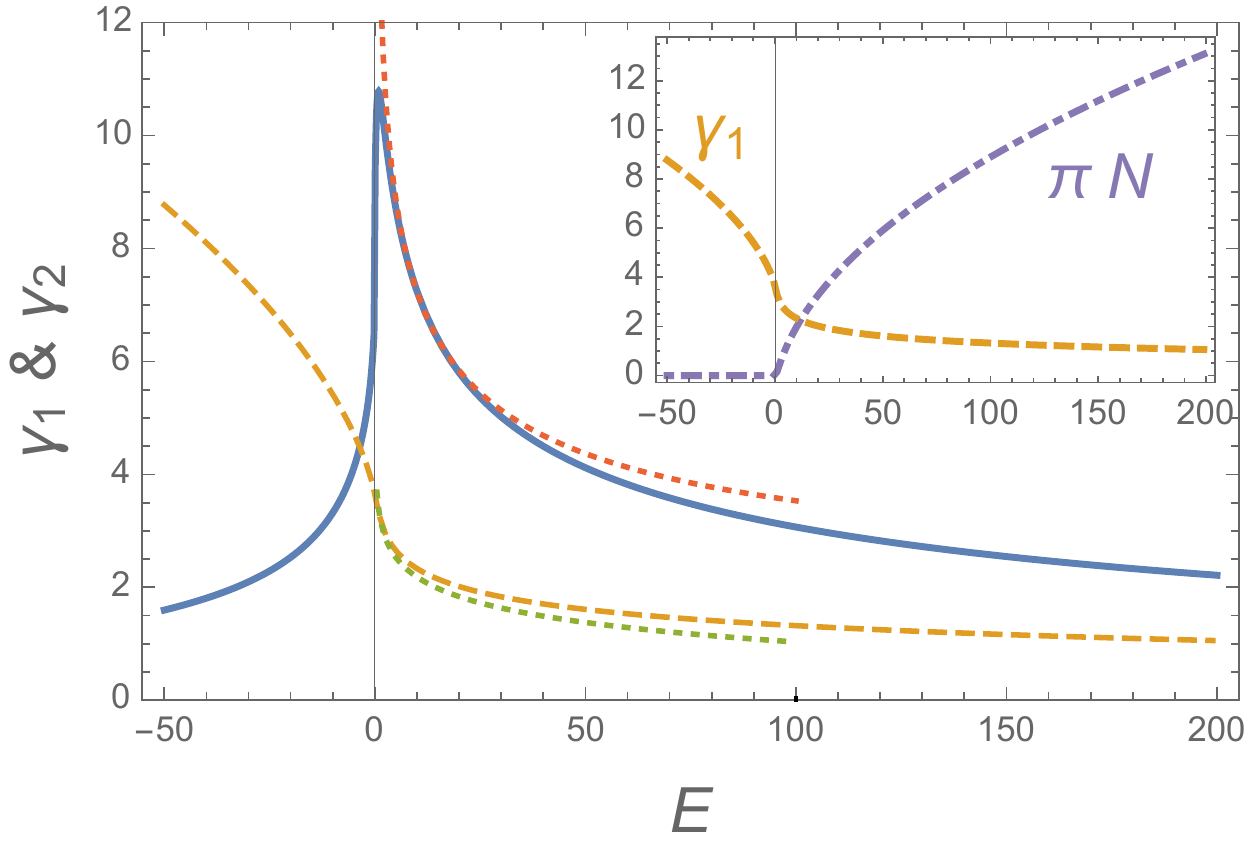}
\caption{\it  The Lyapunov exponent $\gamma_1$ (dashed orange line) and the variance $\gamma_2$ (blue line) for the Frisch-Lloyd model for $\rho=1$ with exponentially distributed weights with $\mean{v_n}=100$, from Eqs.~(\ref{eq:PhiFrischLloyd},\ref{eq:Lyapunov},\ref{eq:FinalGamma2}).
Dotted lines are discussed in the text.
Inset show $\gamma_1$ and the IDoS $\IDoS$ on a larger scale.
}
\label{fig:FLld}
\end{figure}

\paragraph{Low density of impurity.---}

Next we consider the low density regime by setting $\bv=\mean{v_n}=100$ (Fig.~\ref{fig:FLld}).
We also see a tendancy for the two curves to converge at high energy, although this occurs on a larger scale, not visible on the figure (see Fig.~\ref{fig:RatioDeych}).
More interestingly, we have compared $\gamma_2$ obtained from \eqref{eq:FinalGamma2} (and also $\gamma_1$) with the approximate expressions (dotted lines) obtained with a different approach in Appendix~\ref{app:PhForm}
\begin{equation}
  \label{eq:Gamma1Gamma2log}
  \begin{cases}
  \gamma_1   
  \simeq
  \rho\,\ln\left(\frac{\bv\,\EXP{-\mathrm{C}}}{2k}\right)
  \\[0.25cm]
  \gamma_2 
  \simeq
  \rho  \, \frac{\pi^2}{4} + \rho\,\ln^2\left(\frac{\bv\,\EXP{-\mathrm{C}}}{2k}\right)
  \end{cases}
  \hspace{0.5cm}\mbox{for }\rho\ll k=\sqrt{E} \ll \bv
  \:,
\end{equation}
where $\mathrm{C}\simeq0.577$ is the Euler-Mascheroni constant.
The agreement is excellent.
For $k\ll\rho$, by continuity, we expect that $\gamma_1$ and $\gamma_2$ saturate to values given by replacing $k$ by $\rho$ in \eqref{eq:Gamma1Gamma2log}~:
$\gamma_1 \sim  \rho\,\ln\left(\bv/\rho\right)$ 
and 
$\gamma_2 \sim  \rho\,\ln^2\left(\bv/\rho\right)$ 
(see \cite{GraTexTou14} for a discussion of the saturation value of~$\gamma_1$).

\subsection{A critical remark on Deych et al.'s criterion for SPS}

The limits of validity of single parameter scaling (SPS) have been the subject of intensive discussions which have motivated the search of a precise analytical criterion.
In a series of papers, Deych, Lisyansky, Altshuler and others
\cite{DeyLisAlt00,DeyLisAlt01,DeyEreLisAlt03} 
have strongly criticized the role played by phase randomization \cite{AndThoAbrFis80,StoAllJoa83,CohRotSha88} in the emergence of SPS (i.e. the requirement of a flat distribution of the phase).~\footnote{
  A more precise study of phase randomization for the 1D Anderson model was provided in Ref.~\cite{BarLuc90}.
} 
Instead, they have proposed a \og universal criterion \fg{} based on the comparison of the localisation length $\xi=1/\gamma_1$ and a characteristic scale $l_s=1/(\pi\IDoS)$, where $\IDoS$ is the integrated density of states per unit length~:
they have stated that SPS holds for $\xi\gg l_s$, i.e. $\gamma_1\ll\IDoS$ \cite{DeyLisAlt00,DeyLisAlt01}.

\begin{figure}[!ht]
\centering
\includegraphics[scale=0.45]{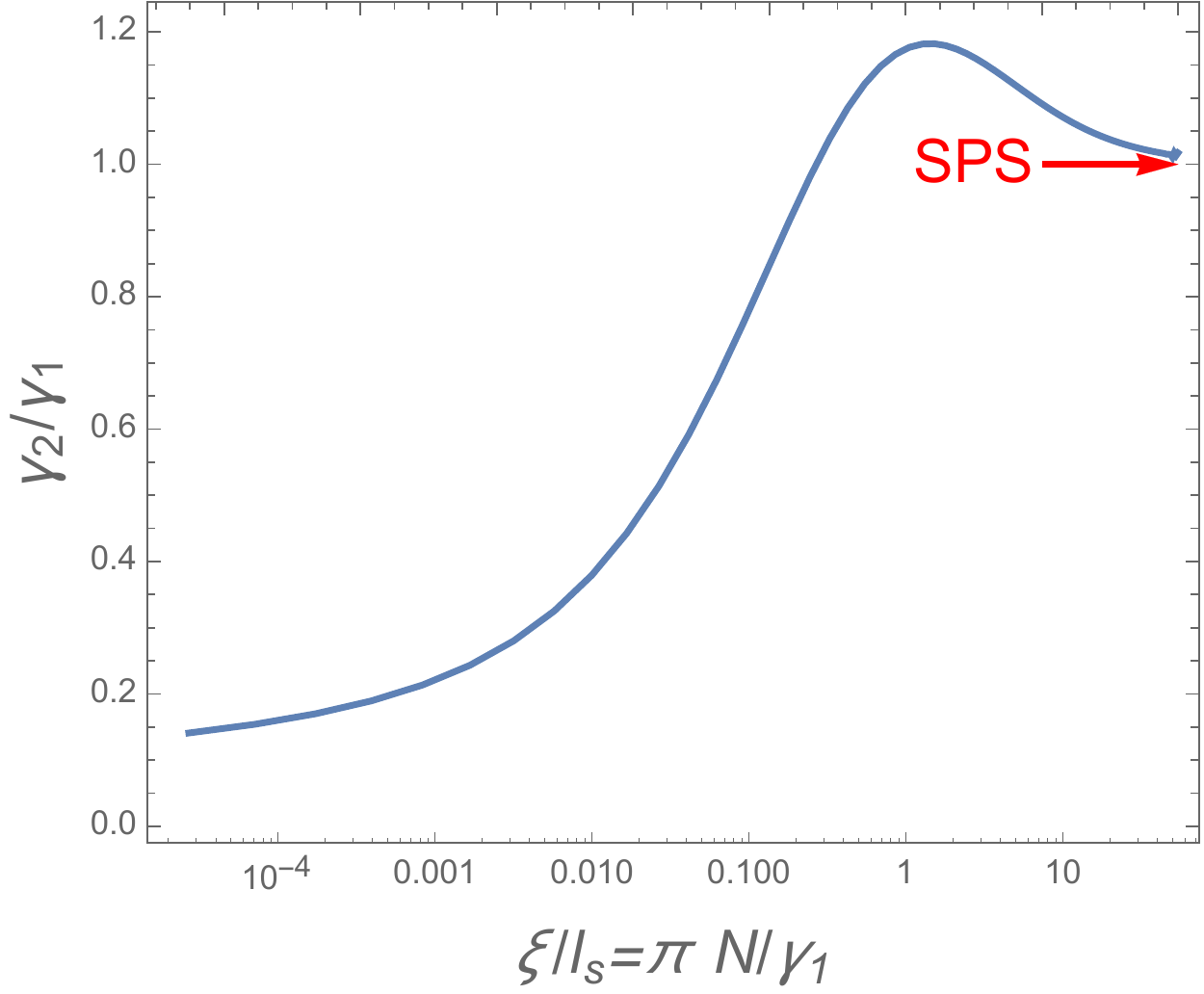}
\hfill
\includegraphics[scale=0.45]{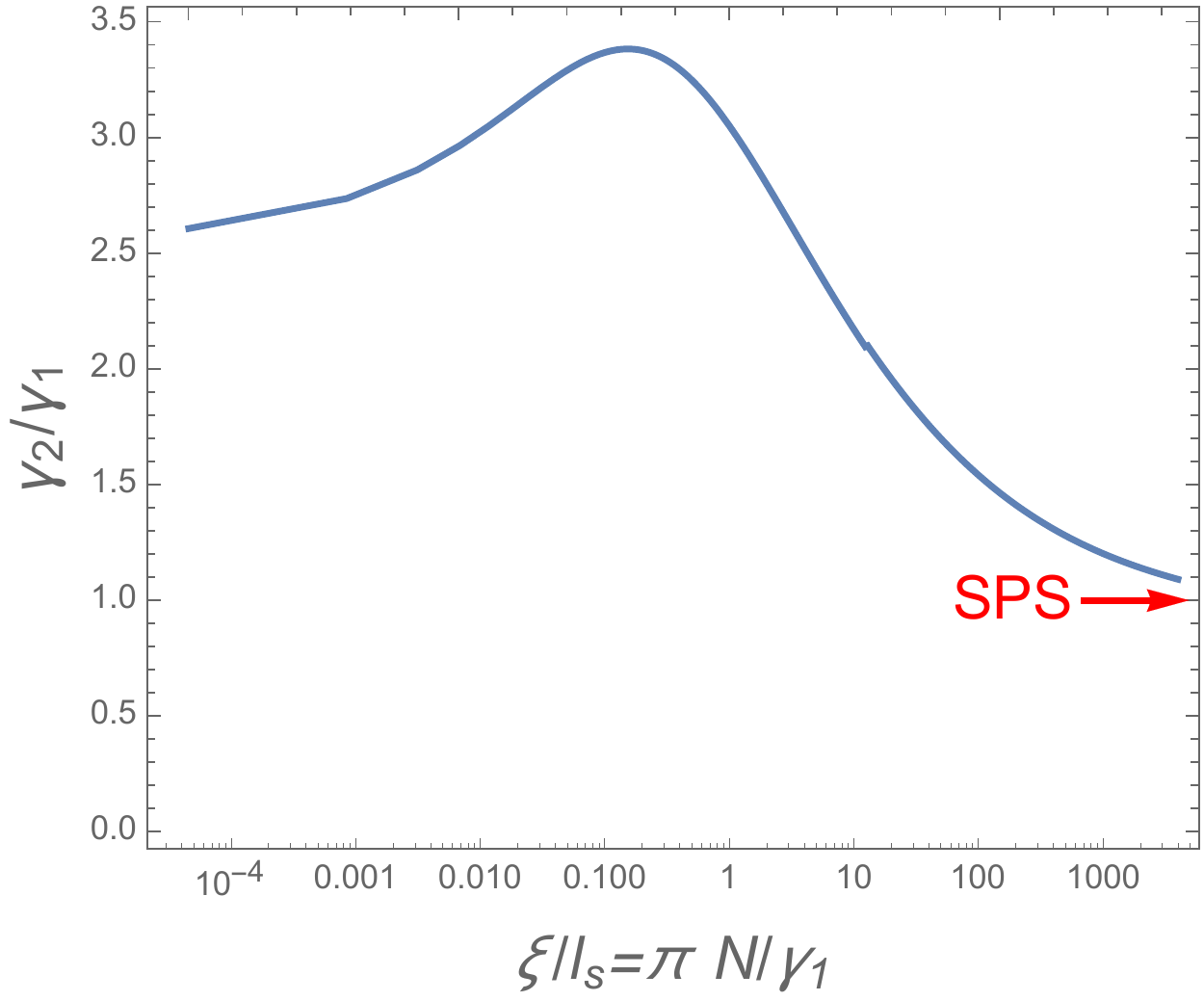}
\caption{\it 
  The ratio $\gamma_2/\gamma_1$ as a function of Deych et al.'s parameter $\xi/l_s$ for high (left) and low impurity density (right).
  We observe that $\xi/l_s\sim1$ does not correspond to the boundary of the SPS regime in general.}
\label{fig:RatioDeych}
\end{figure}

The Frisch \& Lloyd model studied here offers a suitable ground to test Deych et al.'s statement.

In the high density regime, corresponding to the Halperin model, the model has only \textit{one} relevant characteristic energy scale $\sigma^{2/3}=(\rho\mean{v_n^2})^{2/3}$. 
Thus all transitions (strong/weak disorder, peaked/broad phase distribution, etc) occur at this energy scale, in particular the transition from non-SPS to SPS regime~:
cf. left part of Fig.~\ref{fig:FLld} (this is made clear by plotting in left part of Fig.~\ref{fig:RatioDeych} the ratio $\gamma_2/\gamma_1$ as a function of Deych et al.'s scaling parameter $\xi/l_s$)~: 
we observe that $\gamma_2/\gamma_1\sim1$ for $\xi/l_s\gtrsim1$ in this case.

The weak density regime $\rho\ll \bv$ is more interesting as the problem is now controlled by \textit{two} energy scales, thus we have two transitions at the characteristic energies $\rho^2$ and $\bv^2$.
The transition at $k\sim\rho$ corresponds to phase randomization (see chapter 4 of \cite{Tex99} and Appendix \ref{app:PhForm}), however the perturbative regime is expected only for $k\gtrsim\bv$.
In particular, we see in Fig.~\ref{fig:FLld} that, in the intermediate regime $\rho\ll k\ll \bv$ such that $\gamma_1\sim\rho\ll\IDoS\sim k$, SPS does not hold as the two cumulants involve two different scales
$\gamma_1 \sim  \rho\,\ln\left(\bv/\rho\right)$ 
and 
$\gamma_2 \sim  \rho\,\ln^2\left(\bv/\rho\right)$.
In order to emphasize this point, we have plotted in Fig.~\ref{fig:RatioDeych} the ratio $\gamma_2/\gamma_1$ as a function of Deych et al.'s parameter $\xi/l_s=\pi\IDoS/\gamma_1$. 
In this case, the crossover between SPS and non-SPS regime takes place for $\xi/l_s$ much larger than one. 
We conclude that $\xi/l_s\gtrsim1$ \textit{is not a sufficient condition for SPS}.


This discussion thus corroborates the analysis of Schrader, Schulz-Baldes and Sedrakyan \cite{SchSchSed04}~: 
SPS only holds to lowest order perturbation theory in the disorder (see Appendix~\ref{app:PhForm} or chapter 4 of Ref.~\cite{Tex99}).

\section{Continuum limit}
\label{sec:ContinuumLimit}
  
By considering the continuum limit of random matrix products, i.e. matrices $M_n\in\mathrm{SL}(2,\mathbb{R})$ close to the identity matrix, 
it was shown in Ref.~\cite{ComLucTexTou13} that a systematic calculation of the Lyapunov exponent is possible, leading to a full classification of solutions. 
In the Mathematics literature, the stochastic differential equation limits of certain matrix products were also established in several papers \cite{ValVir14,GenGiaGre17,ComGiaGre19}.
In this section, we follow the same strategy  as in Ref.~\cite{ComLucTexTou13}, but for the GLE, although the general resolution of the spectral problem remains out of the scope of the present paper.

At first, we do not assume a specific choice of decomposition.
The group $\mathrm{SL}(2,\mathbb{R})$ is a three parameter group, characterised by infinitesimal generators $\Gamma_i$. 
The index $i$ can label the generators of the Iwasawa decomposition or those of any other decomposition. 
The matrices are decomposed as 
\begin{equation}
  M = M_1(t_1)M_2(t_2)M_3(t_3)
  \simeq 
  \boldsymbol{1}_2 + \sum_{i=1}^3t_i\Gamma_i
  \:,
\end{equation}
where $t_1,\,t_2,\,t_3\to0$ parametrize the elements. 
%
We now reformulate the spectral problem~\eqref{eq:TheSpectralPb}.

\subsection{The spectral problem in the continuum limit}
\label{subsec:SPcontinuumlimit}

The starting point is to expand the transfer operator \eqref{eq:Representation4Elements} up to second order in the parameters
\begin{align}
  \mathscr{T}_{M}(q)
  \simeq 
  1 + \sum_{i=1}^3t_i \, \mathscr{D}_i(q)
  + \frac{1}{2}  \sum_{i=1}^3t_i^2 \, \mathscr{D}_i(q)^2
  + \sum_{i<j}t_it_j \, \mathscr{D}_i(q)\mathscr{D}_j(q)
  \:.
\end{align}
Writing
$
2\,\mathscr{D}_i(q)\mathscr{D}_j(q)
=
\big\{\mathscr{D}_i(q)\,,\,\mathscr{D}_j(q)\big\}
+\big[\mathscr{D}_i(q)\,,\,\mathscr{D}_j(q)\big]
$ 
(sum of the anticommutator and the commutator),
we finally get the useful representation
\begin{equation}
  \mathscr{T}_{M}(q)
  \simeq 
  1 + \sum_{i=1}^3t_i \, \mathscr{D}_i(q)
  + \frac{1}{2} \sum_{i, j}t_it_j \, \mathscr{D}_i(q)\mathscr{D}_j(q)
  + \frac{1}{2} \sum_{i<j}\sum_k t_it_j \, c_{ijk}\mathscr{D}_k(q)
  \:,
\end{equation}
where $c_{ijk}$ are the structure constants of the group, defined by \eqref{eq:DefStructureCstes}.

The continuum limit corresponds to let the parameters $t_i$'s go to zero, assuming that their mean values  and covariances all scale in the same way \cite{ComLucTexTou13},
while properly rescaling the GLE.
This assumption can be rewritten by introducing a small parameter $\epsilon$ and by splitting each parameter as 
\begin{equation}
  t_i = \epsilon\, \tilde\mu_i + \sqrt{\epsilon}\,\tau_i
  \:,
\end{equation}
where $\overline{t_i}\equiv\mean{t_i}=\epsilon\, \tilde\mu_i$ corresponds to the mean value, and $\tau_i$ corresponds to the rescaled fluctuating part, with $\mu_i\sim\mathcal{O}(1)$ and $\tau_i\sim\mathcal{O}(1)$.
Accordingly, it will become clear that the proper rescaling of the GLE corresponds to 
\begin{equation}
  \lambda(q) =\EXP{-\epsilon\,\ell(q) + \mathcal{O}(\epsilon^{3/2})} 
  \hspace{0.5cm}\mbox{i.e.}\hspace{0.5cm}
  \widetilde{\Lambda}(q) = \epsilon\,\ell(q) + \mathcal{O}(\epsilon^{3/2})
  \:.
\end{equation}
Rewriting \eqref{eq:TheSpectralPb}, we get the differential equation
\begin{align}
  \left\{
  \epsilon\,\tvmu  \cdot \underline{\mathscr{D}}(q)
  +
   \frac{\epsilon}{2} \underline{\mathscr{D}}(q)\cdot\tcovma^2\cdot\underline{\mathscr{D}}(q)
   + \frac{\epsilon}{2} \sum_{i<j}\sum_k (\tcovma)_{ij} \, c_{ijk}\mathscr{D}_k(q)
   + \mathcal{O}(\epsilon^{3/2})
  \right\}
  \Phi_q^\mathrm{R}(z)
  \nonumber\\
  = \left[\epsilon\,\ell(q) + \mathcal{O}(\epsilon^{3/2})\right]\, \Phi_q^\mathrm{R}(z)
\end{align}
(we do not make explicit the $\epsilon$-dependence of the eigenfunction for simplicity),
where we have introduced the vectors
\begin{align}
  \tvmu 
  =
  \lim_{\epsilon\to0}
  \frac{1}{\epsilon}
  \begin{pmatrix}
    \overline{t_1} \\ 
    \overline{t_2} \\ 
    \overline{t_3} 
  \end{pmatrix}
  \hspace{0.5cm}\mbox{and}\hspace{0.5cm}
  \underline{\mathscr{D}}(q)
  = \begin{pmatrix}
   \mathscr{D}_1(q) \\
   \mathscr{D}_2(q) \\
   \mathscr{D}_3(q)
  \end{pmatrix}
  = \deriv{}{z}\vec{g}(z) + q\,\vec{h}(z) 
\end{align}
and the covariance matrix with elements 
$(\tcovma^2)_{ij} 
=\lim_{\epsilon\to0}
  \frac{1}{\epsilon}\mathrm{Cov}(t_i,t_j)
$.
In the limit $\epsilon\to0$, we get 
\begin{equation}
  \label{eq:SpectralPbContinuum}
  \left\{
   \frac{1}{2} \underline{\mathscr{D}}(q)\cdot\tcovma^2\cdot\underline{\mathscr{D}}(q)
   + \frac{1}{2} \vec{\tilde{c}}\cdot
     \Big(\underline{\mathscr{D}}(q)\times\underline{\mathscr{D}}(q)\Big)
   +    \tvmu  \cdot \underline{\mathscr{D}}(q)
  \right\}
  \Phi_q^\mathrm{R}(z)
  = \ell(q)\, \Phi_q^\mathrm{R}(z)
\end{equation}
where $\vec{\tilde c}=((\tcovma^2)_{23},-(\tcovma^2)_{13},(\tcovma^2)_{12})$.
Taking the limit $\epsilon\to0$, the information on higher cumulants of the three parameters $t_i$'s have disappeared, as we are considering the Brownian motion limit of the random walk in the group. 

The existence of a non-trivial continuum limit means that the rescaled GLE $\ell(q)$ can be expressed in terms of a scaling function $G_q$ of nine arguments, the three mean values and the six elements of the covariance matrix. Thus
\begin{equation}
  \widetilde{\Lambda}(q) \underset{\epsilon\to0}{\simeq}
  \epsilon\,\ell(q)
  =
  \epsilon\:
  G_q\!\left(
    \frac{\overline{t_1}}{\epsilon}, \cdots, \frac{\overline{t_3}}{\epsilon},\frac{\mean{t_1^2}}{\epsilon}, \frac{\mean{t_1t_2}}{\epsilon}, \cdots,\frac{\mean{t_3^2}}{\epsilon}
  \right)
  \:.
\end{equation}
The function $G_q$ characterizes the limiting process obtained in the continuum limit. 
Furthermore, one can reduce the number of arguments by one, since one can always choose $\epsilon$ equal to one of the nine parameters, e.g. $\epsilon=\overline{t_1}$.

Below, we prefer to use the simplest notations of Ref.~\cite{ComLucTexTou13} and deal with the (non rescaled) small parameters $\mu_i=\overline{t_i}$ and $(\covma^2)_{i,j}=\mathrm{Cov}(t_i,t_j)$,
 so that we can simply write the spectral problem
\begin{equation}
  \label{eq:SpectralPbContinuum3}
  \boxed{
  \left\{
   \frac{1}{2} \underline{\mathscr{D}}(q)\cdot\covma^2\cdot\underline{\mathscr{D}}(q)
   + \frac{1}{2} \vec{c}\cdot
     \Big(\underline{\mathscr{D}}(q)\times\underline{\mathscr{D}}(q)\Big)
   +    \vec{\mu}  \cdot \underline{\mathscr{D}}(q)
  \right\}
  \Phi_q^\mathrm{R}(z)
  \simeq \widetilde{\Lambda}(q)\, \Phi_q^\mathrm{R}(z)
}
\end{equation}
 where
\begin{equation}
  \vec{c}
  =\begin{pmatrix}
    \big(\covma^2\big)_{23} \\[0.125cm]
    -\big(\covma^2\big)_{13} \\[0.125cm] 
    \big(\covma^2\big)_{12} 
  \end{pmatrix}
  \:.
\end{equation}
In a more explicit form, Eq.~\eqref{eq:SpectralPbContinuum3} reads~:
\begin{align}
  \label{eq:SpectralPbContinuum2}
  \deriv{}{z}
  &\left\{
      \frac{1}{2}\vec{g}(z)\cdot \covma^2\cdot \deriv{}{z}\vec{g}(z)
    + \frac{1}{2}\vec{c} \cdot \left[\vec{g}(z)\times\vec{g}'(z)\right] + \vec{\mu}  \cdot \vec{g}(z)
  \right\}\Phi_q^\mathrm{R}(z)
  =\bigg\{
    \widetilde{\Lambda}(q)
  \nonumber
  \\
   &\hspace{-.5cm}
     - q \left(
           \vec{h}(z) \cdot\covma^2 \cdot\deriv{}{z}\vec{g}(z) 
           +\frac{1}{2}\vec{h}'(z)\cdot \covma^2 \cdot\vec{g}(z) 
           + \frac{1}{2}\vec{c} \cdot \left[\vec{g}(z)\times\vec{h}'(z)\right]
           +  \vec{\mu}   \cdot \vec{h}(z)
         \right)
  \nonumber
  \\
   &\hspace{.5cm}
     - \frac{q^2}{2}    \vec{h}(z)\cdot \covma^2 \cdot\vec{h}(z) 
  \bigg\}   \Phi_q^\mathrm{R}(z)
\end{align}
Setting $q=0$, we recover Eq.~1.34 of Ref.~\cite{ComLucTexTou13} for the invariant density
\footnote{
In Ref.~\cite{ComLucTexTou13}, a general formula for the Lyapunov exponent was used:
\begin{equation}
  \label{eq:Eq2-46CLTTcorrected}
   \lambda_1 = -\overline{w}
   + \int\D z\, z \left\{
     \overline{\theta} 
     + \deriv{}{z} \left(
       \frac{D_{\theta\theta}}{2}(1+z^2) - 2D_{\theta w} \, z - D_{\theta u} 
     \right)
   \right\}
   f(z)
\end{equation}
which is the limit of \eqref{eq:GeneralFormulaLyapunov} for small parameters.
In Eq.~2.46 of Ref.~\cite{ComLucTexTou13}, the integral was (incorrectly) splitted, which is not always possible. A simple example where it cannot be splitted is the case with disorder on angles, $D_{\theta\theta}\neq0$ with finite mean values $\btheta$, $\bw$ and $\bu$: the invariant density $f$ is easy to find and one can see that when splitted, the integrals do not converge, even in principal part, while \eqref{eq:Eq2-46CLTTcorrected} is convergent. 
}
 $\Phi_0^\mathrm{R}=f$. 

If we consider the Iwasawa decomposition, the vectors collecting the three infinitesimal generators are $\vec{g}(z)= ( g_K \, , \, g_A \,,\, g_N )$
and $\vec{h}(z)= ( h_K \, , \, h_A \,,\, h_N )$, where the functions are given in Table~\ref{tab:InfGenerators}.


\subsection{Illustration 1 : the Halperin model (matrices $KN$ and $\widetilde{K}N$)}
\label{subsec:Halperin}

We first consider the spectral problem \eqref{eq:SpectralPbContinuum2} for fixed angles $\theta_n=\btheta$ with disorder on $u_n$'s only~:
\begin{equation}
   \vec{\mu} 
   = 
   \begin{pmatrix}
     \btheta 
     \\
     0
     \\
     0
   \end{pmatrix}
   \hspace{0.5cm}\mbox{and}\hspace{0.5cm}
   \covma^2
   = 
   \begin{pmatrix}
      0 & 0 & 0 \\
      0 & 0 & 0 \\
      0 & 0 & D_{uu}
   \end{pmatrix}
   \:.
\end{equation}
This corresponds to transfer matrices for the Schr\"odinger equation with $\delta$-impurities with random weights on a lattice. 
The continuum limit corresponds to send the lattice spacing $\propto\btheta$ to zero, while scaling the disorder $D_{uu}$ to zero in the same way.

Choosing the measure $J=J_N$, we see that the components of the two vectors $\vec{g}(z)$ and $\vec{h}(z)$ are polynomials, which allows to take the Fourier transform of Eq.~\eqref{eq:SpectralPbContinuum2}, $z\to\I\partial_s$ and $\partial_z\to\I s$, leading to 
\begin{equation}
  \I s \left[
     \frac{\I}{2}D_{uu}\, s  + \btheta \left(1 - \deriv{^2}{s^2} \right) 
  \right]
  \widehat{\Phi}^\mathrm{R}_q(s)
  = 
  \left[
     \widetilde{\Lambda}(q) - \I q \btheta \deriv{}{s}
  \right]  
  \widehat{\Phi}^\mathrm{R}_q(s)
  \:.
\end{equation}
We introduce the notations
\begin{equation}
  \frac{D_{uu}}{\btheta} = \frac{\sigma}{k^3}  
  \hspace{0.5cm}\mbox{and}\hspace{0.5cm}
  \frac{\widetilde{\Lambda}(q)}{\btheta} = \frac{\Lambda(q) }{k}
\end{equation}
which now allow to take the limit $\btheta\to0$ and $D_{uu}\to0$.
Rescaling $s$ as $s\to ks$, we get
\begin{equation}
  \label{eq:SPHalperinFourier}
  \I s \left[
    - \deriv{^2}{s^2}+ E + \I\frac{\sigma}{2}\, s 
  \right]
  \widehat{\Phi}^\mathrm{R}_q(s)
  = 
    \left[
     \Lambda(q) - \I q \deriv{}{s}
  \right]  
  \widehat{\Phi}^\mathrm{R}_q(s)
  \:,
\end{equation}
where $E=+k^2$ (the case $E=-k^2$ corresponds to matrices $\widetilde{K}N$).
We have precisely recovered the equation \eqref{eq:LocalisationEqForPhiR} for the Halperin model corresponding to the Schr\"odinger equation \eqref{eq:Schrodinger} for a Gaussian white noise potential $\mean{V(x)V(x')}=\sigma\,\delta(x-x')$, corresponding to the L\'evy exponent $\levy(s)=\sigma s^2/2$. This was expected since $\btheta\to0$ corresponds to take the limit of high density of impurities (see also Appendix~\ref{app:Halperin}).

\begin{figure}[!ht]
\centering
\includegraphics[scale=0.6]{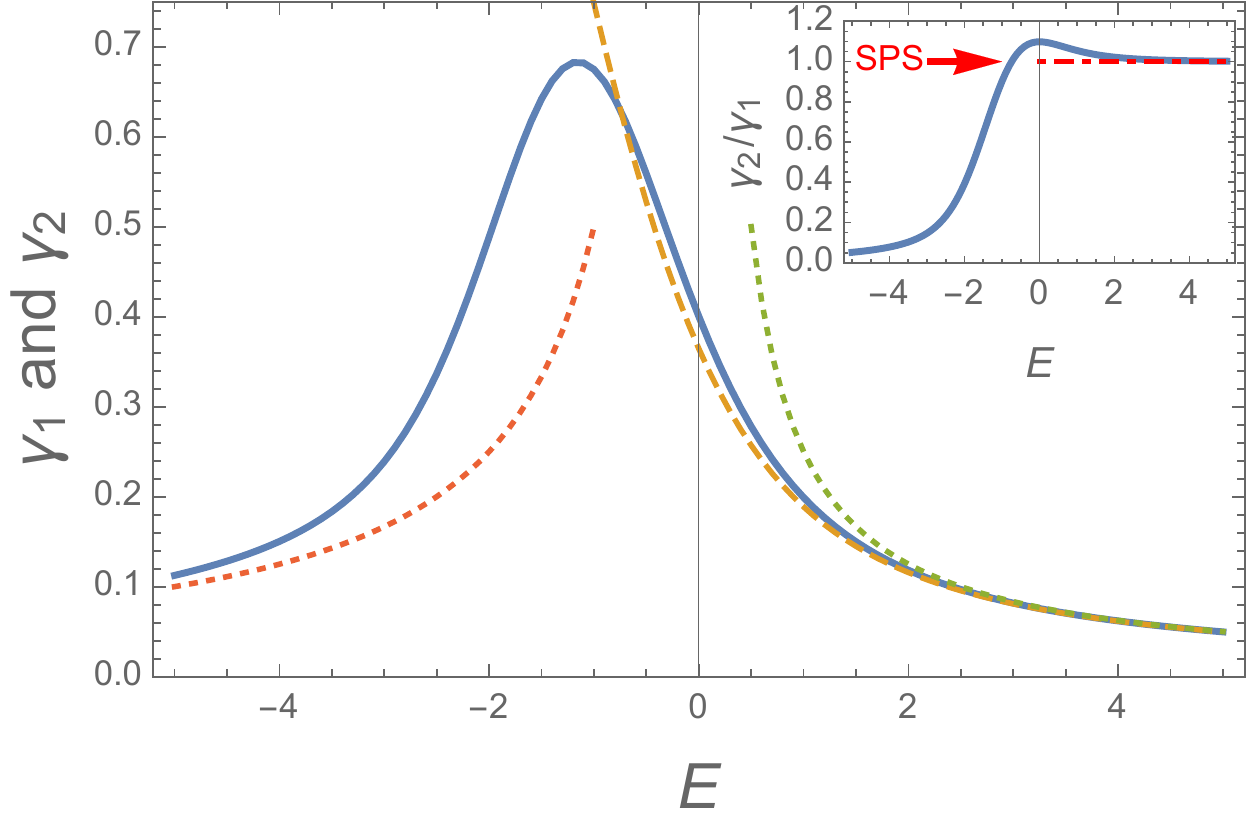}
\caption{\it  The Lyapunov exponent $\gamma_1$ (dashed orange line) and the variance $\gamma_2$ (blue line) from (\ref{eq:PhiHalperin},\ref{eq:FinalGamma2}) for the Halperin model with $\sigma=2$.
  The dotted lines are the two asymptotics 
   \eqref{eq:HalperinHighEnergy} and \eqref{eq:HalperinLowEnergy}.  
  Inset~: the rapid convergence of the ratio $\gamma_2/\gamma_1\to1$ for $E\to+\infty$ corresponds to SPS.
}
\label{fig:Halperin}
\end{figure}

\paragraph{The variance.---}

As it corresponds to \eqref{eq:LocalisationEqForPhiR} for $\levy(s)=\sigma\,s^2/2$, Eq.~\eqref{eq:SPHalperinFourier} is solved along the lines of Section~\ref{sec:Perturbation}. 
we get \eqref{eq:FinalGamma2} where $\hat{f}(s)$ is now expressed in terms of Airy functions \eqref{eq:PhiHalperin}.~\footnote{
  Equivalently we can obtain the variance from \eqref{eq:FinalLambda2} by taking the continuum limit, $\gamma_2=-\lim_{\rho\to\infty}\rho\lambda_2$ (with $\mean{v_n}=0$ and $\rho\mean{v_n^2}=\sigma$ fixed, cf.~Appendix~\ref{app:Halperin}).
}
Therefore, here we have obtained an explicit form of the variance in terms a \textit{single} integral, instead of the multiple integrals of Refs.~\cite{SchTit02,RamTex14}.
The result is plotted in Fig.~\ref{fig:Halperin} and perfectly agrees with these papers. 
We also remark that the curves are similar to the one obtained for the Frisch-Lloyd model for a high density of impurities, $\rho\gg v$ (cf. Fig.~\ref{fig:FLhd}).
 
We recall the limiting behaviours.
In the perturbative regime one gets \cite{AntPasSly81,Tex99,RamTex14} 
\begin{equation}
  \label{eq:HalperinHighEnergy}
     \gamma_1\simeq\gamma_2 \simeq \frac{\sigma}{8E} 
     \hspace{0.5cm}\mbox{for }
     E\to+\infty
\end{equation} 
(this property expresses to \og single parameter scaling \fg{}, discussed in Subsec.~\ref{subsec:RSO}). Furthermore, 
by analyzing the first few cumulants,
 Schomerus and Titov \cite{SchTit02} also showed that the GLE is quadratic in this limit, $\Lambda(q)\simeq\gamma_1\,\big[q+\frac12q^2\big]$, i.e. that higher cumulants decay faster with $E$. 
  The same conclusion was reached in \cite{Tex19b} by a direct analysis of the GLE.
In the other limit, one has \cite{RamTex14,FyoLeDRosTex18}  
\begin{equation}
  \label{eq:HalperinLowEnergy}
     \gamma_1\simeq \sqrt{-E} + \frac{\sigma}{8E} 
     \hspace{0.5cm}\mbox{and}\hspace{0.5cm}
     \gamma_2 \simeq \frac{\sigma}{4(-E)} 
     \hspace{0.5cm}\mbox{for }
     E\to-\infty
     \:.
\end{equation}


\paragraph{Remark on analyticity~:}

Analytic properties are known to be central for the analysis of the Lyapunov exponent $\gamma_1$~: 
see the general discussion of the book \cite{Luc92} or the specific discussions in \cite{ComTexTou13,ComLucTexTou13}.
More precisely, analyticity arises from the fact that the Lyapunov exponent $\gamma_1$ and the integrated density of states $\IDoS $ are real and imaginary parts of a single analytic function of the energy, respectively.
In particular, the limiting behaviours $\gamma_1\simeq\sqrt{-E}$ for $E\to-\infty$ and $\pi \IDoS \simeq\sqrt{E}$ for $E\to+\infty$ mentioned above have the same origin.

One could ask whether a similar property holds for the variance $\gamma_2$. The results obtained for the Halperin case suggests that the answer is negative. Indeed, the inspection of the limiting behaviours 
$\gamma_2\simeq\sigma/(-4E)$ for $E\to-\infty$ and $\gamma_2\simeq\sigma/(8E)$ for $E\to+\infty$ 
shows that $\gamma_2$ cannot be the real part of an analytic function of~$E$.

\paragraph{Large deviations~:}

the large deviations, i.e. the large $q$ behaviour of the GLE $\Lambda(q)$ was considered in \cite{FyoLeDRosTex18} (and earlier for even $q$ by the replica trick in \cite{BouGeoLeD86}), where it was shown that $\Lambda(q)\simeq\frac{3}{4}(\sigma/2)^{1/3}\,|q|^{4/3}$ for $q\to+\infty$ (replica trick was also applied to RMP in \cite{BouGeoHanLeDMai86}).~\footnote{
  note that the exponent $4/3$ is compatible with the numerics of Ref.~\cite{ZilPik03}.
}
Using the symmetry \eqref{eq:Vanneste2010} discussed in the conclusion, we conclude that this is also the behaviour for $q\to-\infty$.
Accordingly, the distribution of the wave function $\Upsilon(x)=\ln|\psi(x)|$ takes the form
$
  \mathscr{P}_x(\Upsilon) \sim \exp\big\{-\Upsilon^4/(2\sigma x^3)\big\}
$
for $x\to\infty$ and $\Upsilon\to\pm\infty$ (note that this characterizes the solution $\psi(x)$ of the Cauchy initial value problem and not the normalised wave function).


\subsection{Illustration 2 : the Dirac/supersymmetric case (matrices $\widetilde{K}A$)}
\label{sec:ContinuumSusy}

The case of matrices of type $\widetilde{K}A$ in the continuum limit has been considered in the literature~: 
choosing
\begin{equation}
   \vec{\mu} 
   = 
   \begin{pmatrix}
     \btheta 
     \\
     \bw
     \\
     0
   \end{pmatrix}
   \hspace{0.5cm}\mbox{and}\hspace{0.5cm}
   \covma^2
   = 
   \begin{pmatrix}
      0 & 0 & 0 \\
      0 & D_{ww} & 0 \\
      0 & 0 & 0
   \end{pmatrix}
\end{equation}
corresponds to the Dirac equation \eqref{eq:Dirac} when the mass is a Gaussian white noise or equivalently the supersymmetric Schr\"odinger equation. The Lyapunov exponent was first derived in \cite{BouComGeoLeD90} (see also \cite{ComTexTou11,ComLucTexTou13,GraTexTou14}).
The variance $\gamma_2$ was computed in \cite{RamTex14}.
This case is also related to a model of random matrix products introduced by Derrida and Hilhorst \cite{DerHil83} which has attracted a lot of attention (see for example \cite{Luc93,FigMosKno98,GenGiaGre17}).
Note however that the matrices $T_n$ considered by Derrida and Hilhorst do not have a unit determinant~:
we can write $T_n=M_n\,\sqrt{\det T_n}$ where $M_n=\widetilde{K}_nA_n$.
Hence, the Lyapunov exponents for matrices $T_n$ and $M_n$ only differ by a trivial constant, however the difference between the variances of $\ln||\big(\prod_nT_n\big)\vec{x}_0||$, studied in \cite{ComGiaGre19}, and $\ln||\big(\prod_nM_n\big)\vec{x}_0||$, studied in \cite{RamTex14} and here, involves the correlation between $\ln||\big(\prod_nM_n\big)\vec{x}_0||$ and $\sum_n\ln\det T_n$.

Here we simply write the main equation \eqref{eq:SpectralPbContinuum2}
\begin{align}
  \label{eq:9.18}
  &\deriv{}{z}
  \left\{  2 D_{ww} \, z\deriv{}{z} z + \btheta\, (-1+z^2) - 2\bw \, z \right\}\Phi_q^\mathrm{R}(z)
\nonumber
\\
&  =
  \bigg\{ \widetilde{\Lambda}(q)  
   + q \left[
      D_{ww} \left( 2\,h_A(z)\deriv{}{z} z  + h_A'(z)\,z \right) - \btheta\,h_{\widetilde{K}}(z)
      - \bw\,h_A(z) \right]
\nonumber
\\  
&\hspace{1.5cm}
       -q^2\,\frac{D_{ww}}{2}h_A(z)^2\bigg\}\Phi_q^\mathrm{R}(z)
      \:.
\end{align}
Inspection of Table~\ref{tab:InfGenerators} makes clear that the choice $J=J_A$ greatly simplifies the problem since $h_A=0$. 
In order to make contact with the notations of Refs.~\cite{ComTexTou11,GraTexTou14,RamTex14}, we reintroduce $k$ in \eqref{eq:9.18} by performing the following substitutions~:
$\bw/\btheta=-\mu g/k$,
$D_{ww}/\btheta=g/k$,
$\widetilde{\Lambda}(q)/\btheta=\Lambda(q)/k$ and $z\to z/k$. We get
\begin{align}
  \label{eq:SpectralPbForSUSYContinuum}
  \deriv{}{z}
  \left\{  2 g \, z\deriv{}{z} z -k^2+z^2 +2\mu\,g \, z \right\}\Phi_q^\mathrm{R}(z)
  =
  \bigg\{ \Lambda(q)  
   - \frac{q}{2} \, \left( z+ \frac{k^2}{z}\right)
   \bigg\}\Phi_q^\mathrm{R}(z)
   \:.
\end{align}
To proceed with the perturbative approach we consider the adjoint problem
$
\mathscr{G}_z\Phi_q^\mathrm{L}(z)
=
\big\{ 
  \Lambda(q) - \frac{q}{2} \, \big( z+ \frac{k^2}{z}\big)
\big\}\Phi_q^\mathrm{L}(z)
$.
This corresponds to follow the little discussion at the end of Section~\ref{sec:Intro}.
We recover some of the formulae of Ref.~\cite{RamTex14}~:
$
  \gamma_1 = \frac{1}{2}\int\D z\, \left( z+ {k^2}/{z^2} \right)\, f(z)
$
and  
$
  \gamma_2 = \int\D z\,L_1(z)\, \left( z+ {k^2}/{z^2} \right)\, f(z)
$
where $\mathscr{G}_zL_1(z)=\gamma_1-\frac{1}{2} \, \big( z+ \frac{k^2}{z}\big)$ with the condition $\int\D z\,L_1(z)\,f(z)=0$.
It is interesting to point that, while the combination $z+k^2/z$ has resulted from a trick in Ref.~\cite{RamTex14}, the new presentation gives a group theoretical interpretation of this trick.


Another convenient way to study the differential equation \eqref{eq:SpectralPbForSUSYContinuum} is to use Mellin transform, leading to the continuum limit of \eqref{eq:SPsusyMellin}.

  
\subsection{Illustration 3 : a two parameter subgroup (matrices $AN$)}
\label{subsec:CaseAN}

Finally, we consider matrices of the form $M_n=A(w_n)N(u_n)$.
Contrary to the previous examples (matrices $KN$ or $\widetilde{K}A$), such matrices form a two parameter subgroup of $\mathrm{SL}(2,\mathbb{R})$. 
In the continuum limit, the problem is characterised by five parameters~: 
\begin{equation}
  \vec{\mu} 
  =\begin{pmatrix}
    0 \\ 
    \bw \\ 
    \bu
  \end{pmatrix}
    \hspace{0.5cm}\mbox{and}\hspace{0.5cm}
    \covma^2 =
   \begin{pmatrix}
       0 & 0 & 0 \\
       0 & D_{ww} & D_{wu} \\
       0 & D_{wu} & D_{uu}
   \end{pmatrix}
   \:.
\end{equation}
Two natural choices for the measure which lead to simplifications are $J=J_A$ or $J=J_N$.
Inspection of Table~\ref{tab:InfGenerators} shows that the second choice, with $h_A=-1$ and $h_N=0$, further simplifies the analysis as it leads to a differential equation of hypergeometric type (which is \textit{a priori} not the case for $J=J_A$).
Some algebra shows that \eqref{eq:SpectralPbContinuum2} takes the form
\begin{align}
  \label{eq:SPsubgroupAN}
  \bigg\{
   & \left(2D_{ww}z^2+2D_{wu}\,z+\frac12D_{uu}\right)\deriv{^2}{z^2}
  \nonumber\\
  &+ \Big( 2\left[ (3+q)D_{ww} - \bw \right] z + (2+q)D_{wu} -\bu \Big) \deriv{}{z}
  \nonumber\\
  &+ D_{ww}\left(2+2q+\frac12q^2\right)-(q+2) \bw - \widetilde{\Lambda}(q) 
  \bigg\}
  \Phi_q^\mathrm{R}(z)
  =0
  \:,
\end{align}
which is of the hypergeometric form
\begin{equation}
  \label{eq:Hypergeom}
  x\,(1-x)\, y''(x) + \big[ c - (a+b+1)\,x \big]\, y'(x) - ab\,y(x) = 0
  \:.
\end{equation}
The two roots of the polynomial $4D_{ww}z^2+4D_{wu}\,z+D_{uu}$ are $\omega$ and $\omega^*$ with 
\begin{equation}
  \omega = \frac{\I\sqrt{D_{ww}D_{uu}-D_{wu}^2}-D_{wu}}{2D_{ww}}
  \:. 
\end{equation}
Two independent solutions of \eqref{eq:SPsubgroupAN} are for instance 
\begin{equation}
   _2F_1\!\left( a, b ; c ; \frac{z-\omega^*}{\omega-\omega^*} \right)
   \hspace{0.5cm}\mbox{and} \hspace{0.5cm}
   _2F_1\!\left( a, b ; c ; \frac{z-\omega}{\omega^*-\omega} \right)
   \:.
\end{equation}

If we set $\bu=0$ and $D_{wu}=0$, the expression simplifies and we get 
\begin{align}
   _2F_1\!\left( a, b ; c ;\frac12 -  \I\sqrt{\frac{D_{ww}}{D_{uu}}} \,z   \right)
\hspace{0.25cm}\mbox{with }
\label{eq:abc}
\begin{cases}
  ab = 1+q+\frac{q^2}{4} - \left(1+\frac{q}{2}\right)\frac{\bw}{D_{ww}} - \frac{\widetilde{\Lambda}(q)}{2D_{ww}}
  \\
  a+b = 2 + q -\frac{\bw}{D_{ww}}
  \\
  c = \frac{1}{2}\left(3 + q -\frac{\bw}{D_{ww}}\right)
\end{cases}  
\end{align}
In order to solve the spectral problem \eqref{eq:SPsubgroupAN}, the next step is to express $\Phi_q^\mathrm{R}(z)$ as a linear combination of two such solutions. 
The condition that it vanishes both at $+\infty$ and $-\infty$ provides the GLE.

\subsubsection{A simple case}

We further simplify the analysis by setting $D_{uu}=4D_{ww}=1$ and $D_{wu}=0$, with $\bw<0$.
This case is related to the diffusion studied in \cite{BalTarFigYor01} (see also Section~8 of \cite{ComLucTexTou13}).
Eq.~\eqref{eq:SPsubgroupAN} gives
\begin{align}
  \label{eq:SPyor}
  \left[
    \frac{1}{2}\left(\deriv{}{z}z + \frac{q}{2}\right)^2 + \frac{1}{2} \deriv{^2}{z^2}
    -2 \bw \left(\deriv{}{z}z + \frac{q}{2}\right)
    - \bu \deriv{}{z}
  \right]
  \Phi_q^\mathrm{R}(z)
  =\widetilde{\Lambda}(q)\, \Phi_q^\mathrm{R}(z)
  \:.
\end{align}  
As we have already seen above, it is useful to analyse first the asymptotic behaviours of the solution. Keeping the terms dominant at infinity we have
\begin{equation}
  \left[
      \left(\deriv{}{z}z\right)^2 
    + q\, \deriv{}{z}z + \frac{q^2}{4} 
    - 4\bw \deriv{}{z}z  -2\bw\,q - 2 \widetilde{\Lambda}(q)
  \right]
  \Phi_q^\mathrm{R}(z)
  \simeq 0
\end{equation}
leading to 
\begin{equation}
  \label{eq:AsympPhiSimpleCase}
  \Phi_q^\mathrm{R}(z) \sim |z|^{-1-\frac{q}{2}+2\bw-\sqrt{4\bw^2+2\widetilde{\Lambda}(q)}}
  \hspace{0.5cm}\mbox{for }
  z\to\pm\infty
  \:.
\end{equation}
With this observation in mind, we now apply the perturbative approach to \eqref{eq:SPyor}.

\paragraph{Order $q^0$.---}

At order $q=0$, we get the differential equation for the invariant density $\Phi_0^\mathrm{R}(z)=f(z)$~:
\begin{equation}
   \left[ (1+z^2)\deriv{}{z} + (1-4\bw)\,z  - 2\bu  \right] f(z)=0
\end{equation}
where we have assumed a vanishing diffusion current (this is imposed by normalisability).
We deduce 
\begin{equation}
  \label{eq:InvariantDensityAN}
  f(z) = \frac{\mathcal{A}_0}{(1+z^2)^{\frac12-2\bw}} \,\EXP{2\bu\,\arctan z}
\end{equation}
where $\mathcal{A}_0$ is a normalisation constant.
This is precisely the hitting distribution obtained in \cite{BalTarFigYor01}.
The solution is normalisable only for 
$\bw<0$.

\paragraph{Order $q^1$.---}

We now identify the order $q^1$ terms of \eqref{eq:SPyor}~:
\begin{equation}
  \label{eq:Order1simpleCase}
     \deriv{}{z}\left[ (1+z^2)\deriv{}{z} + (1-4\bw)\,z  - 2\bu  \right] R_1(z)=
     2\left( \gamma_1 + \bw - \frac{1}{2}\deriv{}{z}z  \right)f(z)
     \:.
\end{equation}
Asymptotic behaviour \eqref{eq:AsympPhiSimpleCase} shows that the dominant term of the $q^n$ term is $R_n(z)\sim|z|^{-1+4\bw}\ln^n|z|$ so that $z^2R_1'(z)$ vanishes at infinity for $\bw<0$. Hence, integration of \eqref{eq:Order1simpleCase} simply gives 
\begin{equation}
  \gamma_1 = -\bw > 0
  \:.
\end{equation}

\paragraph{Order $q^2$.---}

We get from \eqref{eq:SPyor}~:
\begin{align}
  \label{eq:Order2simpleCase}
     &\deriv{}{z}\left[ (1+z^2)\deriv{}{z} + (1-4\bw)\,z  - 2\bu  \right] R_2(z)
     \nonumber\\
     =
     &\left(\gamma_2 - \frac{1}{4}\right) f(z)
     + 2\left( \gamma_1 + \bw - \frac{1}{2}\deriv{}{z}z  \right)R_1(z)
     \:.
\end{align}
Once again, after integration we get rid of boundary terms which all vanish at infinity, leading to 
$\gamma_2 - \frac{1}{4}+2\left( \gamma_1 + \bw\right)=0$, thus 
\begin{equation}
  \gamma_2 = \frac{1}{4}
  \:.
\end{equation}
We see that the parameter $\bu$ plays no role.

\paragraph{Exact solution of the spectral problem.---}

We now show that it is possible to solve exactly the spectral problem \eqref{eq:SPyor}.
Since $\bu$ has played no role in the above analysis, we set $\bu=0$ and introduce the notation $\epsilon=-2\bw>0$ for convenience.
Equations \eqref{eq:abc} have solution
\begin{align}
\begin{cases}
   a = 1 + \epsilon + \frac{q}{2} + \sqrt{\epsilon^2+2\widetilde{\Lambda}}
   \\
   b = 1 + \epsilon + \frac{q}{2} - \sqrt{\epsilon^2+2\widetilde{\Lambda}}
   \\
   c = \frac{3+q}{2} + \epsilon 
\end{cases}
\end{align}
Seeking a solution vanishing at infinity, it is convenient to use the solution of \eqref{eq:Hypergeom}  
\begin{equation}
   x^{-a}\: _2F_1\!\left( a, a-c+1 ; a-b+1 ; 1/x \right) 
   \hspace{0.5cm}\mbox{with }
   x = \frac{1-\I z}{2}
   \:.
\end{equation}
Setting $2\widetilde{\Lambda}=\epsilon q+q^2/4$, we get 
\begin{equation}
   a = 2(a-c+1) = a-b+1 = 1 + q + 2\epsilon
   \:.
\end{equation}
We recognize the function $_2F_1\!\left( a, a/2 ; a ; 1/x \right)=(1-1/x)^{-a/2}$.
Thus, $\Phi_q^\mathrm{R}(z)\propto x^{-a}\,(1-1/x)^{-a/2}\propto (1+z^2)^{-a/2}$ 
vanishes for $z\to\pm\infty$. 
We conclude that the exact solution of the spectral problem \eqref{eq:SPyor} (for $\bu=0$) is 
\begin{equation}
  \label{eq:PhiRexactCasSimple}
  \Phi_q^\mathrm{R}(z) 
  = \mathcal{A}_q\, \left( 1+z^2 \right)^{-\frac{1}{2}-\epsilon-\frac{q}{2}}
  = \mathcal{A}_q\, \left( 1+z^2 \right)^{-\frac{1}{2}+2\bw-\frac{q}{2}}
\end{equation}
with 
\begin{equation}
  \label{eq:GLEcaseAN}
  \widetilde{\Lambda}(q) 
  = \frac{1}{2}\left(q\epsilon + \frac{q^2}{4}\right) 
  = -q\,\bw + \frac{q^2}{8}
  \:,
\end{equation}
(this result coincides with Eq.~8.25 of \cite{ComLucTexTou13}, which was obtained by a direct study of the diffusion in the hyperbolic plane).
The constant $\mathcal{A}_q$ is a normalisation (for $q=0$, we have $\int\Phi_0^\mathrm{R}=\int f=1$ and thus $\mathcal{A}_0=B(1/2,\epsilon)^{-1}$, where $B(x,y)$ is the Euler beta function). 
The GLE is quadratic and we have simply recovered the two leading terms given by the perturbative approach.
Thus, we have found that, for products of matrices of the form $M_n=A_nN_n$, in the continuum limit the distribution of $\ln||\Pi_n\vec{x}_0||$ is exactly Gaussian~: 
this is a rather unusual feature and large deviations are in general non-Gaussian (see for example the discussion of large deviations for the Halperin model in Ref.~\cite{FyoLeDRosTex18}, or Subsection~\ref{subsec:Halperin}).

\paragraph{Remark~:}

Equation \eqref{eq:GLEcaseAN} is a particular instance of the general formula for the GLE of matrices $M_n=A(w_n)N(u_n)$:
\begin{equation}
  \label{eq:TrivialGLE}
  \widetilde{\Lambda}(q)  = \ln\mean{\EXP{\pm qw_n}}
  \hspace{0.5cm}\mbox{where }
  \pm=\mathrm{sign}(\bw)
  \:,
\end{equation}
valid beyond the continuum limit for $\bw\neq0$.
This formula can be proven by making use of the upper triangular form of the matrices.~\footnote{Note that for $\bw>0$, the invariant density is trivial as the inverse of the Riccati variable is distributed according to a delta function. For $\bw<0$, the invariant density is a non trivial distribution, as we have seen above in the continuum limit, Eq.~\eqref{eq:InvariantDensityAN}.}
In this case, the GLE is independent of the randomness on the $N$ part.


\section{Conclusion}
\label{sec:Conclusion}

In this article we have studied the cumulant generating function for the logarithm of products of random matrices of the group $\mathrm{SL}(2,\mathbb{R})$~:
\begin{equation}
    Q_n(z_0;q) = \mean{||\Pi_n\vec{x}_0||^q} 
    = 
    \mean{ \EXP{-\frac{q}{2}\sum_{i=1}^n\sigma(M_i,z_{i-1})} }
    \:,
\end{equation}
where $\sigma(M,z)$ is the additive cocycle \eqref{eq:DefCocycle} and $\{z_i\}$ the random sequence~\eqref{eq:RiccatiRecursion}.
Following Tutubalin \cite{Tut65}, we have shown how ideas from group theory can be used to complete this program, which has led us to introduce a family of representations of the group with multipliers, each family being labeled by the parameter $q$
\begin{equation}
    \left[
     \mathscr{T}_M(q) f
  \right](z) 
  =
  \EXP{\frac{q}{2}\sigma(M^{-1},z) }
  \,
  \deriv{\mathcal{M}^{-1}(z)}{z}\,
  f\!\left( \mathcal{M}^{-1}(z) \right)
  \:.
\end{equation}
The average $\smean{\mathscr{T}_M(q)}_M=\overline{\mathscr{T}}(q)$ plays the same role as the exponential of the differential operator $\mathscr{G}_z^\dagger+q\,\phi(z)$ in the example at the end of Section~\ref{sec:Intro}.
As a result, the generalized Lyapunov exponent (GLE) $\widetilde{\Lambda}(q)$ of the random matrix product can be obtained by solving the spectral problem \eqref{eq:TheSpectralPb}. 
We have emphasized that the spectral problem is independent of the choice of Jacobian (multiplier), which corresponds to choose different \textit{realizations}, for the same \textit{representation} (same $q$).
This can be used to simplify the spectral problem through the transformation \eqref{eq:TransformationTbar}, by an appropriate choice.
We have illustrated the concepts and methods on specific cases of random matrices motivated by the theory of Anderson localisation.
Furthermore, we have concentrated ourselves on products of random transfer matrices for the Schr\"odinger equation with a random potential. In this context, we have derived new formulae for the variance of the logarithm of the random matrix product (RMP), Eqs.~\eqref{eq:FinalLambda2} and \eqref{eq:FinalGamma2}.
We have also discussed the continuum limit of random matrix products~: then the spectral problem involves a differential operator, Eq.~\eqref{eq:SpectralPbContinuum3} or \eqref{eq:SpectralPbContinuum2}, which has allowed us to make some progresses.

Table~\ref{tab:conclu} gives an overview of the different cases discussed.
%
%

\begin{table}[!ht]
\centering
\begin{tabular}{lllll}
Matrices & distribution   & Jacobian                        & Transform & Sections
\\
\hline
\hline
$K(\theta)N(u)\phantom{\overset{|}{=}}\hspace{-0.5cm}$  
  & $\theta$: exponential \& $u$: arbitrary
  & $J=J_N$ 
  & Fourier        
  & \ref{subsec:FL1}, \ref{subsec:IntTransfSchrod}, \ref{sec:PeturbFL}
\\
  & $\theta$, $u$: $\ll1$, Gaussian  & $J=J_N$  &   no      & \ref{subsec:Halperin}
\\   
\hline
$\widetilde{K}(\theta)N(u)\phantom{\overset{|}{=}}\hspace{-0.5cm}$ 
  & $\theta$: exponential  \& $u$: arbitrary
  & $J=J_N$ 
  & Fourier        
  &  \ref{subsec:IntTransfSchrod},  \ref{subsubsec:CaseNzero}
\\
  & $\theta$, $u$: $\ll1$, Gaussian  & $J=J_N$  &   no      & \ref{subsec:Halperin}
\\   
\hline
$\widetilde{K}(\theta)A(w)\phantom{\overset{|}{=}}\hspace{-0.5cm}$  
  & $\theta$: exponential \&  $w$: arbitrary 
  & $J=J_A$ 
  & Mellin         
  & \ref{subsec:SuSy1}, \ref{subsec:MellinTransfForSusy}, \ref{sec:PerturbativeSusy}
\\
  & $\theta$, $w$: $\ll1$, Gaussian  &  $J=J_A$ &   no      & \ref{sec:ContinuumSusy}
\\   
\hline
$K(\theta)A(w)N(u)\phantom{\overset{|}{=}}\hspace{-0.5cm}$  
  & $\theta$, $w$, $u$: $\ll1$, Gaussian  &   &   no      & \ref{subsec:SPcontinuumlimit}
\\   
\hline
$A(w)N(u)\phantom{\overset{|}{=}}\hspace{-0.5cm}$  
  & $w$, $u$: $\ll1$, Gaussian  & $J=J_N$   &    no     & \ref{subsec:CaseAN}
\\   
\hline
\end{tabular}
\caption{\it Summary of the different cases studied with the appropriate transformations.
  The case of small parameters ($\theta$, $w$, $u$: $\ll1$) corresponds to the continuum limit.}
\label{tab:conclu}
\end{table}

This study opens several challenging problems.
\begin{enumerate}[label=({\it\roman*}),leftmargin=*,align=left,itemsep=+0.1cm]



\item
The perturbative approach has been successfully applied to the spectral problem \eqref{eq:SpectralPbIntro} for matrices of type $KN$ or $\widetilde{K}N$ in \S~\ref{sec:PeturbFL}, \S~\ref{sec:GLElocalisation} and \S~\ref{subsec:Halperin}. 
This has led to a general formula for the variance, Eq.~\eqref{eq:FinalLambda2} or Eq.~\eqref{eq:FinalGamma2} (the latter simplifies the multiple integrals representations of Refs.~\cite{SchTit02,RamTex14}).
An interesting question would be to extend such formulae to other types of matrices~; of specific interest is the case of matrices of type $KA$ or $\widetilde{K}A$ which have been widely studied and for which formulae involving multiple integrals have been obtained in Ref.~\cite{RamTex14} in the continuum limit. 
The goal would be to derive an alternative (more simple and more general) formula, as we did here for matrices $KN$.

\item
The search for compact formulae for the variance of the logarithm of the RMP also suggests to investigate other models~:
in particular, group theoretical considerations could help identifying new interesting solvable cases, which is always of great interest given the relative scarcity of solvable disordered models.
We stress that, here, we have only considered models involving pairs of matrices of the Iwasawa decomposition~: $KN$, $KA$, or $AN$ (note that only the third case corresponds to a two-parameter subgroup). 
Identifying a solvable case involving the three matrices together still seems extremely challenging (the mixed models studied in \cite{HagTex08,TexHag09,GraTexTou14} could be good candidates).

Another possibility could be to consider the case where the three parameters have exponential distributions~: then the spectral problem \eqref{eq:TheSpectralPb} takes the form of a third order differential equation 
\begin{equation}
\left[ 1 - \bu\,\mathscr{D}_N(q)  \right]
\left[ 1 - \bw\,\mathscr{D}_A(q)  \right]
\left[ 1 - \btheta\,\mathscr{D}_K(q)  \right] \Phi^\mathrm{R}_q(z)
=
\lambda(q) \, \Phi^\mathrm{R}_q(z)
\end{equation}
(note that one $q$-dependence among the three generators might be removed by appropriate choice of the Jacobian $J$).

\item
In Ref.~\cite{ComLucTexTou13}, a classification of the possible solutions for the Lyapunov exponent for RMP in the continuum limit was possible through a Hilbert transform.
An open question is to provide such a general analysis of Eq.~\eqref{eq:SpectralPbContinuum3} or Eq.~\eqref{eq:SpectralPbContinuum2}.

\item
The perturbative approach was carried out up to second order and has led to the two first cumulants of the logarithm of the RMP.
Only very few analytical results are available for higher cumulants~:
the third and fourth cumulants for the Schr\"odinger equation~\cite{SchTit02},
the third cumulant for the lattice model with Cauchy disorder~\cite{TitSch03} and the high energy behaviour of all cumulants for power law disorder~\cite{Tex19b}.
The study of higher cumulants would certainly be of interest.

\item
In Subsection~\ref{subsec:CaseAN}, we have studied products of random matrices of type $AN$, for which the spectral problem is exactly solvable.
It would be interesting to clarify whether the fact that $AN$ forms a two-parameter subgroup is related to the existence of simple solutions in terms of hypergeometric functions, or not.
Our analysis has led to a quadratic GLE, i.e. purely Gaussian fluctuations, a rather unusual feature [this comes from the fact that the GLE is somehow trivial in this case, cf. Eq.~\eqref{eq:TrivialGLE}].
Identifying another model leading to some analytic expression of the GLE, with non-Gaussian large deviations, seems extremely challenging.


\item
In the paper, the emphasis was put on the $q\to0$ expansion of the GLE $\widetilde{\Lambda}(q)$ and $\Lambda(q)$, with the main motivation being in the characterization of the variance of the logarithm of the RMP, i.e. in \textit{typical} fluctuations.
Besides the intrinsic interest for \textit{atypical} fluctuations, the GLE of the Schr\"odinger equation with Gaussian white noise potential (related to products of matrices $KN$ or $\widetilde{K}N$ in the continuum limit) was recently shown to control the number of equilibria of a polymer in a random medium~\cite{FyoLeDRosTex18}.
In this study, the relevant quantity is $\Lambda(1)$. 
In Ref.~\cite{FyoLeDRosTex18}, the analysis of the spectral problem has revealed some remarkable features, which has played a crucial role~: 
the vanishing of all perturbative contributions (in the disorder) for $q=1$, whereas it was demonstrated that $\Lambda(2n)$ is purely analytic in the disorder strength. 
This observation seems to be related to the resurgent expansions discussed in 
Ref.~\cite{JenZin04}  
for a similar spectral problem (see Ref.~\cite{KozSulTanUns18} for more references on resurgent series).
This point requires a better understanding and to clarify whether these features are accidental, related to the specific model, or more generic within the study of GLE, and unveil their group theory interpretation.

\item
\textit{A symmetry of the GLE~:}
Finally, let us raise some questions in relation with a remarkable symmetry relation for the GLE \cite{Van10}, which relies on the symplectic nature of the group.~\footnote{  
A group of $2m\times2m$ real matrices is symplectic if matrices satisfy 
$
  M^\mathrm{T}\,J\,M = J 
$
where $J=\I\sigma_2\otimes\boldsymbol{1}_m$.
It follows that the $2m$ eigenvalues come by pairs 
$\mathrm{Spec}(M)=\{\lambda_1,\cdots,\lambda_m,\lambda_1^{-1},\cdots,\lambda_m^{-1}\}$.
}
A first consequence is that products of i.i.d. matrices $M_i$ and $M_i^{-1}$ have the same GLE~:
indeed, because the product $\Pi_n=M_n\cdots M_2M_1$ is also symplectic, $\Pi_n$ and $\Pi_n^{-1}$ have the same eigenvalues, hence, using that the matrices are i.i.d., we can write
\begin{equation}
  \label{eq:GLEsymplectic}
  \widetilde{\Lambda}(q)
  = \lim_{n\to\infty} \frac{ \ln\mean{||M_n\cdots M_2M_1\vec{x}_0||^q} }{n}
  = \lim_{n\to\infty} \frac{ \ln\mean{||M_n^{-1}\cdots M_2^{-1}M_1^{-1}\vec{x}_0||^q} }{n}
  \:.
\end{equation}
These considerations apply in the case of $\mathrm{SL}(2,\mathbb{R})$ of interest in the paper since the group is also the symplectic group $\mathrm{Sp}(2,\mathbb{R})$.
Thus $1/\lambda(q)=\EXP{\widetilde{\Lambda}(q)}$ is the largest eigenvalue of the two transfer operators
$
  \smean{ \mathscr{T}_M(q) } 
$
and 
$
  \smean{ \mathscr{T}_{M^{-1}}(q) } 
$.   
For simplicity, we choose the Jacobian $J=J_N$ so that we can use \eqref{eq:SymTdaggerT}, thus $1/\lambda(q)$ is the largest eigenvalue of 
$  \smean{ \mathscr{T}_M(q) } $ and $ \smean{ \mathscr{T}_{M}^\dagger(-q-2) } $.
The operator and its adjoint having the same spectrum, we deduce $\lambda(q) = \lambda(-q-2)$, i.e.~\footnote{ 
  For $2m\times2m$ symplectic real matrices, Vanneste has obtained~\cite{Van10}
  $\widetilde{\Lambda}(q)=\widetilde{\Lambda}(-2m-q)$.
 }
\begin{equation}
  \label{eq:Vanneste2010}
  \widetilde{\Lambda}(q)=\widetilde{\Lambda}(-2-q)
  \:.
\end{equation}
Starting from the definition (\ref{eq:FluctPsiX},\ref{eq:DefLambdaLocalisation}) of \S~\ref{sec:GLElocalisation}, it is possible to prove a similar symmetry relation, by manipulations of Eq.~\eqref{eq:SpectralPbExpRL} and its adjoint~:
\begin{equation}
  \label{eq:Vanneste2010bis}
   \Lambda(q) = \Lambda(-2-q)
  \:.
 \end{equation} 
The identities (\ref{eq:Vanneste2010},\ref{eq:Vanneste2010bis}) hold if the search of the largest eigenvalue of the two operators $  \smean{ \mathscr{T}_M(q) } $ and $ \smean{ \mathscr{T}_{M^{-1}}(q) }=\smean{ \mathscr{T}_{M}^\dagger(-q-2) } $ is the same spectral problem, i.e. assuming that the two operators act in the same space of functions.

This symmetry relation has been introduced and tested numerically in Ref.~\cite{Van10} in a specific case. 
It has also been verified analytically for products of matrices of the form $M=KN$ in the continuum limit in Ref.~\cite{Tex19b}. 
Although the argument seems quite general, it is possible to find counter examples showing that the symmetry is not always satisfied~: 
Sturman and Thiffeault~\cite{StuThi19} have recently obtained some bounds for the GLE of matrices $N(\theta)^\mathrm{T}$ and $N(u)$ chosen with equal probability, demonstrating a monoteneous behaviour of $\widetilde{\Lambda}(q)$ as a function of~$q$.
Eqs.~\eqref{eq:TrivialGLESubgroupKtilde} and \eqref{eq:GLEcaseAN} derived above furnish other cases.
This last examples could provide a hint to track the origin of the violation of the symmetry \eqref{eq:Vanneste2010}~: the matrices in both cases clearly do not satisfy the irreducibility condition (see \cite{BurMen16}), since the two cases correspond to subgroups of $\mathrm{SL}(2,\mathbb{R})$.

In order to get a hint of the possible origin of the violation of the symmetry \eqref{eq:Vanneste2010}, let us first stress that the relation corresponds to a symmetry of the distribution of $\Upsilon_N=\ln||\Pi_N||$. Its distribution is expected to present the scaling form
$\mathscr{P}_N(\Upsilon)\sim\exp\big\{-N\,\Phi(\Upsilon/N)\big\}$ for $N\to\infty$, where the large deviation function is related to the GLE through a Legendre transform
\begin{equation}
   \Phi(y)=\underset{q}{\mathrm{min}}\big\{q\,y-\widetilde{\Lambda}(q)\big\}
   \:.
\end{equation}
Hence, the symmetry \eqref{eq:Vanneste2010} corresponds to 
\begin{equation}
  \Phi(-y) = \Phi(y) + 2\,y
  \:,
\end{equation}
i.e. relates the asymptotic behaviours of the distribution of the norm of the matrix product for $||\Pi_N||\to0$ and $||\Pi_N||\to\infty$.
To close the discussion, one can make few remarks on the case of matrices $M=N(\theta)^\mathrm{T}N(u)$~:
depending on the choice for the distribution of parameters, the symmetry holds or not. 
We first consider the case where the two random parameters are positive (this is equivalent to the case considered by Sturman and Thiffeault~\cite{StuThi19}) and compare the action of a sequence of random matrices $M=N(\theta)^\mathrm{T}N(u)$ and $M^{-1}=N(-u)N(-\theta)^\mathrm{T}$ on an initial unit vector~:
the results are represented on the two plots of Fig~\ref{fig:GeomVannesteSym}.
The fact that the parameters have a fixed sign generates a drift and the vector can never come back in the vicinity of the origin.
The comparison of the two figures also shows that the support of the distribution $f(z)$ of the Riccati variable   $z=x/y$ is not the same in the two cases (the support is in $\mathbb{R}^+$ for $\btheta=\bu>0$, while it is in $\mathbb{R}^-$ for $\btheta=\bu<0$).
This suggests that the eigenvectors of the two operators $\smean{ \mathscr{T}_M(q) }$ and $\smean{ \mathscr{T}_{M}^{-1}(q) }$ belong to different functional spaces.

\begin{figure}[!ht]
\centering
\includegraphics[scale=0.35]{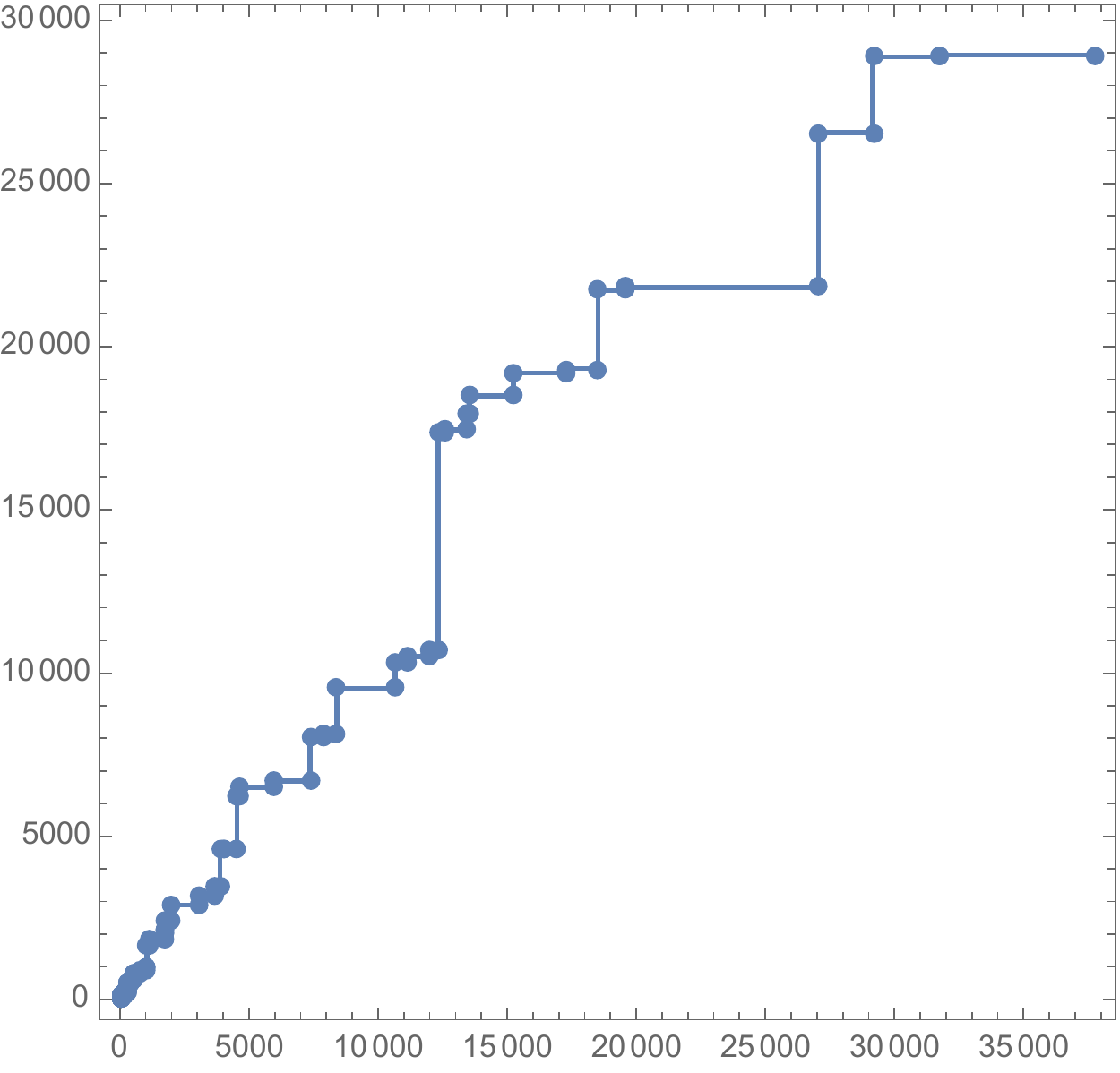}
\hspace{1cm}
\includegraphics[scale=0.35]{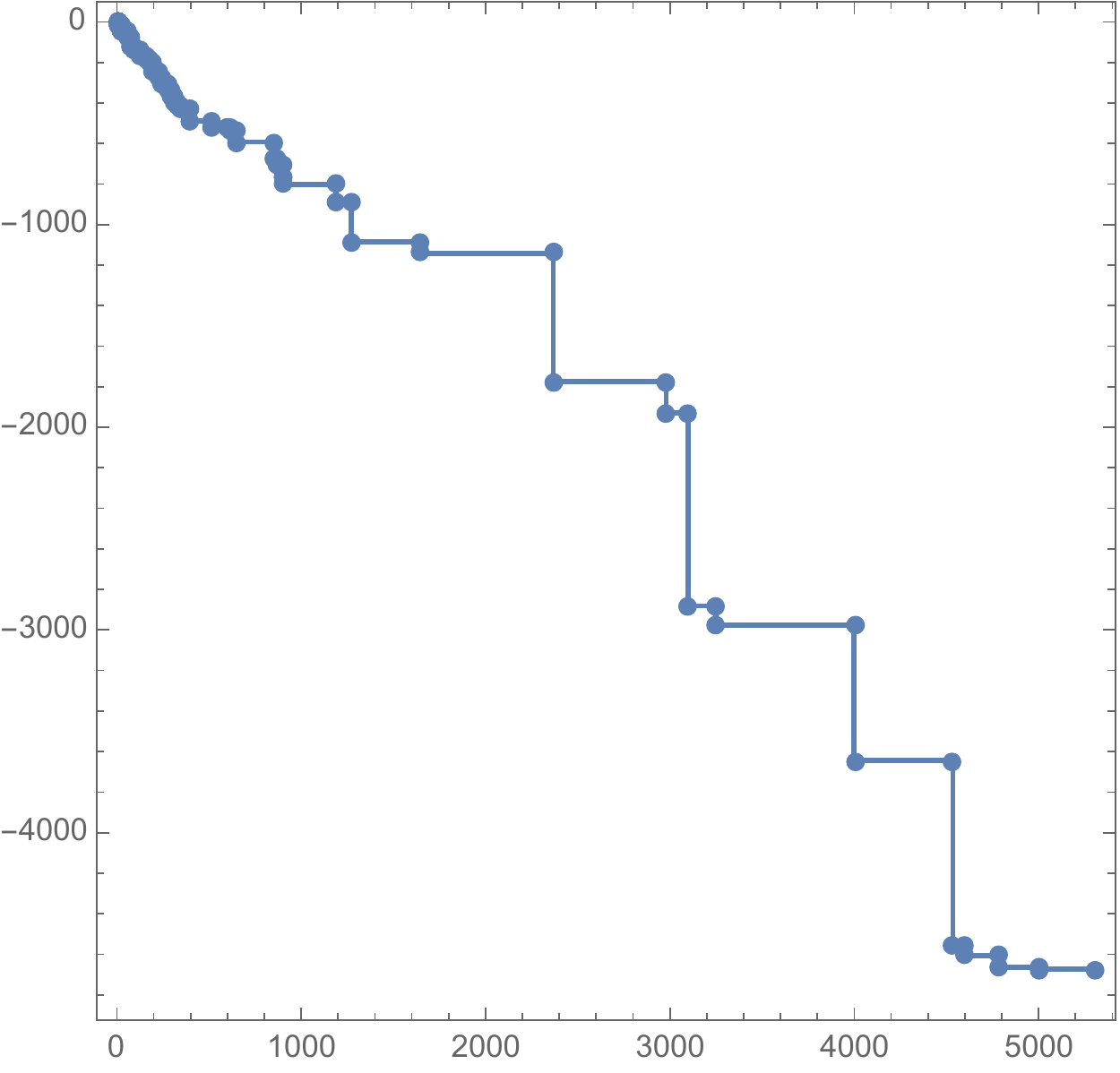}
\caption{\it Action of matrices $M=N(\theta)^\mathrm{T}N(u)$ on an initial unit vector. Parameters are exponentially distributed with $\btheta=\bu=0.1$ (left). The figure on the right corresponds to $\btheta=\bu=-0.1$, which is equivalent to consider $\big[N(\theta)^\mathrm{T}\big]^{-1}N(u)^{-1}$ for positive coefficients.}
\label{fig:GeomVannesteSym}
\end{figure}

On the contrary, considering $\theta$ and $u$ with random signs (corresponding to the case studied by Vanneste \cite{Van10}) allows for a better exploration of the plane under the action of sequences of matrices $M$ or $M^{-1}$ (see Fig.~\ref{fig:GeomVannesteSym2}).
In this case, the support of the distribution $f(z)$ is $\mathbb{R}$, both for matrices $M$ or $M^{-1}$.

\begin{figure}[!ht]
\centering
\includegraphics[scale=0.35]{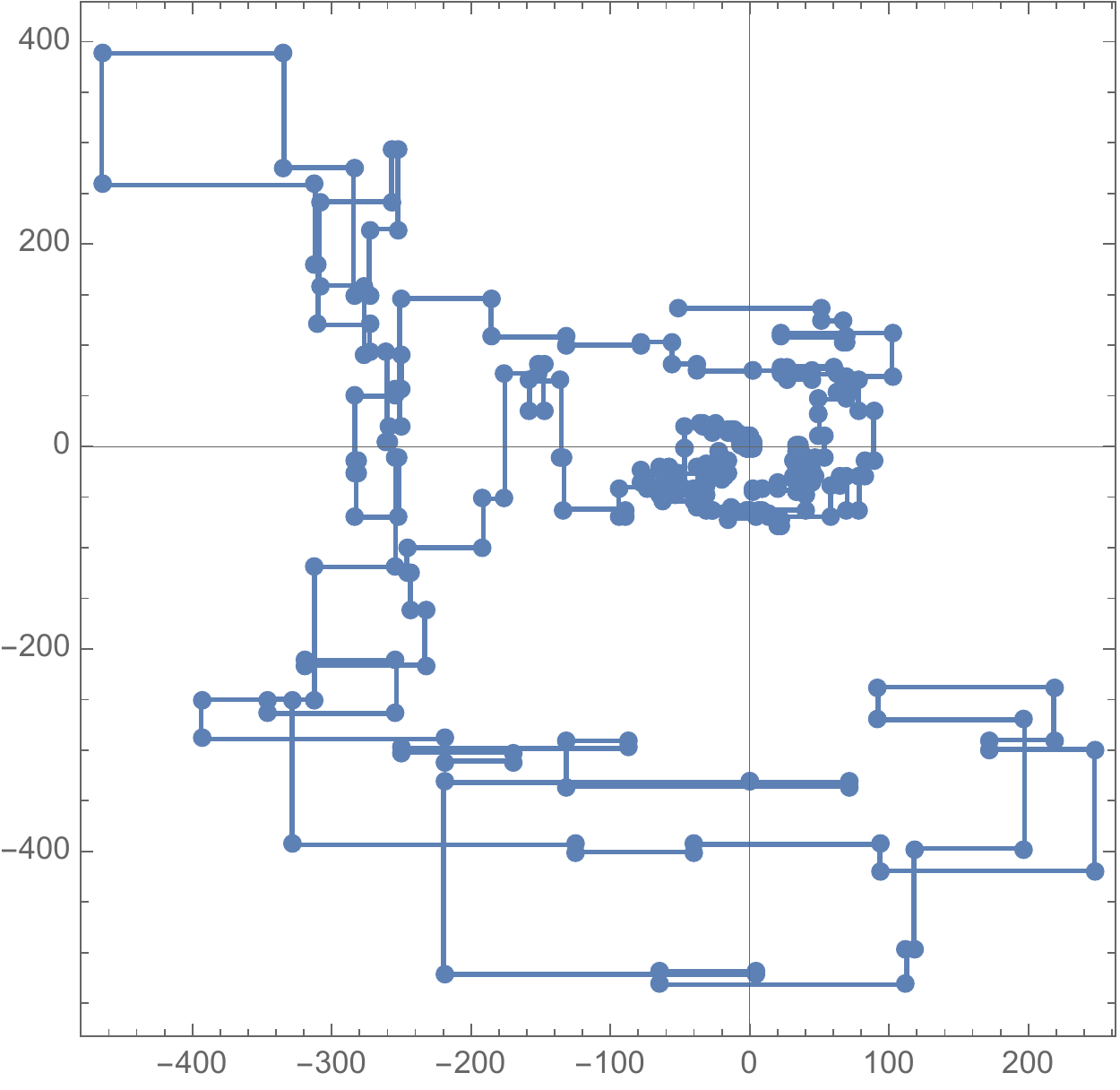}
\hspace{1cm}
\includegraphics[scale=0.35]{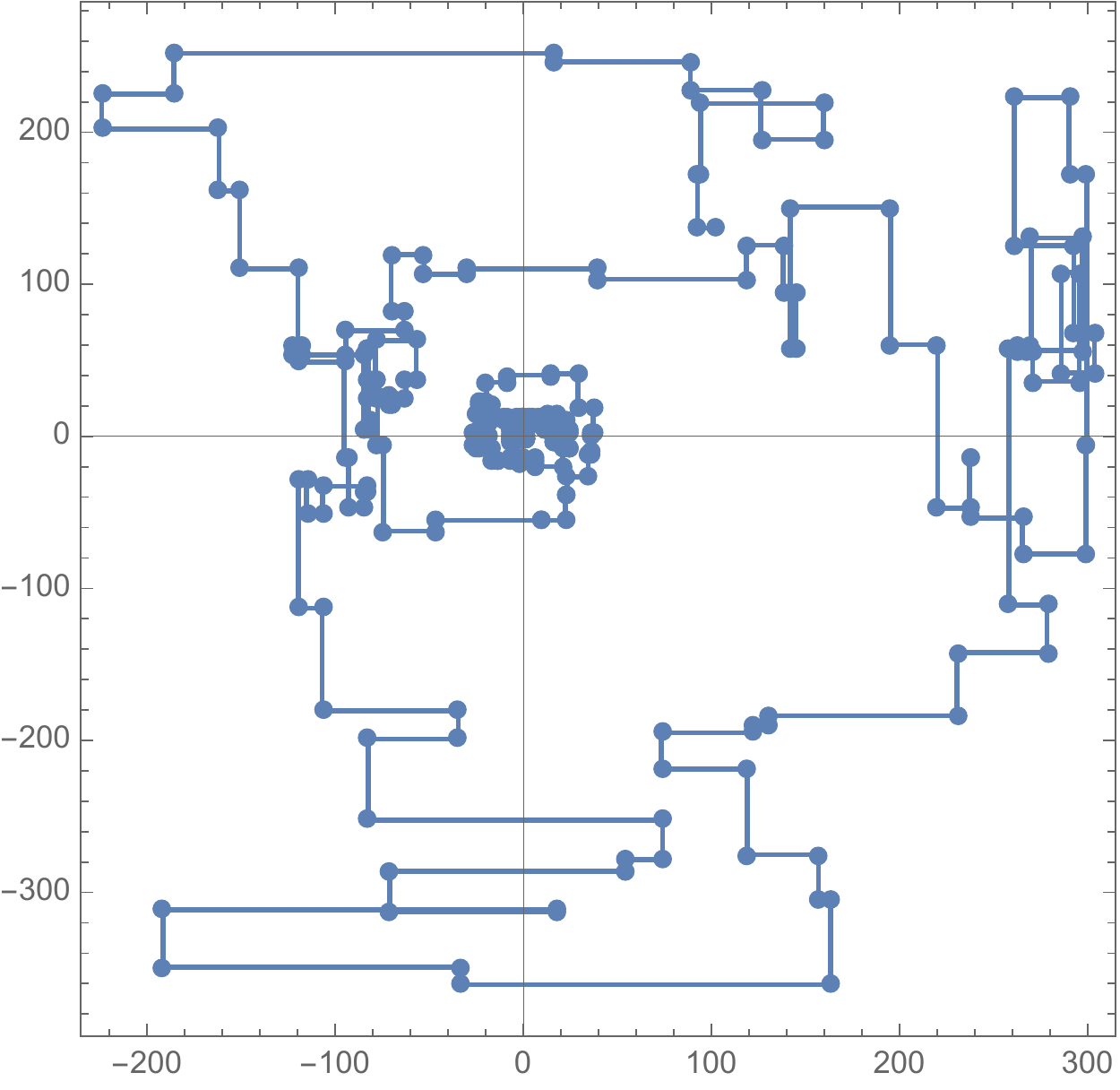}
\caption{\it Action of matrices $M=N(\theta)^\mathrm{T}N(u)$ for weights with Gaussian distribution for variances $\Dtt=\Duu=0.1$ and means $\btheta=-\bu=0.1$ (left). 
The figure on the right shows the action of the inverse matrices. By changing the sign of the means, one has reversed the current around the origin.}
\label{fig:GeomVannesteSym2}
\end{figure}

Further investigations for the conditions of validity of the relation \eqref{eq:Vanneste2010} are however required.

\item
  Finally, the most challenging issue is the study of the GLE for other groups of matrices.
  Here, the one dimensional nature of the projective space has been used early in the exposition.
  The article \cite{ComTexTou19} gives some hint on how some of the ideas presented here can be extended to study the more general case.

\end{enumerate}

\section*{Acknowledgments}

This paper has benefited from a long term collaboration with Alain Comtet and Yves Tourigny, to whom I express my gratitude. In particular, the introduction of the representation with multipliers and the role of Jacobians owes them a lot.  
The connection with representation theory is discussed in another paper \cite{ComTexTou19}.
I am grateful to Jean-Marc Luck for several stimulating discussions. 
I thank Olivier Giraud for a helpful remark and Maxime Allemand. 
I thank Alain Comtet, Satya Majumdar and Yves Tourigny for remarks on the manuscript.
I am grateful to Jean-Luc Thiffeault for pointing to my attention Ref.~\cite{StuThi19}. 
Finally I am indebted to the two heroic anonymous referees for having carefully examined this long manuscript and for their helpful remarks.
This work has benefited from the financial support \og Investissements d'Avenir du LabEx PALM \fg{} (ANR-10-LABX-0039-PALM), project ProMAFluM.


\appendix


\section{Representation with multipliers: another convention}
\label{app:RepresMatheux}

In this appendix, the connection is made with the convention of Ref.~\cite{ComTexTou19} for the construction of the representation with multipliers. 
This second convention, used in the mathematical literature, is probably more appropriate for generalization to groups of matrices of larger dimensions.

The starting point is to introduce the adjoint of the transfer operator, with same definition as in the paper,
\begin{equation}
  \big[
     \mathscr{T}_M^\dagger(q) \psi
  \big](z) 
  \eqdef
  J^{-q/2}(M,z) 
  \,
  \psi\!\left( \mathcal{M}(z) \right)
  \:.
\end{equation}
We now define the scalar product as 
\begin{equation}
  \label{eq:DefScalarProductYves}
    \braketYves{\psi}{\tilde{f}} \eqdef \int\D z \,\rho(z)\, \psi(z)^*\, \tilde{f}(z)
  \:.
\end{equation}
The relation with the scalar product~\eqref{eq:DefinitionScalarProduct} can be simply established by writing 
\begin{equation}
  \label{eq:RelationBetweenScalarProducts}
  \braketYves{\psi}{\tilde{f}} = \braket{\psi}{f}
  \hspace{1cm}\mbox{with }
  f(z)= \rho(z)\, \tilde{f}(z)\:.
\end{equation}

With this new scalar product in hand, we can obtain the new transfer operator as follows
\begin{align}
  &\braketYves{\mathscr{T}_M^\dagger(q)\psi}{\tilde{f}}
  = \int \D y\, \rho(y) \, J^{-q/2}(M,y)\, \psi\left( \mathcal{M}(y) \right)^*\, \tilde{f}(y)
  \\
  \nonumber
  &= \int \D z\,  \rho\left( \mathcal{M}^{-1}(z)\right)\, 
  \deriv{\mathcal{M}^{-1}(z)}{z}\, 
    J^{-q/2}(M,\mathcal{M}^{-1}(z))\, 
    \psi(z)^*\, \tilde{f}\left( \mathcal{M}^{-1}(z) \right)
   = \braketYves{\psi}{\widetilde{\mathscr{T}}_M(q)\tilde{f}}
\end{align}
with 
\begin{equation}
  \label{eq:Repres4Yves}
  \left[
     \widetilde{\mathscr{T}}_M(q) f
  \right](z) 
  =
  J^{1+q/2}(M^{-1},z) 
  \:
  f\!\left( \mathcal{M}^{-1}(z) \right)
   \:.
\end{equation}
From \eqref{eq:RelationBetweenScalarProducts}, we see that the two representations (\ref{eq:Repres4},\ref{eq:Repres4Yves}) are related by 
\begin{equation}
  \frac{1}{\rho(z)}\, \mathscr{T}_M(q)\, \rho(z)= \widetilde{\mathscr{T}}_M(q)
  \:.
\end{equation}

Contrary to $\mathscr{T}_M={\mathscr{T}}_M(q=0)$, which preserves the norm  $\int\D z\,f(z)$, the operator $\widetilde{\mathscr{T}}_M(q=0)$ conserves $\int\D z\,\rho(z)\,\tilde{f}(z)$~:
$
  \int \D z\, \rho(z)\,\big[
     \widetilde{\mathscr{T}}_M(0) \tilde{f}
  \big](z) = \int \D z\, \rho(z)\,\tilde{f}(z) 
  \:.
$
This is one of the reasons why we have prefered the representation \eqref{eq:Repres4} in the paper.

Finally, note that the transfer operators \eqref{eq:Repres4Yves} are related to infinitesimal generators of the form
\begin{equation}
   \widetilde{\mathscr{D}}_i(q)
   = \frac{1}{\rho(z)}\, \mathscr{D}_i(q)\, \rho(z) 
   = g_i(z)\,\deriv{}{z} + (q+2)\, h_i(z)
  \hspace{0.5cm}
  \mbox{for } 
  i\in\{ K, \, A, \, N  \}
  \:.
\end{equation}
It is also easy to check that they realise the Lie algebra of $\mathrm{SL}(2,\mathbb{R})$ and that they are adjoint to the infinitesimal generators \eqref{eq:AdjointInfinGene}, with respect with the scalar product~\eqref{eq:DefScalarProductYves}.

%
%


\section{Irreducible representations of $\mathrm{SL}(2,\mathbb{R})$}
\label{app:RepresentationTheory}


We recall basic properties of the representation theory of real unimodular matrices, taken from the monograph~\cite{GelGraVil66}.
For $M\in\mathrm{SL}(2,\mathbb{R})$, let us start with the group action in the space $\mathscr{F}(\mathbb{R}^2)$ of infinitely differentiable functions defined in the plane
\begin{equation}
  \label{eq:TransfFR2}
  (T_M F)(\vec{x}) \eqdef F(M^{-1}\cdot \vec{x})
\end{equation}
Classification of the irreducible representations of the group requires to identify the smallest proper invariant subspaces of functions.
For this purpose, homogeneous functions play an important role.
A function satisfying the property
\begin{equation}
  \label{eq:DefHomogeneous}
  F(\lambda\,\vec{x}) = \left(\sign(\lambda) \right)^{1-\epsilon} \,|\lambda|^{-\eta}\,F(\vec{x})
  \hspace{0.5cm}\forall\: \lambda\in\mathbb{R}
\end{equation}
is called a homogeneous function of degree $\eta\in\mathbb{C}$, where $\epsilon\in\{+1,\,-1\}$ defines even and odd sectors.
We denote by $\mathscr{E}_{\eta}^+$ the space of homogeneous functions of degree $\eta$ with even parity, and $\mathscr{E}_{\eta}^-$ with odd parity. 
Clearly, the property \eqref{eq:DefHomogeneous} is preserved by the transformation \eqref{eq:TransfFR2}, hence we have identified subspaces of $\mathscr{F}(\mathbb{R}^2)$ invariant under the action of the group.
The space $\mathscr{E}_{\eta}^\pm$ for noninteger $\eta$ is known to generate an irreducible representation of $\mathrm{SL}(2,\mathbb{R})$~\cite{GelGraVil66}.

A key observation is that each homogeneous function of given parity is uniquely determined by its value on the projective line~:
for any $F\in\mathscr{E}_{\eta}^+$, we have $F(x,y)=|y|^{-\eta}F(x/y,1)=|x|^{-\eta}F(1,y/x)$, hence each homogeneous function of degree $\eta$ can be represented by an infinitely differentiable function defined on the projective line
\begin{equation}
    f(z) = F(z,1) = |z|^{-\eta}\, F(1,1/z)
    \:.
\end{equation} 
This shows that, on the projective line, the elements of $\mathscr{E}_{\eta}^+$ are represented by functions such that the two limits
\begin{equation}
  \label{eq:PropertyIrreducibleRepres}
  \lim_{z\to-\infty} (-z)^{\eta}\,f(z) 
  \hspace{0.25cm}\mbox{and}\hspace{0.25cm}
  \lim_{z\to+\infty} z^{\eta}\,f(z) 
  \hspace{0.25cm}\mbox{exist and are equal.}
\end{equation}


In order to apply these considerations to our case we must relate the degree of the representation to the parameter $q$.
Given a matrix \eqref{eq:MatrixOfSL2R},  we consider the transformation $\tilde{f}=\mathscr{T}_M(q)f$ defined by  \eqref{eq:Repres4}. It is straighforward to get the asymptotic behaviour
\begin{equation}
  \label{eq:ftilde}
  \tilde{f}(z) 
  \underset{z\to\pm\infty}{\simeq}
  \begin{cases}
    \displaystyle
    \left( \frac{\sqrt{\rho(-d/c)}}{|c|}\right)^q\frac{f(-d/c)}{c^2}\,\frac{1}{z^2|z\sqrt{\rho(z)}|^q}
    & \mbox{for } c\neq0
    \\[0.25cm]
    \displaystyle
    |a|^{-2-q}\left(\frac{\rho(z/a^2)}{\rho(z)}\right)^{q/2}\, f(z/a^2)
    & \mbox{for } c=0
  \end{cases}
\end{equation}
In the paper, we have considered Jacobians involving symmetric densities $\rho(z)=\rho(-z)$ with power law tail $\rho(z)\sim|z|^{-2\omega}$ where
\begin{equation}
  \begin{cases}
     \omega=0 & \mbox{for } \rho_N
     \\
     \omega=1/2 & \mbox{for } \rho_A
     \\
     \omega=1 & \mbox{for } \rho_K
  \end{cases}  
\end{equation}
It is now clear from \eqref{eq:ftilde} that the operators $\mathscr{T}_M(q)$ preserve the property \eqref{eq:PropertyIrreducibleRepres}, hence these operators form an irreducible representation of the unimodular group of degree
\begin{equation}
  \eta=2+(1-\omega)\,q
  \:.
\end{equation}
For example, for the choice of Jacobian $J=J_N$, we have simply $\eta=2+q$ (Subsection~\ref{subsec:FL1}).
For  $J=J_A$, we have $\eta=2+q/2$ (Subsection~\ref{subsec:SuSy1}).

\medskip

A remark~:
\begin{itemize}
\item 
the case $c=0$ corresponds to the subgroup of matrices $AN$. 
As it is clear from \eqref{eq:ftilde}, the exponent $\eta$ cannot be related to $q$ in this case.


\end{itemize}

\subsection{Eigenvectors of the Casimir operator and $\mathscr{D}_K(q)$}
\label{app:DiagDK}

Irreducible representations of $\mathrm{SL}(2,\mathbb{R})$ can also be analysed by the algebraic method, as it well known for $\mathrm{SU}(2)$.
For this purpose, we find more convenient to consider the decomposition of the group in terms of matrices of the subgroups $\mathrm{K}$, $\widetilde{\mathrm{K}}$ and $\mathrm{A}$, i.e. as $M=K(\theta)\widetilde{K}(\varphi)A(w)$, according to the notations of Section~\ref{sec:RepresentationsSL2R}.
With this choice, the three infinitesimal generators obey the algebra
\begin{equation}
  \left[\Gamma_K\,\,,\Gamma_{\widetilde{K}}\right] = 2\Gamma_A \:, \hspace{0.25cm}
  \left[\Gamma_K\,\,,\Gamma_A\right] = -2\Gamma_{\widetilde{K}} \:, \hspace{0.25cm}
  \left[\Gamma_{\widetilde{K}}\,\,,\Gamma_A\right] = -2\Gamma_K 
\end{equation}
which is the usual form for the Lie algebra of $\mathrm{SO}(2,1)$, the Lorentz group in $2+1$ dimensions, studied by Bargmann \cite{Bar47} (this is also the Lie algebra for $\mathrm{SU}(1,1)$ as expected, cf. \cite{ComTexTou10}). 

The connection to the Lorentz group makes easy to identify the Casimir operator as $-\Gamma_K^2+\Gamma_{\widetilde{K}}^2+\Gamma_A^2=3\,\mathbf{1}_2$, and, for the representation of interest in the paper~:
\begin{equation}
  \Casimir = 
  -\mathscr{D}_{K}(q)^2 + \mathscr{D}_{\widetilde{K}}(q)^2 + \mathscr{D}_{A}(q)^2 = q\,(q+2)
  \:,
\end{equation}
where we have used the expressions \eqref{eq:GeneratorsDdeQ} derived in Subsection~\ref{subsubsec:InfGenDq}.
This makes clear that irreducible representations are classified by the index $q$ (this is similar to the Bargmann index).~\footnote{
  Note that Bargmann has considered irreducible unitary representations, while the representaions of interest here are non-unitary.
}
This explains the origin of 
the symmetry $q\leftrightarrow-q-2$ discussed in Section~\ref{sec:Conclusion}.

Irreducible representations can be further characterised by diagonalization of one of the generator.
We consider the operator $\mathscr{D}_K(q)=\deriv{}{z}(1+z^2)+q\,z$, acting on functions $\phi(z)$ defined on the projective line. We have chosen here the measure $\rho=\rho_N$.

In a first step, it is more clear to map the projective line on the half circle, $z=\cotan\theta$. 
Correspondingly, we introduce the transformed operator 
\begin{equation}
  \mathscr{B}_K(q) = \deriv{z}{\theta} \, \mathscr{D}_K(q) \, \deriv{\theta}{z}
  = -\deriv{}{\theta} + q\, \cotan\theta
\end{equation}
acting on periodic functions $\psi(\theta)=\psi(\theta+\pi)$ (the projective space is the space of \textit{directions}, hence the period $\pi$). 
Here, $\phi(z)$ and  $\psi(\theta)$ play the same role as densities, related by $\phi(z)\,\D z=\psi(\theta)\,\D\theta$.
It is now straightforward to get the eigenvector of $\mathscr{B}_K(q)$~:
\begin{equation}
  \psi_n^\mathrm{R}(\theta) = \frac{1}{\sqrt{\pi}}\,\EXP{2\I n\theta}\,(\sin\theta)^q
  \hspace{0.5cm}\mbox{ for eigenvalue }\hspace{0.5cm}
  \lambda_n=-2\I n 
  \:,
  \hspace{0.25cm}\mbox{with }
  n\in\mathbb{Z}
\:.
\end{equation}
Similarly, it is instructive to derive the eigenvector of the adjoint operator $\mathscr{B}^\dagger_K(q) =\deriv{}{\theta} + q^*\, \cotan\theta$ (we consider $q\in\mathbb{C}$ in this paragraph)~:
\begin{equation}
  \psi_n^\mathrm{L}(\theta) = \frac{1}{\sqrt{\pi}}\,\EXP{2\I n\theta}\,(\sin\theta)^{-q^*}
  \:.
\end{equation}
The orthonormalisation reads
$\int_0^\pi\D\theta\,\psi_n^\mathrm{L}(\theta)^*\psi_m^\mathrm{R}(\theta)=\delta_{n,m}$.

We now come back on the projective line.  
The right and left eigenvectors of $\mathscr{D}_K(q)$ are $\phi_n^\mathrm{R}(z)=\psi_n^\mathrm{R}(\theta)\deriv{\theta}{z}$ and $\phi_n^\mathrm{L}(z)=\psi_n^\mathrm{L}(\theta)$~:
\begin{align}
   \phi_n^\mathrm{R}(z) = \frac{1}{\sqrt{\pi}}\,\left(\frac{z+\I}{z-\I}\right)^n\,(1+z^2)^{-1-q/2}
  \hspace{0.25cm}\mbox{and}\hspace{0.25cm}
   \phi_n^\mathrm{L}(z) = \frac{1}{\sqrt{\pi}}\,\left(\frac{z+\I}{z-\I}\right)^n\,(1+z^2)^{q^*/2}   
\end{align}
for eigenvalue $\lambda_n=-2\I n$. They satisfy the orthonormalisation condition
\begin{equation}
\int_\mathbb{R}\D z\,\phi_n^\mathrm{L}(z)^*\phi_m^\mathrm{R}(z)=\delta_{n,m}
\:.
\end{equation}
The vectors $\phi_n^\mathrm{R}(z)$, which are labelled by the two numbers $(q,n)$, play the same role as spherical harmonics for the group $\mathrm{SO}(3)$.

To complete the analysis we introduce the ladder operators 
\begin{equation}
  \mathscr{D}_\pm(q) = \mathscr{D}_{\widetilde{K}}(q) \pm\,\I\, \mathscr{D}_{A}(q)
  =\pm\I\left( \deriv{}{z} (z\pm\I)^2 + q \,(z\pm\I)\right)
  \:.
\end{equation}
In summary we have 
\begin{align}
  \mathscr{C}\, \phi_n^\mathrm{R}(z) &= q\,(q+2)\,\phi_n^\mathrm{R}(z) 
  \\
  \mathscr{D}_K(q)\, \phi_n^\mathrm{R}(z) &= -2\I n\,\phi_n^\mathrm{R}(z)
  \\
  \mathscr{D}_\pm(q)\, \phi_n^\mathrm{R}(z) &=
  -2\left( n \pm\frac{q+2}{2} \right) \phi_{n\pm1}^\mathrm{R}(z)
  \:.
\end{align}
Starting from the vector $\phi_0^\mathrm{R}(z)$, the ladder operators allow to construct an infinite number of eigenvectors of $\mathscr{D}_K(q)$, hence the irreducible representation has an infinite dimension, unless $(q+2)/2=-N$ is a negative integer, leading to only $2N+1$ vectors, i.e. irreducible representation of finite dimension.

More can be found about group theoretical aspects of the problem in Ref.~\cite{ComTexTou19}.


\section{Boundary conditions for the spectral problems \eqref{eq:SPFrischLloyd} and \eqref{eq:LocalisationEqForPhiR}~: Frisch-Lloyd case (matrices $KN$ or $\widetilde{K}N$)}
\label{app:SmallSbehaviours}

\subsection{Behaviour of $\widehat{\Phi}^\mathrm{R}_q(s)$ for $s\to0$}
\label{subsec:PhiRSmallSbehaviour}

Let us discuss the $s\to0^+$ behaviour of the solution of Eq.~\eqref{eq:LocalisationEqForPhiR} (the analysis for Eq.~\eqref{eq:SPFrischLloyd} is similar).
We expect that the solution presents analytic terms in $s$ and also non analytic terms.
Assume that the first terms of the $s\to0$ expansion are $\widehat{\Phi}^\mathrm{R}_q(s)\simeq1+c\,s + \omega\,s^{a+1}$. 
Injecting this expansion in \eqref{eq:LocalisationEqForPhiR} and assuming $\levy(s)\sim s$ for $s\to0$ (i.e. finite first moment of the weight $v_n$), we obtain  
$-\I\,\omega\,a(a+1)\,s^{a}+\mathcal{O}(s)=\Lambda(q)-\I\,q\big(c+\omega\,(a+1)s^{a}\big)+\mathcal{O}(s)$, thus 
$c=-\I\Lambda(q)/q$ and $a=q$.
Hence we conclude that 
\begin{equation}
  \label{eq:LimitingBehaviourPhiR}
  \boxed{
  \widehat{\Phi}^\mathrm{R}_q(s)
   \underset{s\to0}{=} 
   1 - \frac{\I\,\Lambda(q)}{q}\, s + \omega_q\, |s|^{q+1} +\mathcal{O}(s^2)
   }
\end{equation}
(these are the first terms when $0\leq q<1$).
The crucial point is that $\omega_q$ is real as we now discuss.
The $s\to0$ behaviour of the Fourier transform selects the asymptotic behaviour \eqref{eq:AsymptoticBehaviourForPhiR}~:
 writing
\begin{align}
  \widehat{\Phi}^\mathrm{R}_q(s)
  &= \int\D z\, \Phi^\mathrm{R}_q(z) 
  - \int\D z\, \Phi^\mathrm{R}_q(z) \left(1-\EXP{-\I sz}\right)
\\
  \label{eq:C3}
  &\underset{s\to0}{\simeq} 
  \int\D z\, \Phi^\mathrm{R}_q(z)
  - 2\mathcal{A}_q\,\Gamma(-1-q)\,\sin\left(\frac{\pi q}{2}\right)\,|s|^{q+1}
  +\mathrm{regular\:terms}
  \:.
\end{align}
The coefficient controlling the power law tail \eqref{eq:AsymptoticBehaviourForPhiR}, is proportional to $\omega_q$ controlling the non analytic behaviour in \eqref{eq:LimitingBehaviourPhiR}~:
 $\omega_q=-2\Gamma(-1-q)\,\sin\left(\frac{\pi q}{2}\right)\mathcal{A}_q$. Hence $\omega_q$ is \textit{real}.

For $q\to0$, we get $\omega_0=-\pi\,\mathcal{A}_0=-\pi\,\IDoS$, where $\IDoS$ is the integrated DoS of the disordered model.
Thus \eqref{eq:AsymptoticBehaviourForPhiR} corresponds with the well-known expansion of the Fourier transform of the invariant density (see \cite{GraTexTou14} and references therein)
$
  \hat{f}(s) = \widehat{\Phi}^\mathrm{R}_0(s) 
   =  1 - \pi \,\IDoS\, |s| - \I\,\gamma_1\,s + \mathcal{O}(s^2)
$, 
as it should.

\subsection{Expansion in powers of $q$}

Let us discuss the consequences for the $s\to0$ behaviour of the functions in the expansion \eqref{eq:ExpansionPhiR}.
The term $\mathcal{O}(q^n)$ of \eqref{eq:LimitingBehaviourPhiR} has the form
%
%
%
%
%
%
%
\begin{equation}
  \widehat{R}_n(s) 
  \underset{s\to0}{\simeq} 
  |s|\sum_{m=0}^n \beta_{n,m}\,\ln^m|s|
  +\mathrm{regular\:terms} 
    \:.
\end{equation}
The coefficients $\beta_{n,m}$ are \textit{real} since $\omega_q$ is real.
In particular, since $\omega_0=-\pi\IDoS $, we find $\beta_{n,n}=-\pi \IDoS /n!$.
Let us apply these considerations to $\widehat{R}_1(s)$ and $\widehat{R}_2(s)$.
We find
\begin{equation}
  \widehat{R}_1'' (s) \underset{s\to0}{\simeq}  \frac{\beta_{1,1}}{|s|} + 2\beta_{1,0}\,\delta(s)
 +\mathrm{regular\:terms}  
\end{equation}
so that 
$\I s\widehat{R}_1'' (s)\simeq\I\beta_{1,1}\sign(s)$ for $s\to0$~;
this shows that $\re[\I s\widehat{R}_1'' (s)]\to0$ in this limit.

We have also
\begin{equation}
  \widehat{R}_2'' (s) \underset{s\to0}{\simeq}  
  2\beta_{2,2}\frac{\ln|s|+1}{|s|} + \frac{\beta_{2,1}}{|s|} +2\beta_{2,0}\,\delta(s)
 +\mathrm{regular\:terms}  
\end{equation}
As a result, $\I s\widehat{R}_2'' (s)$ is logarithmically divergent for $s\to0$, however 
$\re[\I s\widehat{R}_2'' (s)]\to0$.
These remarks were used in \S~\ref{sec:PeturbFL} and \S~\ref{sec:GLElocalisation}.

%


\section{The distribution $\mathrm{Pf}(1/|x|)$}
\label{app:Pf}

Consider a regular function $\psi(x)$, decaying at infinity.
We define the distribution $\mathrm{Pf}(1/|x|)$ as 
\begin{equation}
  \int_{-\infty}^{+\infty} \D x\, \psi(x) \, \mathrm{Pf}\frac{1}{|x|}
  \eqdef 
  \lim_{\epsilon\to0^+}
  \left[
    \left( \int_{-\infty}^{-\epsilon} + \int_{+\epsilon}^{+\infty}\right)\D x\,\frac{\psi(x)}{|x|}
    +2\,\psi(0)\,\ln\epsilon
  \right]
  \:,
\end{equation}
in the same spirit as Hadamard's regularization of the integral $\int\D x\,\psi(x)/x^2$.
Equivalently, using $\Big(\int_{-1}^{-\epsilon}+\int_\epsilon^1\Big)\D x/|x|=-2\ln\epsilon$, we can avoid the regulator and define the distribution as
\begin{equation}
  \int_{-\infty}^{+\infty} \D x\, \psi(x) \, \mathrm{Pf}\frac{1}{|x|}
  \eqdef \left( \int_{-\infty}^{-1} + \int_{+1}^{+\infty}\right)\D x\,\frac{\psi(x)}{|x|}
  + \int_{-1}^{+1}\D x\,\frac{\psi(x)-\psi(0)}{|x|}
  \:.
\end{equation}
We can check two useful properties~:
\begin{equation}
  \big[ \mathrm{sign}(x)\,\ln|x| \big] ' = \mathrm{Pf}\frac{1}{|x|}
  \hspace{0.5cm}\mbox{and}\hspace{0.5cm}
  \big[ |x|\,(\ln|x|-1) \big] '' = \mathrm{Pf}\frac{1}{|x|}
\end{equation}
where derivation is understood here in the distributional sense.

\paragraph{Application~:}

The distribution can be used in order to write the solution of the equation 
\begin{equation}
  \left( -\deriv{^2}{x^2} + k^2 \right) \varphi(x) = \frac{1}{|x|}
  \:,
\end{equation}
which is a simplified version of Eq.~\eqref{eq:111}.
The solution is expected to be continuous at $x=0$, but non differentiable as $ \varphi(x) \simeq  \varphi(0) -|x|\,\big(\ln|x|-1\big)$ for $x\to0$.
The finite part allows to write the solution under the integral form
\begin{equation}
  \varphi(x) = \int_{-\infty}^{+\infty} \D y\, G(x-y) \, \mathrm{Pf}\frac{1}{|y|}
  \hspace{0.5cm}\mbox{for }
  x\neq0
  \:,
\end{equation}
where $G(x)=(2k)^{-1}\EXP{-k|x|}$ is the Green's function.
The integral is well defined thanks to the finite part.


\section{Continuum limit of the Frisch-Lloyd model~: the Halperin model}
\label{app:Halperin}

The disordered model \eqref{eq:Schrodinger} for a Gaussian white noise potential 
$
  \mean{V(x)V(x')}=\sigma\,\delta(x-x')
$
corresponds to the so-called Halperin model \cite{Hal65,LifGrePas88}.
It can be recovered from the impurity models in two different manners.
As discussed in Subsection~\ref{subsec:Halperin}, the continuum limit for fixed angles (regular lattice of impurities) leads to the Halperin model.
The high impurity density limit of the Frisch-Lloyd model, $\rho\to\infty$, with $v_n\to0$, also leads to the Halperin model (this limit was already studied by Frisch and Lloyd in~\cite{FriLlo60}).
The Halperin model has been extensively studied in the literature. In particular the fluctuations have been analysed in \cite{SchTit02,RamTex14} and the generalized Lyapunov exponent in \cite{FyoLeDRosTex18}. Hence it is a good case to benchmark the method of the present article.
In this appendix, we establish the correspondence with the equations studied in these papers.
Considering the limit $\rho\to\infty$ and $v_n\to0$ in Eqs.~(\ref{eq:SPFrischLloyd},\ref{eq:SPFrischLloydLoc}), keeping $\rho\mean{v_n}=0$ and $\rho\mean{v_n^2}=\sigma$ fixed, leads to 
\begin{equation}
  \label{eq:SPHalperin}
  \left[
     \mathscr{G}^\dagger 
     + q\, z 
  \right]\Phi^\mathrm{R}_q(z)
  =
  \Lambda(q)\,  \Phi^\mathrm{R}_q(z)
  \hspace{0.5cm}\mbox{where }
  \mathscr{G}^\dagger = \frac{\sigma}{2}\deriv{^2}{z^2} + \deriv{}{z}( E + z^2 )
\end{equation}
is the (forward) generator of the diffusion for the Riccati variable $z(x)=\psi'(x)/\psi(x)$ involved in the localisation problem \cite{FriLlo60,Hal65,Luc92,ComTexTou13}. 
Eq.~\eqref{eq:SPHalperin} is precisely the equation given in Refs.~\cite{RamTex14,FyoLeDRosTex18}. 

It is interesting to discuss the origin of \eqref{eq:Gamma2} without using Fourier transform.
A convenient starting point is the equation in real space for the $q^n$ order contribution to $\Phi^\mathrm{R}_q(z)$~:
\begin{equation}
  \label{eq:Recursion}
  \mathscr{G}^\dagger R_n(z)=(\gamma_1-z)R_{n-1}(z) + \sum_{m=2}^n\frac{\gamma_m}{m!}R_{n-m}(z)
  \:.
\end{equation}
In order to get $\gamma_1$, one can integrate the perturbation equation at order $q^1$~:
\begin{equation}
\lim_{X\to+\infty}\int_{-X}^{+X}\D z\,
\left[
  \mathscr{G}^\dagger R_1(z)+(z-\gamma_1)R_0(z)
\right] = 0
\end{equation}
thus, using the normalisation $\int R_0=\int f=1$, 
\begin{equation}
\gamma_1 = \dashint\D z\, z\, R_0(z) = \int\D z\, z\, \frac{R_0(z)-R_0(-z)}{2}
         = \I\, \frac{\widehat{R}_0'(0^+)+\widehat{R}_0'(0^-)}{2} 
         = - \im\left[\hat{f}'(0^+)\right]
\end{equation}
where the principal part is important as $\mathscr{G}^\dagger R_1(z)\simeq z \, R_0(z) \simeq \IDoS /z$ at infinity.  
This corresponds to \eqref{eq:Gamma1}.
Similarly, starting from the equation at order $q^2$, 
$\mathscr{G}^\dagger R_2(z)=(\gamma_1-z)R_1(z)+(1/2)\gamma_2f(z)$, we integrate and get  
$0=\dashint\D z\,(\gamma_1-z)\,R_1(z)+\gamma_2/2$
  (note that $ R_2(z) \sim \ln^2|z|/z^2$ for $z\to\infty$, cf. Appendix~\ref{app:SmallSbehaviours}, hence the importance of the principal part).
This leads to  
\begin{equation}
   \gamma_2 = \lim_{s\to0}\I\, \left[\widehat{R}_1'(s)+\widehat{R}_1'(-s)\right] -2\, \gamma_1\,\widehat{R}_1(0)
\end{equation}
or more conveniently 
\eqref{eq:Gamma2}. 
The final result is given in the text, Eq.~\eqref{eq:FinalGamma2} where $\hat f(s)=\varphi(s)/\varphi(0)$ is given by~\eqref{eq:PhiHalperin}.

\paragraph{The adjoint spectral problem.---}

In the paper, we have considered the spectral problem \eqref{eq:TheSpectralPb} for $\Phi^\mathrm{R}_q(z)$.
We can equivalently work with the adjoint problem, which here takes the form
\begin{equation}
  \left[
     \mathscr{G} + q\, z 
  \right]\Phi^\mathrm{L}_q(z)
  =
  \Lambda(q)\,  \Phi^\mathrm{L}_q(z)
  \hspace{0.5cm}\mbox{where }
  \mathscr{G}  = \frac{\sigma}{2}\deriv{^2}{z^2} -(E+z^2) \deriv{}{z}
  \:.
\end{equation}
We expand the vector as $\Phi^\mathrm{L}_q(z)=\sum_{n=0}^\infty q^n\,L_n(z)$ with $L_0(z)=1$.
In particular, at order $q^1$ we get
\begin{equation}
\mathscr{G} L_1(z)=\gamma_1-z
\:.
\end{equation}
Multiplication by $f(z)$ and integration gives $\gamma_1=\dashint\D z\, f(z)\, z$ (using 
$\dashint\D z\, f(z)\big[\mathscr{G} L_1(z)\big]=\dashint\D z\, \big[\mathscr{G}^\dagger f(z)\big]L_1(z)=0$).
Similarly, 
imposing the condition $\int\D z\,L_1(z)\,f(z)=0$,
we get 
$\gamma_2=\dashint\D z\, L_1(z)\, z\,f(z)$.
This is analogous to the general discussion for functionals of stochastic processes provided in Section~\ref{sec:Intro}.


\section{Phase formalism for the Frisch-Lloyd model}
\label{app:PhForm}

The Frisch-Lloyd model for impurities at random positions can be conveniently studied with the phase formalism of Refs.~\cite{AntPasSly81,LifGrePas88}. 
This allows to derive some simple asymptotic behaviours. 
Furthermore, we can obtain approximate formulae beyond the perturbative regime.

\subsection{Phase formalism}

The starting point is to parametrize the wave function as $ \psi(x) = \EXP{\xi(x)} \, \sin\theta(x) $ and $\psi'(x) = k\EXP{\xi(x)} \, \cos\theta(x)$ for $E=+k^2$.
The two variables obey the differential equations
$\theta'(x)=k-\big[V(x)/k\big]\,\sin^2\theta$ and $\xi'(x)=\big[V(x)/(2k)\big]\,\sin2\theta$.
The Lyapunov exponent and the variance can be obtained by studying the drift and the diffusion constant of the process $\xi(x)$, corresponding roughly to $\ln|\psi(x)|$ (cf. remark in \cite{ComTexTou10}, p.~442).
This process is constant between two impurities, and make a jump of \cite{Tex99,BieTex08}
\begin{equation}
  \label{eq:DeltaXiN}
  \Delta\xi_n  =\xi(x_n^+) - \xi(x_n^-)
  = \frac{1}{2}\ln\left( 1 + \frac{v_n}{k}\sin2\theta_n^- +  \frac{v_n^2}{k^2}\sin^2\theta_n^- \right)
\end{equation}
through the impurity $n$.
In the $k\gg\rho$ limit, we can assume that the phase $\theta_n^-=\theta(x_n^-)=k\ell_n+\theta(x_n^+)$ modulo $\pi$ is uniformly distributed over $[0,\pi]$.
As a consequence, the increments \eqref{eq:DeltaXiN} are independent and one can easily compute their moments.
Using the independence assumption, we conclude that $\xi(x)$ is a compound Poisson process 
\begin{equation}
  \xi(x) = \sum_{n=1}^{\mathscr{N}(x)} \Delta\xi_n
  \:,
\end{equation}
where $\mathscr{N}(x)$ is a Poisson process of intensity $\rho$.
The drift and the diffusion constant are 
\begin{equation}
  \gamma_1 = \lim_{x\to\infty}\frac{\mean{\xi(x)}}{x} \simeq \rho\, \mean{\Delta\xi_n}   
  \hspace{0.5cm}\mbox{and}\hspace{0.5cm}
  \gamma_2= \lim_{x\to\infty}\frac{\mathrm{Var}(\xi(x))}{x}  \simeq \rho\, \mean{\Delta\xi_n^2}  
\end{equation}
(note that the diffusion constant involves the second \textit{moment} of the increment due to the fact that the number of impurities fluctuates in a fixed interval $[0,x]$~; if the number of impurities would be constant in the interval, $\gamma_2$ would involve the variance of $\Delta\xi_n$).

\subsection{Universal (perturbative) regime}

We fist consider the high energy regime $\rho,\:v_n\ll k=\sqrt{E}$.
In this case we can expand the logarithm in \eqref{eq:DeltaXiN}, and eventually average over the phase.
We see that the two moments are equal~:
\begin{equation}
  \label{eq:PerturbativeUniversalGamma1Gamma2}
  \gamma_1 \simeq \gamma_2 \simeq \frac{\rho\,\mean{v_n^2}}{8E}
  \hspace{0.5cm}\mbox{for }
  E\to+\infty
  \:,
\end{equation}
which is the single parameter scaling property~(as noticed in \cite{Tex99}, this is only true for the leading order term~; see also \cite{SchSchSed04}).

\subsection{Intermediate regime for weak density}

For weak density, in the intermediate (non perturbative) regime  $\rho\ll k\ll v_n$, the phase increment between impurities is $k\ell_n\gg1$, hence we can use that the phase $\theta_n^-$ is uniformly distributed (phase randomization).
We can also simplify the logarithm as 
$
  \Delta\xi_n \simeq \ln\left|\frac{v_n}{k}\sin\theta_n^-\right|
$.
Using that $\int_0^\pi\frac{\D\theta}{\pi}\,\ln\sin\theta=-\ln2$ and 
$\int_0^\pi\frac{\D\theta}{\pi}\,\ln^2\sin\theta=\frac{\pi^2}{12}+\ln^22$ we deduce
\begin{align}
  \gamma_1  \simeq \rho\, \mean{\ln\frac{v_n}{2k} }  
  \hspace{0.5cm}\mbox{and} \hspace{0.5cm}
  \gamma_2  \simeq \rho  \, \frac{\pi^2}{12}
  + \rho\, \mean{\ln^2\frac{v_n}{2k} }
  \:,
\end{align} 
thus we expect $\gamma_2\gg\gamma_1$.
For an exponential distribution of weights $\mathrm{Proba}\{v_n>x\}=\EXP{-x/v}$ we use 
$\int_0^\infty\D x\,\ln x\,\EXP{-x}=-\mathrm{C}$ and 
$\int_0^\infty\D x\,\ln^2 x\,\EXP{-x}=\frac{\pi^2}{6}+\mathrm{C}^2$,
where $\mathrm{C}\simeq0.577$ is the Euler-Mascheroni constant.
Eventually we deduce \eqref{eq:Gamma1Gamma2log}.

\subsection{Large negative energy}

For $E=-k^2$, the process $\xi(x)$ is not constant between the impurities, but grows as 
\begin{equation}
  \label{eq:DeltaXiNcontribEN}
  \xi(x_{n+1}^-) - \xi(x_n^+)
   = k\ell_n 
   + \frac{1}{2}\ln\left(\frac{\tan^2(\theta_n^++\pi/4)+\EXP{-4k\ell_n}}{\tan^2(\theta_n^++\pi/4)+1}\right)
\end{equation}
where $\theta_n^+=\theta(x_n^+)$.
One must add this contribution to \eqref{eq:DeltaXiN}.
Using that the phase is locked $\theta_n^-\to\pi/4$ for $E\to-\infty$, we can expand the total increment $  \Delta\xi_n  =\xi(x_{n+1}^-) - \xi(x_n^-)$ as 
\begin{equation}
  \Delta\xi_n 
  = k\ell_n + \frac{v_n}{2k} + \mathcal{O}\left( \frac{v_n^2}{k^2} \right) 
  + \mathcal{O}\left( \EXP{-4k\ell_n} \right)
\end{equation}
As a result 
\begin{align}
  \gamma_1 \simeq k + \frac{\rho\mean{v_n}}{2k} \simeq \sqrt{-E+\rho\mean{v_n}}
  \hspace{0.5cm}\mbox{and} \hspace{0.5cm}
  \gamma_2 \simeq \frac{\rho\mean{v_n^2}}{4k^2} 
\end{align}
Interestingly, one has recovered the same property as for the Halperin model \cite{RamTex14}
\begin{equation}
  \gamma_2 \big|_{E=-k^2} \simeq 2\,\gamma_2 \big|_{E=+k^2}
  \hspace{0.5cm}\mbox{for }
  k\to\infty
  \:.
\end{equation}


%

\end{document}